\documentclass[prr, twocolumn, superscriptaddress, floatfix]{revtex4-2}
\usepackage[spanish, english]{babel}
\usepackage[usenames]{color}
\usepackage{xcolor}
\usepackage{upgreek}
\usepackage{amsmath}
\usepackage{amssymb} 
\usepackage{graphicx}
\usepackage[utf8]{inputenc}
\usepackage{physics}
\usepackage{empheq}
\usepackage{braket}
\usepackage{siunitx}
\usepackage{float}
\usepackage{tcolorbox}
\usepackage{enumitem}
\usepackage[colorlinks=true,bookmarks=false]{hyperref}
\hypersetup{linkcolor=blue,citecolor=blue,urlcolor=blue}  
\newcommand{\up}{\uparrow}
\newcommand{\down}{\downarrow}
\usepackage[normalem]{ulem}

\newcommand{\bpp}{\hat{b}^{\dagger}}
\newcommand{\bp}{\hat{b}}

\newcommand{\beginsupplement}{%
	\setcounter{table}{0}
	\renewcommand{\thetable}{S\arabic{table}}%
	\setcounter{figure}{0}
	\setcounter{section}{0}
	\setcounter{equation}{0}
	\renewcommand{\thefigure}{S\arabic{figure}}
	\renewcommand{\theHfigure}{S\arabic{figure}}
	\renewcommand{\thesection}{S\Roman{section}}
	\renewcommand{\theHsection}{S\Roman{section}}
	\renewcommand{\theequation}{S\arabic{equation}}
	\hypersetup{linkcolor=black,citecolor=blue,urlcolor=blue}  
	
}%

\makeatletter
\renewcommand*{\@fnsymbol}[1]{\ensuremath{\ifcase#1\or \dagger\or *\or \ddagger\or
		\mathsection\or \mathparagraph\or \|\or **\or \dagger\dagger
		\or \ddagger\ddagger \else\@ctrerr\fi}}
\makeatother

\begin{document}

\title{Phases, instabilities and excitations in a two-component lattice model \\ with photon-mediated interactions}
\author{Leon Carl}
\affiliation{Institute for Quantum Electronics, ETH Zürich, 8093 Zürich, Switzerland}
\author{Rodrigo Rosa-Medina}
\email{rrodrigo@phys.ethz.ch}
\affiliation{Institute for Quantum Electronics, ETH Zürich, 8093 Zürich, Switzerland}
\author{Sebastian D. Huber}
\affiliation{Institute for Theoretical Physics, ETH Zürich, 8093 Zürich, Switzerland}
\author{Tilman Esslinger}
\affiliation{Institute for Quantum Electronics, ETH Zürich, 8093 Zürich, Switzerland}
\author{Nishant Dogra}
\thanks{These authors contributed equally to this work}\email{dubcekt@ethz.ch}
\affiliation{Cavendish Laboratory, University of Cambridge, J. J. Thomson Avenue, Cambridge CB3 0HE, United Kingdom.}
\author{Tena Dubcek}
\thanks{These authors contributed equally to this work}\email{dubcekt@ethz.ch}
\affiliation{Institute for Theoretical Physics, ETH Zürich, 8093 Zürich, Switzerland}

\date{\today}

\begin{abstract}
	
	Engineering long-range interacting spin systems with ultra cold atoms offers the possibility to explore exotic magnetically ordered phases in strongly-correlated scenarios. Quantum gases in optical cavities provide a versatile experimental platform to further engineer photon-mediated interactions and access the underlying microscopic processes by probing the cavity field.
	Here, we study a two-component spin Bose-Hubbard system with cavity-mediated interactions.
	We provide a comprehensive overview of its phase diagram and transitions in experimentally relevant regimes.
	The interplay of different energy scales yields a rich phase diagram with superfluid and insulating phases exhibiting density modulation or spin ordering.
	In particular, the combined effect of contact and global-range interactions gives rise to an antiferromagnetically ordered phase for arbitrarily small spin-dependent light-matter coupling, while long-range and inter-spin contact interactions introduce regions of instability and phase separation in the phase diagram. We further study the  low energy excitations above the antiferrogmagnetic phase. Besides particle-hole branches, it hosts spin-exchange excitations with a tunable energy gap. The studied lattice model can be readily realized in cold-atom experiments with optical cavities.
	
	% Here, we consider a two-component spin Bose-Hubbard system with cavity-mediated global-range interactions. Using complementary mean-field approaches, we show that the interplay of different energy scales yields a rich phase diagram with superfluid and insulating phases exhibiting density modulation or spin ordering.
	% In particular, the combined effect of contact and spin-dependent global-range interactions gives rise to an anti-ferromagnetically ordered phase hosting spin-exchange excitations with tunable energy gaps, while long-range and inter-spin contact interactions introduce regions of instability and phase separation in the phase diagram.
	% The studied extended Bose-Hubbard model can be readily realized in existing cold-atom experiments with optical cavities.
	
\end{abstract}

\maketitle

\section{Introduction}

Experiments with ultacold atoms in optical lattices have substantially extended the scope of quantum simulation of many-body systems~\cite{jaksch1998cold,greiner2002quantum}. Two key strengths are the high-degree of tunability of different energy scales, and the possibility to involve the atomic spin degree of freedom, facilitating the investigation of strongly-correlated phenomena like superfluidity, quantum magnetism, high-temperature superconductivity and complex out-of-equilibrium dynamics~\cite{gross2017quantum,schaefer2020tools}.
While contact interactions naturally occur in ultracold atomic systems~\cite{Vuletic1999,greiner2002quantum}, long-range interactions have been more elusive. Nonetheless, systems that are traditionally used to study long-range interactions, such as dipolar quantum gases, heteronuclear molecules and Rydberg atoms, suffer from small long-range interaction strengths, low densities and short lifetimes, respectively~\cite{Moses2017,Lahaye2009,Browaeys2016}.
Quantum gases coupled to optical cavities thus provide an alternative experimental platform to create photon-mediated long-range interactions, whose strength and sign are controlled by external laser fields~\cite{ritsch2013cold, mivehvar2021cavityqed}. This has facilitated theoretical~\cite{li2013lattice, Bakhtiari2015nonequilibrium, CaballeroBenitez2015Quantum, dogra2016phase,chen2016quantum, sundar2016latticeboson, Flottat2017_phasediagram, liao2018theoretical, Chen2020extended}  and experimental~\cite{Klinder2015observation,landig2016quantum} investigations of lattice supersolid and charge density wave phases in single-component spin systems. The atomic dynamics and many-body excitations can be accessed non-destructively in real time by the light leaking from the cavity~\cite{hruby2018metastability}.
Recently, the inclusion of an internal atomic spin degree of freedom has become feasible in such systems, leading to the observation of density and spin self-organization~\cite{landini2018formation,Kroeze2018spinor,ferri2021emerging}.
Incorporating tunable long-range spin interactions provides a natural path to further enrich the accessible phenomenology ~\cite{Simon2007single, Strack2011Dicke, Gopalakrishnan2011frustration, Buchhold2013dicke, Zhiqiang2017nonequilibrium, Mivehvar2017disorder, LewisSwan2018robust, davis2019photon,  mivehvar2019cavity, Muniz2020exploring, Stitely2020nonlinear}. In combination with optical lattices, this approach will allow the realization of strongly-correlated magnetic phases arising due to the interplay of short- and long-range interactions. Some magnetically ordered phases have been discussed both in bosonic~\cite{Guan2019two,Lozano2020Spin} and fermionic systems~\cite{Fan2018magnetic,CamachoGuardian2017quantum}, but a comprehensive theoretical study of the phase diagram and possible transitions in experimentally accessible regimes is still missing, although it could notably expedite their successful realization.

Here, we investigate an extended two-component Bose-Hubbard (BH) model with cavity-mediated long-range interactions\textemdash the lattice counterpart of the experiment performed in Ref.~\cite{landini2018formation} with a bulk Bose gas. The considered long-range interactions have a `density' and a `spin' contribution, which favor the two atomic components to either occupy a common sublattice or two different ones, each breaking independently a lattice $\mathbb{Z}_2$-symmetry. Their absolute and relative strengths can be tuned via the intensity and the polarization of an external laser field, respectively. Additionally, the two atomic components have different intra- and inter-species contact interactions. We extract the complete phase diagram using a Gutzwiller approach in the case of fixed density at unity filling, and obtain density-modulated and magnetically-ordered phases, both in the superfluid and insulating regimes. Remarkably, the cooperation between short-range and long-range interactions results in the formation of an antiferromagnetic Mott insulator for arbitrarily small spin-dependent coupling strengths. In some regimes, the competing contact-interaction energy scales lead to the separation of the two spin components~\cite{kuklov2003_counterflow,Altman2003_phase_diagram,Lingua2015_demixing_effects}. In addition, long-range interactions can introduce correlated phase-separated states in multicomponent systems~\cite{Bai2020_segregated_quantum,Zhang2022quantumphases}, and induce phase instabilities in systems with only one component~\cite{Batrouni2000_phaseseparation,Flottat2017_phasediagram}. 
To further elucidate the nature of the magnetically ordered phase, we construct an effective Hamiltonian for its low-energy excitations via perturbation theory, and identify spin-exchange branches with a tunable gap.

\section{Description of the System}

\begin{figure}[ht]
	\centering
	\includegraphics[width=0.95\columnwidth]{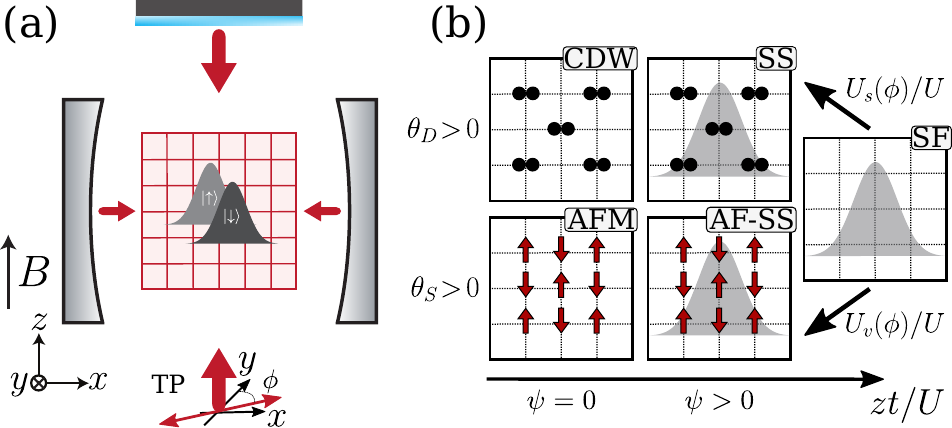}
	\caption{(a) Schematic representation of a two-component BEC ($\ket{\uparrow},\ket{\downarrow}$) confined in a 2D optical lattice inside an optical cavity. The spin-mixture is illuminated by a transverse pump field (TP) with tunable polarization angle $\phi$ in the $(x,y)$-plane. (b) Mean-field order parameters and associated phases of the effective Hamiltonian. The order parameters $\theta_D$, $\theta_S$ and $\psi$ characterize density modulation, spin-order and superfluidity, respectively. For $U_{12}/U=1$, the possible ground state configurations are a superfluid (SF), a charge density wave (CDW), a lattice supersolid (SS), an antiferromagnetic Mott-insulator (AFM) or an antiferromagnetic lattice supersolid (AF-SS). Spin states are represented by red arrows, while the black markers indicate spin-insensitive density configurations.}
	\label{fig:Fig1}
\end{figure}

We consider a balanced spin-mixture of two Bose-Einstein condensates (BECs) coupled to a high-finesse optical cavity. For concreteness, we consider $^{87}$Rb atoms and a two-dimensional (2D) system extending in the $(x,z)$-plane [Fig.~\ref{fig:Fig1}(a)], closely resembling the experiments in Refs.~\cite{landig2016quantum,landini2018formation}. A $\lambda$-periodic quantized cavity mode extends along the $x$-axis, is polarized in $y$-direction and has a resonant frequency $\omega_c$. The two spin components $\ket{\up}=\ket{F=1,m_F=1}$ and $\ket{\down}=\ket{F=1,m_F=-1}$ belong to the total angular momentum $F=1$ manifold, with the quantization axis defined by a magnetic field in the $z$-direction. The mixture is loaded into a 2D $\lambda/2$-periodic square optical lattice. The lattice arm along the $z$-direction has a frequency $\omega_p$  and linear polarization in the $(x,y)$-plane, and  fulfills a dual role as a transverse pump field (TP). It is far red-detuned both from the atomic and cavity resonance, $\Delta_c=\omega_p-\omega_c<0$, thus acting dispersively on the atoms. The light scattered from the TP into the cavity couples the atoms' motional and spin degrees of freedom to the cavity mode. The single-particle Hamiltonian in the rotating frame of the TP reads~\cite{SI}
\begin{equation}
	\begin{split}
		\hat{H}_\text{sp}&=\frac{\hat{\mathbf{p}}^2}{2m}+\hat{V}_{\mathrm{lat}}-\left(\Delta_c-U_0\cos^2\left(\frac{2\pi}{\lambda}\hat{x}\right)\right)\hat{a}^{\dagger}\hat{a}\\
		&+ \cos\left(\frac{2\pi}{\lambda}\hat{x}\right)\cos\left(\frac{2\pi}{\lambda}\hat{z}\right)\left(\eta_{s}\hat{X}+\eta_{v}\hat{P}\hat{F}_z\right),\\
	\end{split}
	\label{eq:H_sp}
\end{equation}
with total momentum $\hat{\mathbf{p}}=(\hat{p}_x+\hat{p}_z)$ and $\hbar=1$ \cite{SI,landini2018formation}.
The first two terms account for the atom moving in the 2D lattice potential ${\hat{V}_{\mathrm{lat}}=-V\left( \cos^2\left(\frac{2\pi}{\lambda}\hat{x}\right)+\cos^2\left(\frac{2\pi}{\lambda}\hat{z}\right)\right).}$ The operator $\hat{a}^{\dagger}$ denotes the creation operator associated to the intra-cavity field.
The presence of the atom dispersively shifts the cavity resonance frequency and leads to an effective detuning ${\tilde{\Delta}_c=\Delta_c-U_0\cos^2(\frac{2\pi}{\lambda}\hat{x})<0}$, where ${U_0<0}$ is the maximal dispersive shift.  The last term describes a self-consistent interference potential that arises due to light scattering between the TP and the cavity mode~\cite{SI}. The scalar component of the atom-light interactions couples the atomic motional degrees of freedom to the real quadrature of the cavity field, $\hat{X}=\left(\hat{a}+\hat{a}^{\dagger}\right)/\sqrt{2},$ giving rise to a $\lambda$-periodic spin-independent density modulation. The vectorial coupling is mediated by the imaginary quadrature, $\hat{P}=i\left(\hat{a}^{\dagger}-\hat{a}\right)/\sqrt{2},$ and gives rise to phase-shifted $\lambda$-periodic modulations for atoms in the two spin states, since the $z$-component of the atomic spin operator $\bold{\hat{F}}$  yields $\hat{F}_z\ket{\uparrow} = +\ket{\uparrow}$ and  $\hat{F}_z\ket{\downarrow} = -\ket{\downarrow}.$ The associated coupling strengths are $\eta_{s}=\eta\cos(\phi)$ and $\eta_v=\eta \xi\sin(\phi),$ with $\xi=\frac{\alpha_v}{2\alpha_s}$ given by the atom-cavity coupling rate $\eta$ and the ratio of the scalar and vectorial polarizabilities ~\cite{kien2013dynamical,cohentannoudji1998atom}.% The ratio of the couplings can be selectively addressed via the TP polarization angle $\phi$.

We adiabatically eliminate the intra-cavity field in a tight-binding approximation \cite{SI,landig2016quantum,dogra2016phase,Zwerger2003mott} and obtain a many-body extended BH Hamiltonian,
\begin{equation}
	\hat{H} =\hat{H}_\text{BH} + \hat{H}_\text{Long},
	\label{eq:H}
\end{equation}
with 
\begin{equation}
	\begin{split}
		\hat{H}_\text{BH} &= - t\sum_{m,<\mathbf{i},\mathbf{j}>}(\hat{b}^\dagger_{\mathbf{i},m}\hat{b}_{\mathbf{j},m}+\text{h.c.}) \\
		&+ \frac{U}{2}\sum_{\mathbf{i},m}\hat{n}_{\mathbf{i},m}\left(\hat{n}_{\mathbf{i},m}-1\right) +U_{12}\sum_{\mathbf{i}}\hat{n}_{\mathbf{i},\up}\hat{n}_{\mathbf{i},\down},
	\end{split}
	\label{eq:H_BH}
\end{equation}
and
\begin{equation}
	\hat{H}_\text{Long} = - \frac{U_s}{K}\hat{\Theta}_D^2 - \frac{U_v}{K}\hat{\Theta}_S^2.
	\label{eq:H_Long}	
\end{equation}
The first term, Eq.~\eqref{eq:H_BH}, constitutes a two-component BH model \cite{fisher1989boson,jaksch1998cold,greiner2002quantum, kuklov2003_counterflow, Altman2003_phase_diagram}, comprising tunneling to the $z$ nearest neighbors, with rate $t>0$ and repulsive inter- and intra-spin contact interactions, $U>0$ and $U_{12}>0.$ We assume identical intra-spin collisional interactions for atoms occupying $\ket{\up}$ and $\ket{\down}$, as is the case for ${}^{87}\mathrm{Rb}$ atoms in the $F=1$ hyperfine groundstate manifold \cite{stamperkurn2013spinorbosegases}.  The operator $\bpp_{\mathbf{i},m}$ ($\bp_{\mathbf{i},m}$) denotes the bosonic creation (annihilation) operator, while  $\hat{n}_{\mathbf{i},m}=\bpp_{\mathbf{i},m}\bp_{\mathbf{i},m}$ counts the total number of atoms with spin $m\in\{\up,\down\}$ at site ${\mathbf{i}=(i_x,i_z)}$. For sufficiently large magnetic fields, both spin-changing collisions and cavity-assisted Raman processes can be neglected \cite{stamperkurn2013spinorbosegases}. The second term, Eq.~\eqref{eq:H_Long}, consists of spin-independent (`scalar') and spin-dependent (`vectorial') global-range interactions that are mediated by the intra-cavity field. 
The scalar long-range interactions are associated with the operator $\hat{\Theta}_D^2=\left(\sum_{\mathbf{i}} (-1)^{|\mathbf{i}|} \hat{n}_{\mathbf{i}}\right)^2,$ where $|\mathbf{i}|=i_x+i_z$ and ${\hat n}_{\bf i}={\hat n}_{{\bf i},\up}+{\hat n}_{{\bf i},\down}$. Its expectation value is maximized for a spin-independent spatial density modulation with all atoms occupying only even or odd sites. The expectation value of the vectorial long-range operator, $\hat{\Theta}_S^2=\left(\sum_{\mathbf{i}} (-1)^{|\mathbf{i}|} \hat{S}_{z,\mathbf{i}}\right)^2$ with $\hat{S}_{z,\mathbf{i}}=\hat{n}_{\mathbf{i},\up}-\hat{n}_{\mathbf{i},\down},$
is maximized for a global antiferromagnetic ordering of the atoms on the lattice, with all atoms in $\ket{\up}$ occupying even sites and all atoms in $\ket{\down}$ occupying odd sites, or vice-versa. The interaction strengths $U_s=U_L\cos^2{\!\phi}$ and $U_v=U_L\xi^2\sin^2{\!\phi}$ can be tuned with respect to each other via the angle $\phi$. The overall interaction strength $U_L>0$ depends on the lattice depth $V$ and the effective detuning $\tilde{\Delta}_c$ ~\cite{SI}. The total number of sites is denoted by $K$. We emphasize that the energy scales of the tunneling, contact and long-range interactions are all independently tunable with respect to each other.
The Hamiltonian, Eq.~\eqref{eq:H}, is invariant under a global spin-flip $\bp_{\mathbf{i},\up}\rightarrow \bp_{\mathbf{i},\down},$ and two global rotations,  $\bp_{\mathbf{i},m}\rightarrow e^{i\phi_m}$ for each $m\in{\up,\down}.$ Furthermore, the scalar and vectorial long-range interaction introduce an additional $\mathbb{Z}_2$-symmetry associated to the two sublattices defined by even and odd sites. Henceforth, the Hamiltonian has a $\mathcal{U}(1)\times \mathcal{U}(1)\times \mathbb{Z}_2\times \mathbb{Z}_2$-symmetry.

\section{Ground State Phase Diagram}

\subsection{Method and Order Parameters}

We explore the zero-temperature phase diagram at unity filling, by using a Gutzwiller mean-field approach \cite{rokhsar1991gutzwiller, altman2002oscillating,huber2007dynamical}. 
We assume a translationally invariant ground state on each of the even $(e)$ and odd $(o)$ sublattices,
\begin{equation}
	\ket{\Psi_{G}}=\prod_{e=0}^{K/2}\prod_{o=0}^{K/2} \ket{\phi_{e}}\ket{\phi_{o}}.
	\label{eq:gutzwiller}
\end{equation}
For each sublattice $i\in \{e,o\}$, the wave function is given by 
\begin{equation}
	%\ket{\phi_i}=\sum_{n=0}^{N_{\up}-1}\sum_{m=0}^{N_{\down}-1} a_i(n,m)\ket{n,m}_i
	\ket{\phi_i}=\sum_{n=0}^{n_{\max}}\sum_{m=0}^{m_{\max}} a_i(n,m)\ket{n,m}_i
\end{equation}
where $\ket{n,m}_i=\frac{(\bpp_{i,\up})^n}{\sqrt{n!}}\frac{(\bpp_{i,\down})^m}{\sqrt{m!}}\ket{0}$ is the local Fock state with $n\leq n_{\max}$ atoms in spin state $\ket{\up}$ and $m\leq m_{\max}$ atoms in state $\ket{\down}$.
The real ground-state coefficients ${\mathbf{a}_{e}} \equiv \left( a_e(n,m)\right)_{n,m}$ and ${\mathbf{a}_{o}} \equiv \left(a_o(n,m)\right)_{n,m}$ are optimized to minimize the effective mean-field energy density 
\begin{equation}
	\mathcal{E}(\mathbf{a}_{e},\mathbf{a}_{o})=\frac{\braket{\Psi_G|\hat{H}|\Psi_G}}{K/2} .
	\label{eq:energy_density}
\end{equation}

The superfluid order parameter $\psi\coloneqq\frac{1}{4}\sum_{i,m}\psi_{i,m}$ with ${\psi_{i,m}=|\braket{\bpp_{i,m}}|}$ (${m\in\{\up,\down\}}$)  signals the transition from an insulating phase (${\psi=0}$) to a phase-coherent superfluid phase exhibiting off-diagonal long-range order (${\psi>0}$). 
The density ${\theta_D=|\braket{\hat{n}_{e}-\hat{n}_{o}}|,}$ and spin ${\theta_S=|\braket{\hat{S}_{z,e}-\hat{S}_{z,o}}|}$ order parameters indicate the degree of global spatial density- and spin-ordering due to long-range interactions [Fig.~\ref{fig:Fig1}(b)].

%Four superfluid order parameters ${\psi_{i,m}=|\braket{\bpp_{i,m}}|}$, ${m\in\{\up,\down\}}$,  signal the transitions from an insulating phase (${\psi\coloneqq\frac{1}{4}\sum_{i,m}\psi_{i,m}=0}$) to a phase-coherent superfluid phase with off-diagonal long-range order (${\psi>0}$). 

\subsection{Phases for a Uniform Mixture}
\begin{figure}[ht]
	\centering
	\includegraphics[width=\columnwidth]{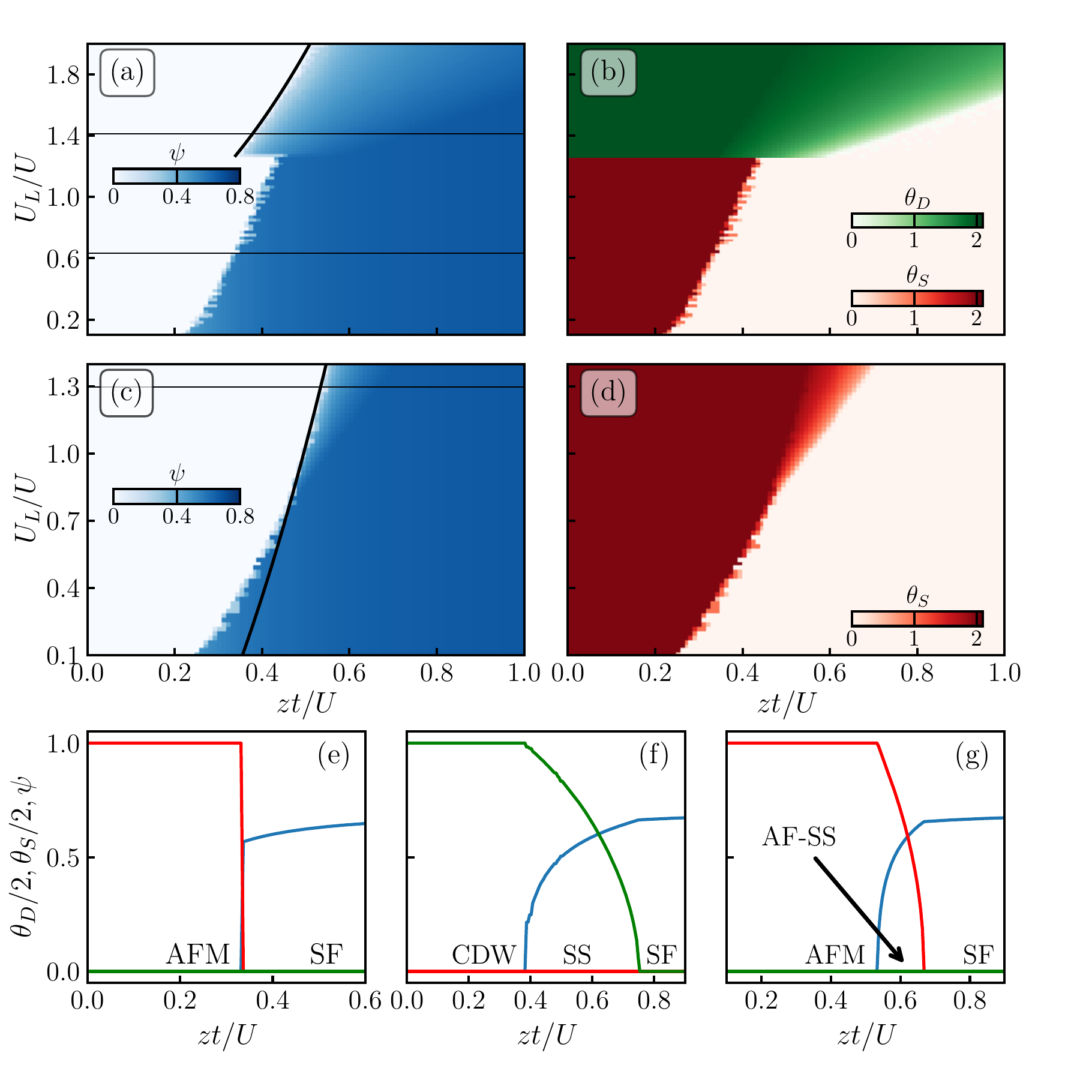}
	\caption{Mean field phase diagrams for a balanced-spin mixture at unity filling, $\rho=2.$ The calculations are performed for $U_s/U_v\approx 4.64$ (a,b) and at $U_s/U_v\approx 0.33$ (c,d). (a,c) Dependence of superfluid order parameter $\psi$. The solid lines are perturbative estimations for the transition. (b,d) Dependence of density and antiferromagnetic order parameters $\theta_D$ and $\theta_S.$ (e,f,g) Cuts along the phase diagrams for $U_s/U_v\approx 4.64$ at constant $U_L/U=0.63$ and $U_L/U=1.41$ (e,f) and for $U_s/U_v\approx 0.33$ at $U_L/U=1.3$ (g), indicated by thin horizontal lines in (a,c). The basis truncation $n_{\max}= m_{\max}=3$ leads to a saturation of the superfluid order parameter at large tunneling.}
	\label{fig:fig2}
\end{figure}

We discuss the case of a balanced spin mixture at unity filling,
\begin{equation}
	\rho_m=\frac{N_m}{(K/2)}=\braket{\phi_e|\hat{n}_{e,m}|\phi_{e}}+\braket{\phi_o|\hat{n}_{o,m}|\phi_{o}}=1
	\label{eq:constraint_balanced mixture}
\end{equation}
for $m=\up,\down$. The choice to work at fixed density is motivated by experiments with ultracold atoms, although a qualitatively similar phase diagram arises in a grand canonical ensemble ~\cite{SI,Guan2019two}.
In this section, we assume ${U_{12}=U}$.
The different order parameters are shown in Fig.~\ref{fig:fig2}. The competition of scalar and vectorial long-range interactions gives rise to two qualitatively different scenarios.

%The phase diagrams corresponding to truncations ${N_{\uparrow}=N_{\downarrow}=3}$ are shown in Fig.~\ref{fig:fig2}

For $U_s>U_v$ [Fig.~\ref{fig:fig2}(a,b)], we observe two distinct insulating phases ($\psi=0$) at low tunneling rates $zt/U$: For large $U_L$, a spin-degenerate charge density wave (CDW), with $\theta_D>0$ and $\theta_S=0.$ For small $U_L$, an antiferromagnetic Mott insulator (AFM), with $\theta_S>0$ and $\theta_D=0.$ Remarkably, the system favors an AFM for arbitrarily small vectorial contributions $U_v$: the contact interaction hinders the formation of a CDW and overcomes the kinetic energy cost to form a unity filling Mott insulator (MI). There, the AFM configuration is favored among all possible MIs by the vectorial long-range interaction. 
The discontinuity in the order parameters $\theta_D$ and $\theta_S$ at constant tunneling as a function of $U_L/U$ signals a first order AFM\textendash~CDW phase transition.
For $t=0$, the boundary between the phases is given by $U_s/U-U_v/U=\frac{1}{2}$ \cite{SI}.
As tunneling increases, the system becomes superfluid $\psi>0$ and can either exhibit spin ($\theta_S>0$) or density ordering ($\theta_D>0$). We denote these phases as antiferromagnetic lattice supersolid (AF-SS) and lattice supersolid (SS), respectively. Meanwhile, a superfluid phase (SF) with $\psi>0$ and $\theta_{D,S}=0$ emerges at even larger tunneling strengths.

In the regime $U_v>U_s$ [Fig.~\ref{fig:fig2}(c,d)], the system exhibits solely spin ordered phases ($\theta_S>0,$ and $\theta_D=0$), as the vectorial long-range and the contact interactions dominate over the scalar long-range interaction.  For small $U_L$, we identify a first-order AFM\textendash~SF phase transition, signaled by a discontinuous jump of $\psi$ and $\theta_S$ [Fig.~\ref{fig:fig2}(e)]. This is in contrast to the second-order MI\textendash~SF transition in the absence of the long-range interactions ($U_L=0$). For larger $U_v/U$ the AFM phase extends towards higher tunneling strengths. AFM\textendash~SF transitions in the context of entanglement properties have recently been studied in three-component BH models with long-range interactions~\cite{Lozano2020Spin}. For larger $U_L$, we observe second-order phase transitions from AFM to AF-SS and from CDW to SS phases, along lines of constant $U_L/U$ [Fig.~\ref{fig:fig2}(f,g)]. The second-order phase transitions from AFM to AF-SS and CDW to SS are supported by perturbative estimations, cf. black lines in Fig.~\ref{fig:fig2}(a,c)~\cite{Oosten2001_quantum_phases,SI}.
% \textcolor{gray}{The transition agrees qualitatively with calculations of the condensate fractions $f_c=\frac{\braket{\hat{n}_{q=0,\up}+\hat{n}_{q=0,\down}}}{N},$ obtained by exact diagonalization (ED) for $6\times 1$ lattices with periodic boundary conditions~\cite{SI,zhang2010exact,ravents2017coldbosons}. This quantity indicates the population in the ${\bf k}=0$ quasi-momentum and captures the degree of superfluidity in the system.  In the insulating regime $f_c$ is small but non-vanishing [Fig.~\ref{fig:fig2}(e-g)], which we attribute to finite-size effects.}
%Our mean-field method does not capture magnetic phenomena such as the supercounterflow occurring in a two-component BH model without long-range interactions~\cite{kuklov2003_counterflow, Altman2003_phase_diagram}. Qualitatively, the vectorial long-range interactions suppresses the spin-exchange mechanism in the insulating regime, thus hindering the supercounterflow. 

\subsection{Phase Diagrams for different $U_{12}/U$}

%Two-component BH models naturally feature phase separation due to repulsive inter-spin interactions~\cite{kuklov2003_counterflow,Altman2003_phase_diagram,Lingua2015_demixing_effects,Chen2003,Bai2020_segregated_quantum,Zhao2014_ferromagnetism,Guan2019two}. Moreover, phase instabilities can arise from long-range interactions~\cite{Batrouni2000_phaseseparation, Flottat2017_phasediagram}. 

%as a function of $U_L/U$ and $zt/U$
We now discuss the phase diagrams for different ratios of inter- and intra-spin interactions $U_{12}/U$ and scalar and vectorial long-range interactions $U_s/U_v$. Besides the homogeneous phases, we also calculate the regions of phase separation between the two spin states (PS), which naturally occur in two-component BH models with repulsive inter-spin interactions~\cite{kuklov2003_counterflow,Altman2003_phase_diagram,Lingua2015_demixing_effects,Chen2003,Bai2020_segregated_quantum,Zhao2014_ferromagnetism,Zhang2022quantumphases,Guan2019two}. Additional regions of phase instability arise from the concurring long-range interactions~\cite{Batrouni2000_phaseseparation, Flottat2017_phasediagram}. To calculate the energy of a phase separated state, the system is divided into halves ($A,B$): one with higher spin-up density ($\rho^{A}_{\uparrow} > \rho^{A}_{\downarrow}$)   and the other with higher spin-down density ($\rho^{B}_{\downarrow} > \rho^{B}_{\uparrow}$), while imposing a density conservation constraint in each of the halves, i.e.,  $\rho\equiv\rho^{A,B}_{\uparrow}+\rho^{A,B}_{\downarrow}=2$ to ensure unity filling. We further assume either $\langle\hat{\Theta}_D\rangle = 0$ or $\langle\hat{\Theta}_S\rangle = 0$. Phase instability, on the other hand, is signaled by a negative compressibility, $\partial_\rho\mu < 0$ ~\cite{Batrouni2000_phaseseparation}, where $\mu(\rho)=\partial_\rho\mathcal{E}(\rho)$ is the chemical potential as a function of the density $\rho$. We calculate the derivative numerically by using the energy densities $\mathcal{E}(\rho)$ extracted from the variational ansatz in Eq.~\eqref{eq:energy_density} \cite{SI}.

\begin{figure}[ht]
	\centering
	\includegraphics[width=0.95\columnwidth]{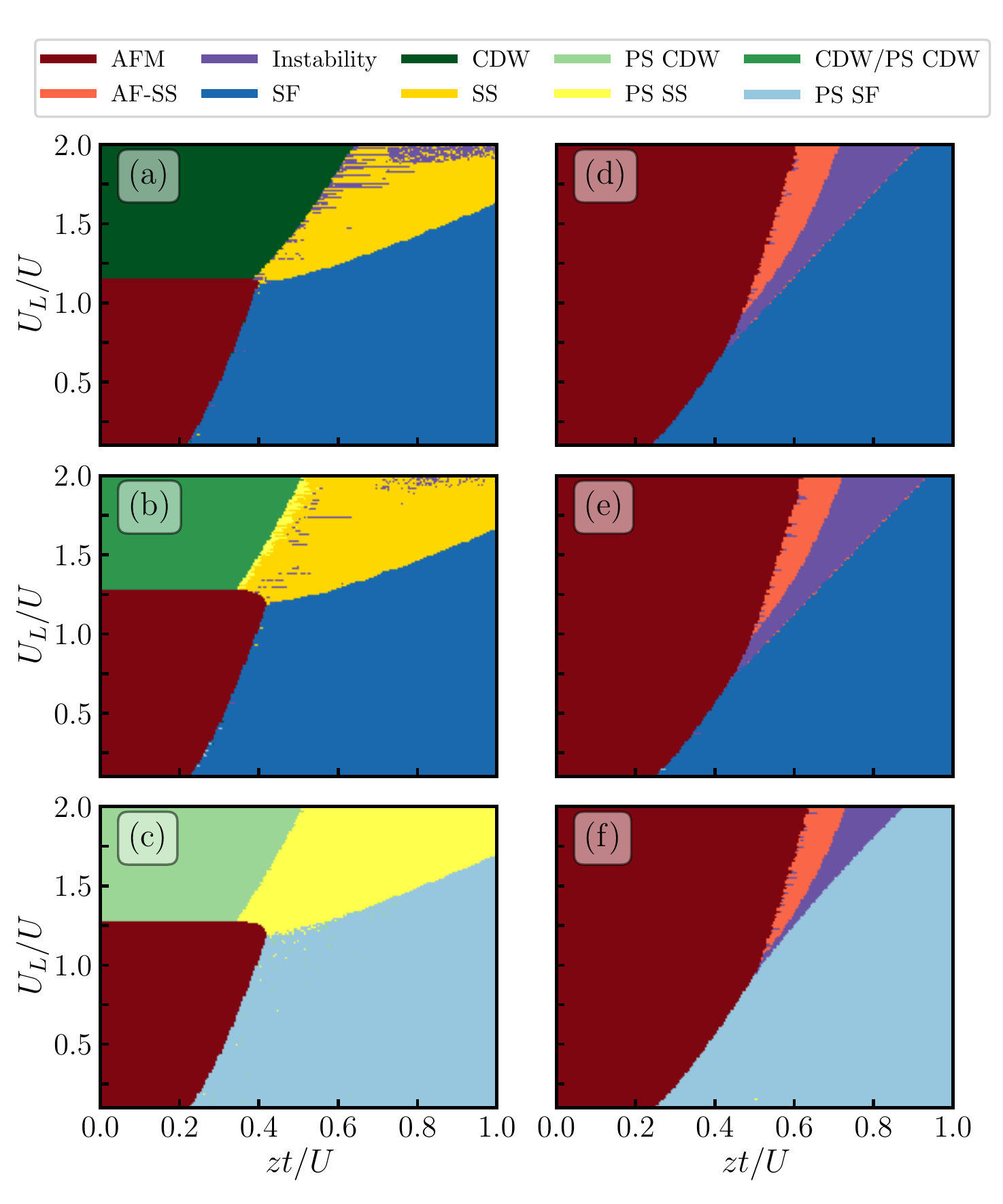}
	\caption{Mean-field phase diagrams for $U_{12}/U=0.9$ (a,d), $~1$ (b,e) and $~1.1$ (c,f) at $U_s/U_v\approx 4.64$ (a-c) and $U_s/U_v\approx 0.33$ (d-f). A total average density of $\rho=2$ is considered.}
	\label{fig:Fig3}
\end{figure}

We first discuss the results for $U_s/U_v>1$, see Fig.~\ref{fig:Fig3}(a-c). For dominating intra-spin interactions [Fig.~\ref{fig:Fig3}(a)], $U_{12}<U$, the phases are identical as those obtained for $U_{12}=U$ assuming a uniform mixture as discussed in the context of Fig.~\ref{fig:fig2}. We additionally find a region of instability in the SS phase for $U_{12}\leq U$. Our observations of phase instability are qualitatively different from the results for spinless systems \cite{Batrouni2000_phaseseparation, Flottat2017_phasediagram}, which predict stable supersolid phases at integer filling in two-dimensional systems. For $U_{12}= U$ [Fig.~\ref{fig:Fig3}(b)], the mixed CDW state  $\ket{\phi_e,\phi_o}=\ket{\up\down,0}$, the entangled state $\ket{\phi_e,\phi_o}=\ket{\up\up,0}+\ket{\down\down,0}$ and the fully phase-separated (PS) configuration with  $\rho^{A}_{\uparrow} = \rho^{B}_{\downarrow} = 2$ are degenerate. Increasing $U_{12}/U$ from 0.9 to 1 shrinks the CDW region: CDW$\rightarrow$AFM transition boundary is shifted towards higher $U_L/U$ and CDW$\rightarrow$AF-SS transition boundary is shifted towards lower $zt/U$. For $U_{12}>U$ [Fig.~\ref{fig:Fig3}(c)], the obtained CDW, SS and SF are fully phase separated.

In the case of dominating vectorial long-rage interactions $U_s/U_v<1$ [Fig.~\ref{fig:Fig3}(d-f)], the phase diagrams for $U_{12}<U$ and $U_{12}=U$  are qualitatively similar: Besides AFM, AF-SS and SF phases, we find extended regions of instability in the AF-SS phase, which is similar to the SS case. For $U_{12}>U$ [Fig.~\ref{fig:Fig3}(f)], the SF is replaced by a fully PS SF, which is compatible with the results for $U_s/U_v>1$. We also find that both the AF-SS and the instability region shrink when $U_{12}/U$ is increased above 1. In contrast, boundaries between insulating regions (PS CDW, AFM) and the phase separated non-insulating state (PS SS, PS SF) do not change with $U_{12}/U,$ as the energy of the PS phases and the AFM phase do not depend on $U_{12}$.

%the transition points to PS non-insulating states (AFM$\rightarrow$PS SF and PS CDW$\rightarrow$PS SS) do not depend on the value of $U_{12}/U$ as the inter-spin interaction strength does not affect the energy of fully phase-separated as well as AFM state.

We note here two limitations of our simulations. First, the identification of different phases in the phase diagrams relies on numerical minimization~\cite{SI} in a high-dimensional landscape. This can lead to spurious solutions, such as the scattered instability points in Fig.~\ref{fig:Fig3}(a,b) and the irregular phase boundaries in Fig.~\ref{fig:fig2}~and~\ref{fig:Fig3}. Second, there is a small region of fully PS SS for $U_{12}=U$ and $U_s/U_v>1$, cf.~Fig.~\ref{fig:Fig3}(b). This is due to the relatively small Hilbert space ($n_{\max}=m_{\max}=3$) used for the simulations. We expect that a larger Hilbert space would lead to degenerate SS and PS SS solutions. Additional information on the different properties of the various phases, the nature of phase transitions and technical details on the identification of phase separation and phase instability is presented in~\cite{SI}. We have also confirmed our results with self-consistent mean-field calculations in a grand canonical ensemble~\cite{dhar2011mean,SI}.

%Within the CE method, we do not observe any changes in the SS, AFM or AFSS phases A comparison with a grand canonical ensemble (GCE) description emphasizes the role of basis truncation. Namely, self-consistent mean-field calculations~\cite{dhar2011mean,SI} with larger local Hilbert space, indicate that the phase-separated SF phase persists up to higher tunneling rates. This is consistent with recent perturbative and DMRG results~\cite{Zhan2014comment}, predicting exclusively a phase-separated SF phase in the absence of the long-range interactions. Including a larger local Hilbert space also suggests the emergence of a phase-separated SS phase.

%We find extended regions of instability (dotted areas in Fig.~\ref{fig:Fig3}c,d) in both the SS and the AFSS phase, and scattered points within the SF regime, which we attribute to numerical inaccuracy. 

\section{Excitations}

\begin{figure}[ht]
	\centering
	\includegraphics[width=1\columnwidth]{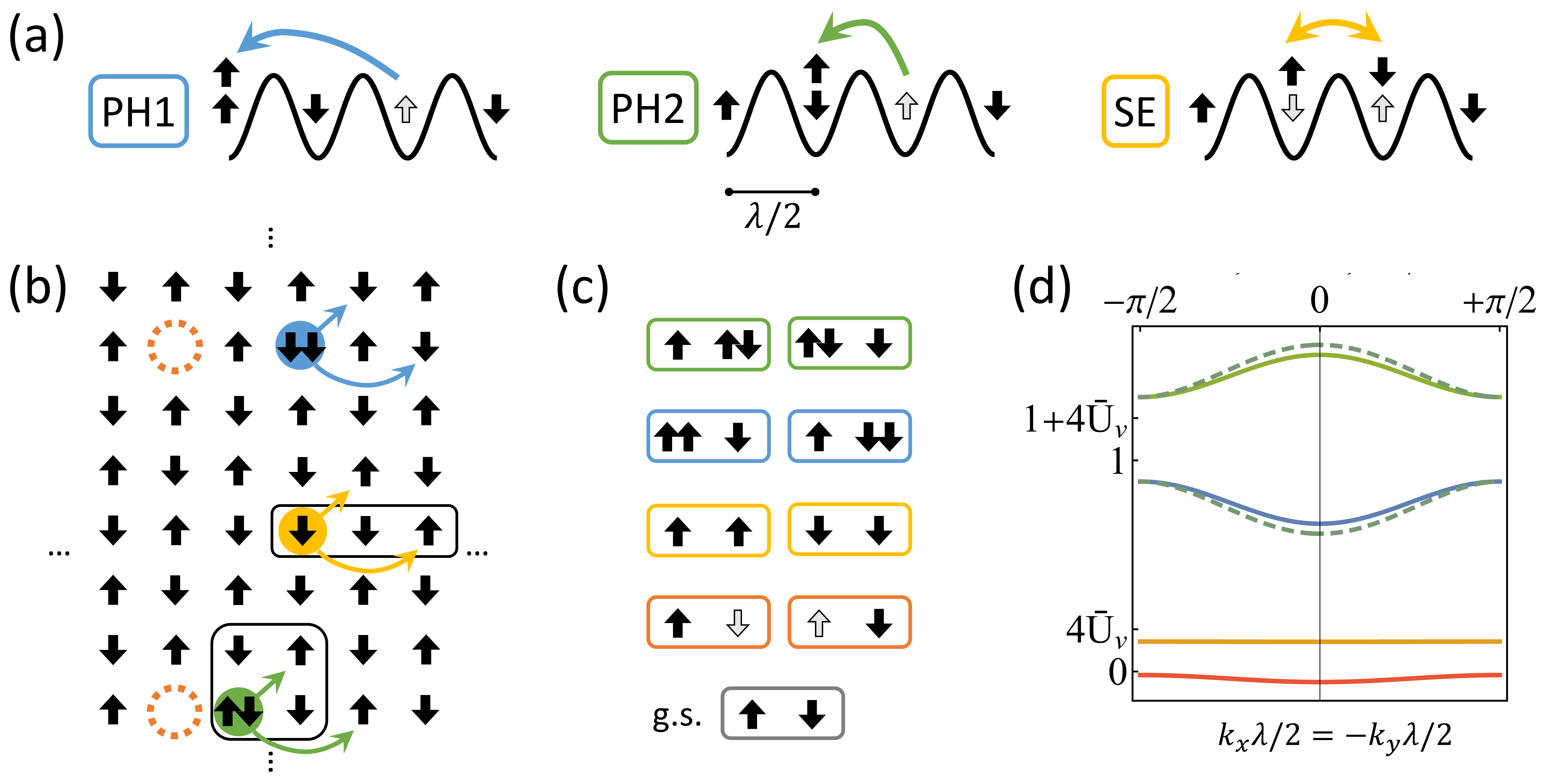}
	\caption{
		(a) Schematic representation of the creation of particle-hole (PH1, PH2) and spin-exchange (SE) excitations in the zero-tunneling limit.
		(b) A finite tunneling rate enables the delocalization of the two components (quasiparticles) of each excitation [color coding as in (a)].
		(c) Sketch of all possible excitation quasiparticles and (d) their band structure for $U_v/U=t/U=0.05$. The dashed green and blue lines denote the bands when mixing of the particle-like components is included in the effective Hamiltonian. The grey box in (c) shows the doublet unit cell when no excitation quasiparticle is present, i.e. the ground state (g.s.).
	}
	\label{fig:Fig4}
\end{figure}

The low-energy excitations provide important information about a given state of the system as they define its dynamical response to external forces and drive the transitions between different macroscopic phases. Here, we study the excitations of the AFM phase, which have no counterpart in single-component BH models or in bulk spinor Bose gases.

We consider vanishing scalar long-range interactions, $U_s=0$. In the $zt/U\rightarrow 0$ limit, the AFM hosts three different low-energy excitations [Fig.~\ref{fig:Fig4}(a)]. The two particle-hole branches PH1 and PH2 correspond to a spin transfer to lattice sites with same and opposite parity, respectively. They yield energy gaps ${\Delta E_{\mathrm{PH1}}/U=1}$ and ${\Delta E_{\mathrm{PH2}}/U=1+4U_v/U}$ above the ground state. The third excitation is a pairwise spin-exchange (SE) involving two atoms with different spin. Its energy gap $\Delta E_{\text{SE}}/U=8U_v/U$  can be arbitrarily tuned with respect to the PH1 and PH2 gaps by adjusting the long-range interaction strength $U_v$.

For a finite $zt/U$, the two components of each excitation (particle and hole for PH1 or PH2; two exchanged spins for SE) delocalize in the lattice [Fig.~\ref{fig:Fig4}(b)].
The effective excitation Hamiltonian,
\begin{equation}
	\label{excitationH}
	H_\text{eff} = \sum_{{\bf n}',{\bf n}}
	\sum_{\beta}
	h_{{\bf n}{\bf n'}}^\beta \left({\hat b}^\beta_{\bf n'}\right)^\dagger {\hat b}^\beta_{\bf n} + \text{H.c.}
\end{equation}
describes quasiparticles $\beta$ [Fig.~\ref{fig:Fig4}(c)], defined on doublets of adjacent spin sites $n$, with creation operators ${\hat b}^\beta_{\bf n}$. The quasiparticles hop on a square superlattice $n_1 {\bf a}_1 + n_2 {\bf a}_2$, with  $n_i \in Z$, ${\bf n} \equiv (n_1,n_2)$, and lattice vectors ${\bf a}_1=({\bf e}_x+{\bf e}_z) \lambda/2$ and ${\bf a}_2= ({\bf e}_x-{\bf e}_z) \lambda/2$. A change in the ground state configuration $|\downarrow,\uparrow\rangle$ of a doublet corresponds to the creation of a quasiparticle at that position.
The coefficients $h_{{\bf n}{\bf n'}}^{\beta}$ characterize the effective tunneling strengths (${\bf n'}\neq {\bf n}$) and energies (${\bf n'}={\bf n}$) of the quasiparticles. They are obtained via first-order perturbation theory, by including up to fourth-order tunneling processes and doubly occupied states~\cite{SI}. We find that the SE excitation energy obtains a second-order leading correction due to tunneling. In the long-wavelength limit, it reads
\begin{equation}
	\frac{\Delta E_{\text{SE}}}{U}
	= 8 \left( \frac{U_v}{U} - \frac{2(3-4U_v/U)}{(1-4U_v/U)(1+4U_v/U)} \left(\frac{t}{U}\right)^2 \right).
	\label{H_eff_exitations}
\end{equation} 

In the resulting band structure~[Fig.~\ref{fig:Fig4}(d)], all the bands are eight times degenerate, four times due to the C4 symmetry invariance of the two excitation components (e.g. particle left, right, above or below the hole), and two times due to two possible configurations of each doublet. The bandwidth of the exchanged-spin branches is $\sim t^4$, and results in an extremely flat band in the considered regime, $t\approx U_v<<U$. The bandwidths of the two particle-like branches are $\sim t^2$, and increases if their weak mixing through a first-order tunneling process is included~\cite{SI}. Additional mixing and level-crossing of the four branches would arise through forth-order (interaction) terms in the Hamiltonian in Eq.~\eqref{excitationH}, which we expect to be pronounced as tunneling is increased and the transition to the AF-SS is approached.

%\textcolor{orange}{The lower PH branch softens and closes the gap at ${\bf k}=0$ as tunneling is increased for any long-range interaction strength, signaling the transition between the insulating and superfluid phases.
% The SE gap, on the other hand, never closes at higher $U_v$~[Fig.~\ref{fig:Fig4}(d)], hinting on the gradual disappearance of spin ordering through a second-order phase transition, and the emergence of an AF-SS between the AFM and SF phase~[Fig.~\ref{fig:Fig3}(c,d)].}
% \text{teal}{The lower particle branch softens and closes the gap at ${\bf k}=0$ as tunneling is increased for any long-range interaction strength, while the behavior of the SE gap strongly depends on it~[Fig.~\ref{fig:Fig4}(d)], resulting in two scenarios: For small $U_v$, the two excitations soften in parallel towards zero as tunneling is increased, signaling an instability of both spin-ordering and insulating properties, in accordance with a first-order AFM\textendash~SF phase transition. For large $U_v$, the SE excitation does not soften for any tunneling strength and always stays gapped, leaving the separate softening of the PH1 branch to be associated with the second-order AFM\textendash~SF phase transition in the standard way.}

\section{Conclusion and Outlook}

We considered an experimentally viable two-component spin BH model, featuring tunable vectorial and scalar cavity-assisted long-range interactions. Its phase diagram at unity filling exhibits both density- and spin-modulated superfluid and insulating phases. We discovered that the insulating spin-modulated phase with global antiferromagnetic order is stabilized for arbitrarily small vectorial long-range interactions, due to their cooperation with repulsive contact interactions. These phases could be readily detected in existing experiments by a combination of cavity field heterodyne detection and time of flight imaging~\cite{landini2018formation,landig2016quantum}, as density and spin-modulated phases couple to orthogonal quadratures of the cavity field.
While dominating intra-spin contact interactions always lead to spatially homogeneous phases, larger inter-spin interactions gives rise to phase separation in the SF and CDW phases. In addition, long-range interactions lead to unstable regions of density- and spin-modulated supersolid phases. 
The low lying excitations above the AFM feature an additional spin-exchange branch, which delocalizes at finite tunneling strengths. A perturbative approach shows that its energy gap can be independently tuned via the vectorial long-range interaction. The inclusion of quasiparticle interactions in the developed effective theory could enable an inspection of the mechanisms by which the interplay between different excitations drives the phase transitions.
% \textcolor{teal}{The low lying excitations above the AFM phase have PH and SE branches, originating from delocalization through tunneling processes. The SE energy gap can be tuned via the vectorial long-range interactions, and its dependence on tunneling strengths (softened or always gapped) hints about the nature of the AFM\textendash~SF phase transition. The inclusion of quasiparticle interactions in the developed effective theory could enable an inspection of the mechanisms in which the dynamics of different excitations drives the phase transitions.}

As a direct extension, it would be interesting to study the first-order phase transition between the density- and the spin-modulated insulating phases. Two key questions there are the lifetime and decay of the metastable states \cite{hruby2018metastability} and the spread of correlations, which might exhibit an unbounded velocity due to the presence of global-range interactions \cite{cheneau2011light-cone,hauke2013spread}. Moreover, regions of phase coexistence could arise when taking into account the harmonic trapping potential \cite{li2013lattice,sundar2016latticeboson}.
In the absence of lattices, cavity dissipation couples density- and spin-modulated states, leading to chiral instabilities and limit cycles \cite{dogra2019dissipation,chiacchio2019dissipation,buca2019dissipation}. The lattice system offers an experimentally tunable access to the boundary between density- and spin-modulated insulating phases, and can help clarify the fate of these instabilities in the Hubbard regime and provide a deeper understanding of the non-Hermitian dynamics in strongly-correlated quantum systems. Finally, the inclusion of spin-changing processes, such as local spin-spin interactions \cite{stamperkurn2013spinorbosegases} or global cavity-mediated  Raman processes \cite{ferri2021emerging,Rosa-Medina2022}, would give rise to another competing scale in the system and could induce novel phases like spin-density waves or chiral states \cite{mivehvar2019cavity, Guan2019two}.
\newline
\begin{acknowledgments}
	We are greatful to Manuele Landini, Francesco Ferri, Fabian Finger and Tobias Donner for inspiring discussions. R.R-M and T.E acknowledge funding from the Swiss National Science Foundation: project numbers IZBRZ2\_186312, 182650 and 175329 (NAQUAS QuantERA) and NCCR QSIT, from EU Horizon2020: ERC advanced grant TransQ (project number 742579).

	\section*{AUTHOR CONTRIBUTIONS}
	R.R.M., S.D.H., T.E., and T.D. conceived the project. L.C. derived the theoretical model and performed the grand-canonical and perturbative calculations of the phase diagrams. N.D. performed the canonical calculations of the phase diagrams. T.D performed the perturbative calculations related to the excitations. L.C., R.R.M., N.D. and T.D. wrote the manuscript. R.R.M., N.D. and T.D. supervised the project. All authors contributed to the discussion and interpretation of the results.
	
\end{acknowledgments}

\newpage
\beginsupplement

%\setstretch{1.25}
\begin{center}
	\large
	\textbf{Supplemental material}
\end{center}
\normalsize

\section{Single-Particle Hamiltonian}
The Hamiltonian of a single atom coupled to a cavity is given by 
\begin{equation}
	\hat{H}_{\mathrm{sp}}=\hat{H}_{\mathrm{atom}} +\hat{H}_{\mathrm{cav}}+\hat{H}_{\mathrm{int}}.
	\label{eq:single_particle_ham}
\end{equation}
The first term describes the bare atomic energy, the second the quantized intra-cavity field and the last term accounts for the atom-light interactions. We work in the rotating frame of the transverse pump field (TP) with frequency $\omega_p$. In the dispersive regime, where the TP is far-detuned from any atomic resonance, excited atomic states can adiabatically be eliminated~\cite{Goldman_2014}. In this regime the bare atomic Hamiltonian reads 
\begin{equation}
	\hat{H}_{\mathrm{atom}}= \frac{\hat{p}^2}{2m}+V_{\mathrm{ext}}(\hat{\mathbf{x}}),
\end{equation}
with $\hbar=1$. The first and second term contain the kinetic energy of the atoms and external confining potentials, respectively. The term accounting for the intra-cavity field is given by 
\begin{equation}
	\hat{H}_{\mathrm{cav}}=-\Delta_c \hat{a}^{\dagger}\hat{a}
\end{equation}
with cavity detuning $\Delta_c=w_p-w_c<0$. The operator $\hat{a}^\dagger$ creates photons in the a cavity mode. The dispersive atom-light interaction can be written as 
\begin{equation}
	\hat{H}_{\mathrm{int}}=\alpha_s\hat{\mathbf{A}}^{\dagger}\cdot\hat{\mathbf{A}}-i\frac{\alpha_v}{2F}\left(\hat{\mathbf{A}}^{\dagger}\times \hat{\mathbf{A}} \right)\cdot\hat{\mathbf{F}},
	\label{eq:Hint}
\end{equation}
where $\hat{\mathbf{A}}$ denotes the co-rotating component of the electromagnetic field and $\hat{\mathbf{F}}=\left(\hat{F}_x,\hat{F}_y,\hat{F}_z\right)^T$ is the total angular momentum operator associated to the $F=1$ ground state manifold of ${}^{87}\mathrm{Rb}$ \cite{kien2013dynamical, landini2018formation}. A static magnetic field $\mathbf{B}=B_0\mathbf{\mathbf{e}}_z$ defines the quantization axis of the spin $\hat{\mathbf{F}}$ along the $z$-direction. Here, we neglect tensorial processes, which is justified for ${}^{87}\mathrm{Rb}$ and TP frequencies close to the experimental realization in Ref. \cite{landini2018formation}. The interaction  Hamiltonian in Eq.~\eqref{eq:Hint} consists of two terms: a scalar contribution $\hat{H}_s=\alpha_s\hat{\mathbf{A}}^{\dagger}\cdot\mathbf{A},$ which does not couple to the spin-degree of  freedom and will give rise to density-modulated phases; and a vectorial contribution $\hat{H}_v=-i\frac{\alpha_v}{2F}\left(\hat{\mathbf{A}}^{\dagger}\times \hat{\mathbf{A}} \right)\cdot\hat{\mathbf{F}},$ which couples to the spin-degree of freedom and will induce spin-modulated phases. The scalar ($\alpha_s$) and vectorial polarizabilities ($\alpha_v$) depend on the electronic structure of the atom and the pump frequency $\omega_p$ and can be tuned relatively to each other \cite{dogra2019phd}.  

The co-rotating part of the light-field is given by
\begin{equation}
	\mathbf{\hat{A}}=E_p\cos(k\hat{z})\left(\cos(\phi)\mathbf{\mathbf{e}}_y+\sin(\phi)\mathbf{\mathbf{e}}_x\right)+E_0\cos(k\hat{x})\mathbf{\mathbf{e}}_y \hat{a}.
	\label{eq:A}
\end{equation}
The first part includes the standing-wave TP field with amplitude $E_p,$ spatial profile $\cos(k\hat{z})$ and linear polarization in the $(x,y)$-plane, which is controlled by the angle $\phi$. The second term accounts for a fundamental quantized mode of the cavity with $E_0$ as the field strength per photon, a spatial mode profile $\cos(k\hat{x})$ and linear polarization along the $y$-direction. Without loss of generality, we assume that both $E_p$ and $E_0$ are real. For a large enough Zeeman splitting, spin-changing processes due to cavity-assisted Raman transitions and atomic collisions, are both suppressed \cite{landini2018formation,stamperkurn2013spinorbosegases}. In this limit the scalar part reads
\begin{equation}
	\begin{split}
		\hat{H}_s&=\underbrace{\alpha_sE_p^2\cos^2(k\hat{z})}_{(1)}+\underbrace{\alpha_sE_0^2\cos^2(k\hat{x})\hat{a}^{\dagger}\hat{a}}_{(2)}\\
		&+\underbrace{\eta\cos(k\hat{x})\cos(k\hat{z})\cos(\phi)\hat{X}}_{(3)}.\\
	\end{split}
	\label{eq:Hs}
\end{equation}
The transverse pump  yields a static lattice for the atoms (1) with lattice depth $V=-\alpha_sE_p^2$ along the $z$-direction of the pump-field. Moreover, the presence of the atoms leads to maximal dispersive shift of the cavity resonanance $U_0=\alpha_sE_0^2$ (2). Finally, the $y$-component of the pump field couples to real quadrature $\hat{X}=(\hat{a}+\hat{a}^\dagger)/\sqrt{2}$ of the cavity mode (3). The spatial modulation of the coupling ($\propto \cos(k\hat{x})\cos(k\hat{z})$) arises from interference of the two fields and scales with $\eta_s=\eta\cos(\phi)$ where 
\begin{equation}
	\eta=\sqrt{2}\alpha_sE_0E_p=\mathrm{sgn}(\alpha_s)\sqrt{2|U_0V|}.
\end{equation}
The vectorial component is given by 
\begin{equation}
	\begin{split}
		\hat{H}_v&=\eta\xi\cos(k\hat{x})\cos(k\hat{z})\sin(\phi)\hat{P}\hat{F}_z\\
	\end{split}
	\label{eq:vectorial_coupling}
\end{equation}
and couples the imaginary quadrature of the cavity field $\hat{P}=i(\hat{a}^{\dagger}-\hat{a})/\sqrt{2}$ to the {$z$-component} of the atomic  spin. The coupling strength is given by  $\eta_v=\eta\xi\sin(\phi)$, with
\begin{equation}
	\xi=\frac{\alpha_v}{2\alpha_sF}.
\end{equation}

In order to obtain a two-dimensional lattice configuration, we consider an additional static optical potential in the $x$-direction 
\begin{equation}
	V_{\mathrm{ext}}(\hat{\mathbf{x}})=-V\cos^2(k\hat{x}). 
\end{equation}
This additional lattice can be created by injecting the cavity-mode with an additional $y$- or $z$-polarized laser beam,  which has a small frequency offset to suppress interference with the transverse pump~\cite{landig2016quantum}. The static lattice potential in the $z$-direction is generated by the transverse pump (see (1) of Eq.~\eqref{eq:Hs}). As in Ref. \cite{landig2016quantum} a third standing wave creates a deep optical lattice in the $y$-direction, which slices the system in an array of effective 2D systems. Altogether, we arrive at the single particle Hamiltonian given in Eq.~(1) of the main text.

\section{Many-Body Hamiltonian}
We derive the many-body Hamiltonian for a mixture of $m_F=\pm 1$ atoms. Here, we identify the Zeeman levels $\ket{F=1,m_F=-1(1)}$ as $\ket{\down(\up)}.$ In second quantization the many-body Hamiltonian reads
\small 
\begin{equation}
	\begin{split}
		&\hat{H}_{\mathrm{mb}}=\int \hat{\Psi}^{\dagger}(x,z)\left(\frac{\hat{\mathbf{p}}^2}{2m}+\hat{V}_{\mathrm{lat}}(x,z)\right)\hat{\Psi}(x,z)\,dxdz \\
		&-\int \hat{\Psi}^{\dagger}(x,z)U_0\cos(kx)^2\hat{a}^{\dagger}\hat{a}\hat{\Psi}(x,z)\,dxdz\\
		&+\int \hat{\Psi}^{\dagger}(x,z)\cos(kx)\cos(kz)\left(\eta_s\hat{X}+\eta_v\hat{P}\hat{F}_z\right)\hat{\Psi}(x,z)\,dxdz\\
		&-\sum_{m\in\{\up,\down\}} \mu_{m}\int\hat{\Psi}_m^{\dagger}(x,z)\hat{\Psi}_m(x,z)\,dxdz \\
		&+\frac{1}{2}\sum_{m\in{\up,\down}} g_{2D,m}\int \Psi^{\dagger}_m(x,z)\Psi^{\dagger}_m(x,z)\Psi_m(x,z)\Psi_m(x,z)\,dxdz\\
		&+g_{2D,\up\down}\int \Psi^{\dagger}_\up(x,z)\Psi_\up(x,z)\Psi^{\dagger}_\down (x,z)\Psi_\down(x,z)\, dxdz\\
		&-\Delta_c\hat{a}^{\dagger}\hat{a},
	\end{split}
\end{equation}
\normalsize
where $\hat{\Psi}=\hat{\Psi}_{\up}+\hat{\Psi}_{\down}$ is the bosonic field operator associated to the mixture. For each  spin state $\ket{\up (\down)}$  there is an associated chemical potential $\mu_{\up(\down)}.$ The last two terms account for on-site contact interactions. The effective interaction strengths for a 2D square lattice are given by ${g_{2D,m}=\frac{4\pi a_m}{M}\int |w^{(y)}_0(y)|^4 \, dy}$ and ${g_{2D,\up\down}=\frac{2\pi a_{\up\down}}{M}\int |w^{(y)}_0(y)|^4 \, dy}$ for intra- and inter-spin species collisions, respectively. Here, $a_{m}$ and $a_{\up\down}$ denote the scattering lengths inter- and intra-spin interactions ($m\in\{\up,\down\}$) and $M$ is the mass of the atoms. The strength of the interactions is modified by the presence of an additional deep lattice in the $y$-direction, which confines the atoms in the $(x,z)$-plane. This is accounted by integrating over the Wannier function $w^{(y)}(y)$  associated to this 3rd lattice direction. For deep enough lattice potentials, the many-body function can be expanded in a localized basis consisting of Wannier functions $w_\mathbf{i}$ in the lowest energy band, i.e., ${\hat{\Psi}_{m}(x,z)=\sum_{\mathbf{i}} w_{\mathbf{i}}(x,z)\bp_{\mathbf{i},m}},$
where $w_{\mathbf{i}}(x,z)$ is the Wannier function localized at site $\mathbf{i}=(i_x,i_z)$ and $\bp_{\mathbf{i},m}$ is the bosonic annihilation operator of a particle in spin $m$ at site $\mathbf{i}.$
Considering only nearest neighbour tunneling, we derive the following lattice Hamiltonian
%We neglect collisional spin-changing processes.
\begin{equation}
	\begin{split}
		\hat{H}=&-t\sum_{m\in \{\up,\down\}}\sum_{\braket{\mathbf{i},\mathbf{j}}}(\bpp_{\mathbf{i},m}\bp_{\mathbf{j},m}+\bpp_{\mathbf{j},m}\bp_{\mathbf{i},m})\\
		&+\frac{1}{2}\negthinspace\sum_{m\in\{\up,\down\}}\negthinspace\negthinspace \negthinspace\negthinspace U_{m}\negthinspace\sum_{\mathbf{i}}\hat{n}_{i,m}(\hat{n}_{\mathbf{i},m}-1)+U_{12}\sum_{\mathbf{i}}\hat{n}_{\mathbf{i},\up}\hat{n}_{\mathbf{i},\down}\\
		&-\sum_{m\in \{\up,\down\}}\mu_{m}\sum_{\mathbf{i}}\hat{n}_{\mathbf{i},m}\\
		&+  M_{0}\left\{\eta_s \hat{\Theta}_D \hat{X}+\eta_v\hat{\Theta}_S\hat{P}\right\}\\
		&-\tilde{\Delta}_c\hat{a}^{\dagger}\hat{a},\\
	\end{split}
	\label{eq:H_mb}
\end{equation}
where $\braket{\mathbf{i},\mathbf{j}}$ denotes a pair of neighbouring sites.
The tunneling strength between neighbouring atoms is given by  
\begin{equation}
	t=\int w_\mathbf{\mathbf{i}}(x,z)\left(-\frac{1}{2m}\left(\frac{\partial^2}{\partial^2_x}+\frac{\partial^2}{\partial^2_y}\right)+V_{\mathrm{lat}}(x,z)\right) w_{\mathbf{j}}(x,z)\ dxdz\\
\end{equation}

and is independent of the spin and tunneling direction as the static lattice is spin-insensitive and equal in strength along $x$ and $z$. 
The intra- and inter-spin contact interaction strengths are given by 
\begin{align}
	U_{m} &=g_{2D,m} \int |w_0(x,z)|^4\ dxdz \ \ \ \text{and}\\
	U_{12}&= g_{2D,\up\down}\int |w_0(x,z)|^4\ dxdz,
\end{align}
respectively. The dynamical interference potential created by photons, which scatter off the atoms into the cavity, yields the overlap integral 
\begin{equation}
	M_{0}=\int w_\mathbf{0}(x,z) \cos(kx)\cos(kz) w_\mathbf{0}(x,z)\ dxdz.
\end{equation}
The scalar component of the atom-light interaction couples the real quadrature $\hat{X}$ of the light-field to the operator
$\hat{\Theta}_D$, as defined in the main text. Meanwhile, the vectorial component couples the imaginary quadrature $\hat{P}$ to the operator $\hat{\Theta}_S.$
Finally, the presence of the atoms in the cavity yields a dispersive shift $\tilde{\Delta}_c=\left(\Delta_c- U_0M_{1}N\right)$ of the cavity resonance, which scales with the overlap integral 
\begin{equation}
	M_{1}=\int w_\mathbf{0}(x,z) \cos(kx)^2 w_\mathbf{0}(x,z)\ dxdz.\\
\end{equation}
Here, $N=\sum_{\mathbf{i},m}n_{\mathbf{i},m}$ denotes the total number of atoms.

\subsection{Adiabatic Elimination of the Cavity Field}
We adiabatically eliminate the cavity-field. This is valid when the energy scales related to the light field, i.e., the cavity losses $\kappa$ and the detuning $\tilde{\Delta}_c$ dominate over the atomic energy scales, such as $t$ and $U_{m,12}$, which is the case for the experiments in Refs.~\cite{landig2016quantum,landini2018formation}. In the presence of cavity dissipation, the corresponding Heisenberg equation of motion is given by 
\begin{equation}
	\frac{d}{dt}\hat{a}=i\tilde{\Delta}_c\hat{a} -i\frac{M_0}{\sqrt{2}}\left(\eta_s\hat{\Theta}_D+i\eta_v\hat{\Theta}_S\right)-\kappa\hat{a}.
	\label{eq:da/dt}
\end{equation}
Setting the right-hand side of Eq.~\eqref{eq:da/dt} to 0 yields 
\begin{equation}
	\hat{a}=\frac{M_0}{\sqrt{2}}\frac{1}{\tilde{\Delta_c}+i\kappa}\left(\eta_s\hat{\Theta}_{D}+i\eta_v\hat{\Theta}_S\right).
	\label{eq:adiabatic_eliminated_a_bh}
\end{equation}
Inserting Eq.~\eqref{eq:adiabatic_eliminated_a_bh} into Eq.~\eqref{eq:H_mb}, we obtain the effective extended Bose-Hubbard model discussed in Eq.(1) in the main text, with $\mu_{\up}=\mu_{\down}=\mu,$  $U_{\up}=U_{\down}=U$ and
\begin{equation}
	\frac{U_L}{K}=-\frac{M_0^2\tilde{\Delta}_c}{\kappa^2+\tilde{\Delta}_c^2}\eta^2.
\end{equation}

\section{Mean-field Methods}
In this section we give details on the different mean-field ansatzes used to construct the full phase diagram: a Gutzwiller ansatz and a self-consistent mean-field approach. The former approach, already outlined in the main text, assumes a translationally invariant ground state of product form. The latter is based on a decoupling of the tunneling and long-range interacting terms, which yields an effective self-consistent two-site two-spin Hamiltonian.   

\subsection{Gutzwiller Ansatz}
We start by discussing the Gutzwiller ansatz used in the main text in detail. We assume that the ground state is of product form (Eq.~(5) of the main text) and the local wave functions for even and odd sites is given by 
\begin{equation}
	\ket{\phi_{e(o)}}=\sum_{m,n=0}^{n_{\max},m_{\max}}a_{e(o)}(n,m)\ket{n,m}
	\label{eq:full_state}
\end{equation}
where we truncated the local occupation of each site (Eq.~(6) of the main text).
As shown in Eq.~(7) of the main text, the energy density to be minimized can be written as
\begin{equation}
	\begin{split}
		\mathcal{E}(\mathbf{a}_{e},\mathbf{a}_{o})&= -2zt\left( \psi_{e,\uparrow}\psi_{o,\uparrow}  + \psi_{e,\downarrow}\psi_{o,\downarrow} \right)\\
		&+  \frac{U}{2}\sum_{m\in\{\up,\down\}}\left(n_{e,m}(n_{e,m}-1\right) + n_{o,m}\left(n_{o,m}-1\right))\\
		& + U_{12}(n_{e,\uparrow}n_{e,\downarrow} + n_{o,\uparrow}n_{o,\downarrow}) -\frac{U_s}{2}\theta_D^2 -\frac{U_v}{2}\theta_S^2.
		\label{eq:energy_density}
	\end{split}
\end{equation}
We repeat the definitions introduced in the main text for completeness here
\[
n_{i,m} = \braket{\hat{n}_{i,m}}, \psi_{i,m} = \braket{\hat{b}^{\dagger}_{i,m}}= \braket{\hat{b}_{i,m}},
\]
with $i$ being an even (e) or odd (o) site and $m~\epsilon \{\uparrow,\downarrow\}.$ The two density order parameters are
\begin{align}
	\theta_D = (n_{e,\uparrow} + n_{e,\downarrow})-(n_{o,\uparrow} + n_{o,\downarrow}), \label{eq:theta_D}\\
	\theta_S = (n_{e,\uparrow} + n_{o,\downarrow})-(n_{o,\uparrow} + n_{e,\downarrow}) \label{eq:theta_S}.
\end{align}

\subsubsection{The limit $t=0$}
In the limit $t=0$ the energy density can be minimized as a function of $\mathbf{n}=\left(n_{e,\up},n_{e,\down},n_{o,\up},n_{o,\down}\right)$ only. To determine the transition from an AFM to a CDW phase for a homogeneous mixture at unity filling, we calculate their respective energies. The AFM phase has two possible configurations $\mathbf{n}_{\mathrm{AFM}}=\left(1,0,0,1\right)$ and $\mathbf{n}_{\mathrm{AFM}}=\left(0,1,1,0\right).$ The corresponding energy density reads
\begin{equation}
	\mathcal{E}_{\mathrm{AFM}}=-2U_v.
\end{equation}
Depending on the relative strength between intra- and inter-species interaction, $U_{12}/U,$ the CDW phase has the following configurations
\begin{equation}
	\begin{cases}
		\mathbf{n}_\text{CDW}=(1,1,0,0), \mathrm{if}\quad \frac{U_{12}}{U}<1\\
		\mathbf{n}_\text{CDW}\in\{ (1,1,0,0),(2,0,0,0),(0,2,0,0)\}, \mathrm{if}\quad \frac{U_{12}}{U}=1\\
		\mathbf{n}_\text{CDW}\in\{(2,0,0,0)(0,2,0,0)\}, \mathrm{if}\quad \frac{U_{12}}{U}> 1\\
	\end{cases}
\end{equation}  
where $(2,0,0,0)$ and $(0,2,0,0)$ denote phase separated solutions, where on one half of the system the even (odd) sites are occupied by two atoms of spin $\up$ ($\down$) and in the other half of the system the even (odd) sites are occupied with two atoms of spin $\down$ ($\up$). For each configuration mentioned above there is also the mirrored configuration with odd sites being occupied instead of the even ones, reflecting the underlying $\mathbb{Z}_2$ of the system.\\
The respective energy density reads  
\begin{equation}
	\mathcal{E}_{\mathrm{CDW}}=\min\{U_{12},U\}-2U_s.
\end{equation}
By comparing $\mathcal{E}_{\mathrm{AFM}}$ and $\mathcal{E}_{\mathrm{CDW}},$ we can infer the 1st order phase transition at $t=0:$
\begin{equation}
	\mathcal{E}_{\mathrm{CDW}}<\mathcal{E}_{\mathrm{AFM}}\iff \frac{\min\{U_{12},U\}}{2}<(U_s-U_v).
	\label{eq:AFM_CDW_transition}
\end{equation}
In particular, under the assumption of unity filling the CDW phase will not be formed if $U_s<U_v$ or equivalently $1/\xi^2<\tan(\phi)^2.$ For the values used in Fig.~2(a,b) of the main text with $|\xi|=0.464$ and $\phi=\pi/4$ the transition occurs at $(U_{L}/U)_c\approx 1.27.$
\
\subsubsection{Energy Minimization} 
We use a series of different ansatzes to construct the full phase diagram of the system. First, we consider the general ansatz given in Eq.~\eqref{eq:full_state} with $n_{\max}=m_{\max}=3.$ Using the MATLAB function \textit{fmincon}, the corresponding energy functional is minimized under the constraint of fixed density for each of the spin states (giving $28$ variational parameters). We find out that density and spin order never coexist which allows us to choose simpler ansatzes for these two orders. For the density order, we assume that the wave functions on every site stay invariant under a spin flip. For the spin order, we assume that the wave functions in odd $|\phi_{e}\rangle$ and even sites $|\phi_{o}\rangle$ are mapped onto each other by a spin flip. Thus, we choose the following ansatzes for uniform density ($|\phi^{D}_{e,o}\rangle$) and spin order ($|\phi^{S}_{e,o}\rangle$)
\begin{align}
	|\phi^{D}_{e(o)}\rangle &=  a_{0e(o)}|0,0\rangle + \frac{a_{1e(o)}}{\sqrt{2}}\left(|1,0\rangle + |0,1\rangle \right) \nonumber\\
	& + \frac{a_{2e(o)}}{\sqrt{2}}\left(|2,0\rangle + |0,2\rangle \right)  
	+ a_{3e(o)}|1,1\rangle \nonumber \\
	&+\frac{a_{4e(o)}}{\sqrt{2}}\left(|2,1\rangle + |1,2\rangle \right) + a_{5e(o)}|2,2\rangle \nonumber \\
	& + \frac{a_{6e(o)}}{\sqrt{2}}\left(|3,0\rangle + |0,3\rangle \right) + \frac{a_{7e(o)}}{\sqrt{2}}\left(|3,1\rangle + |1,3\rangle \right) \nonumber \\
	& + \frac{a_{8e(o)}}{\sqrt{2}}\left(|3,2\rangle + |2,3\rangle \right) + a_{9e(o)}|3,3\rangle ,
\end{align}

\begin{align}
	|\phi^{S}_{e(o)}\rangle &= a_{0(0)}|0,0\rangle + a_{1(2)}|1,0\rangle + a_{2(1)}|0,1\rangle + a_{3(4)}|2,0\rangle + \nonumber \\
	&a_{4(3)}|0,2\rangle + a_{5(5)}|1,1\rangle + a_{6(7)}|2,1\rangle + a_{7(6)}|1,2\rangle + \nonumber \\ 
	&a_{8(8)}|2,2\rangle + a_{9(10)}|3,0\rangle + a_{10(9)}|0,3\rangle \nonumber \\
	&+ a_{11(12)}|3,1\rangle 
	+a_{12(11)}|1,3\rangle + a_{13(14)}|3,2\rangle +\nonumber\\ 
	&a_{14(13)}|2,3\rangle + a_{15(15)}|3,3\rangle. 
\end{align}

%as less parameters allow us to obtain the ground state energy with higher accuracy.
In this way, the number of variables to be optimized is reduced to 18 and 15 for the two cases respectively. The same ansatzes are also used for identifying phase instability. To incorporate phase separation, we use the ansatz as shown in Eq.~\eqref{eq:full_state} with the assumption that there is no coexistence of density and spin order. We thus constrain only the total density (and not the individual spin up and down densities) and find the energy of a phase separated and density (spin) ordered state by setting $U_{s(v)}=0$. Physically, this implies that the whole system is split into two halves: while the one half has higher(lower) density of one(other) spin state, the situation is reversed in the other half in a way that either the density or the spin order vanishes.

\begin{figure}[ht]
	\centering
	\includegraphics[width=\columnwidth]{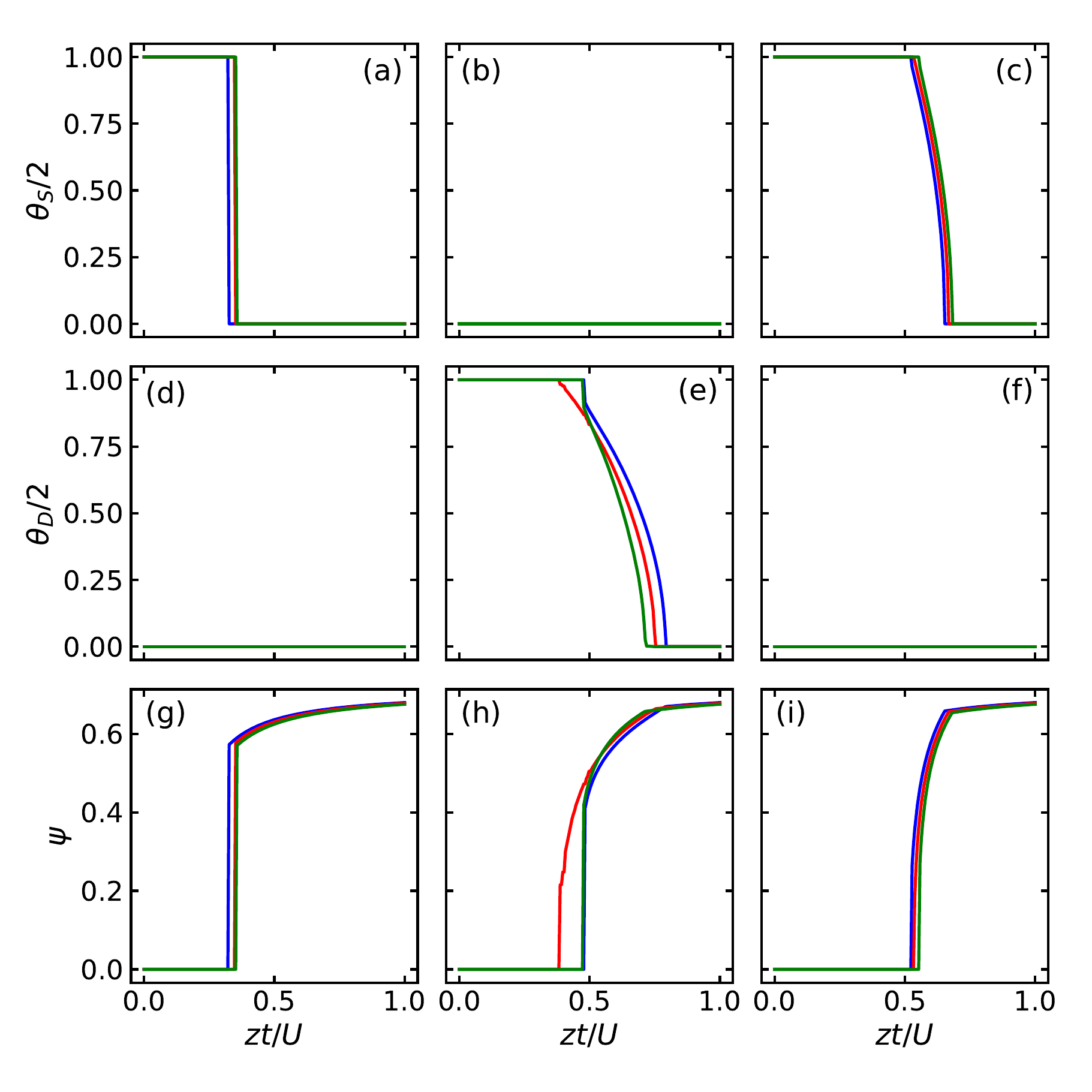}
	\caption{Dependence of spin ($\theta_s$) and density ($\theta_d$) order parameters and superfluid ($\psi$) order parameter as a function of normalized tunnelling strength $zt/U$. Different colours represent different values of $U_{12}/U$: 0.9 (blue), 1(red) and 1.1 (green). (a,d,g) are plotted for $U_s/U_v\approx 4.64$ and $U_L/U=0.63$, (b,e,h) are plotted for $U_s/U_v\approx 4.64$ and $U_L/U=1.41$ and (c,f,i) are plotted for $U_s/U_v\approx 0.33$ and $U_L/U=1.3$.}
	\label{fig:fig1_appendix}
\end{figure}

The energy to be minimized (Eq.~(7) of the main text) can be written as
\begin{align}
	\mathcal{E}=& -2zt\left( \psi_{e,\uparrow}\psi_{o,\uparrow}  + \psi_{e,\downarrow}\psi_{o,\downarrow} \right) +  \frac{U}{2}\left(n_{e,\uparrow}(n_{e,\uparrow}-1\right) + \nonumber \\ &n_{o,\uparrow}\left(n_{o,\uparrow}-1\right)) +  
	\frac{U}{2}\left(n_{e,\downarrow}\left(n_{e,\downarrow}-1\right) + n_{o,\downarrow}\left(n_{o,\downarrow}-1\right)\right)  \nonumber\\
	& + U_{12}(n_{e,\uparrow}n_{e,\downarrow} + n_{o,\uparrow}n_{o,\downarrow}) -\frac{U_s}{2}\theta_D^2 - -\frac{U_v}{2}\theta_S^2
	\label{eq:energy_density2}
\end{align}
where $n_{i,m} = \braket{\hat{n}_{i,m}}$ and $\psi_{i,m} = \braket{\hat{b}^{\dagger}_{i,m}}= \braket{\hat{b}_{i,m}}$, with $i$ being an even (e) or odd (o) site and $m~\epsilon \{\uparrow,\downarrow\}$. %The density and spin order parameters are
%\[
%\theta_D = (n_{e,\uparrow} + n_{e,\downarrow})-(n_{o,\uparrow} + n_{o,\downarrow}),
%\]
%\[
%\theta_S = (n_{e,\uparrow} + n_{o,\downarrow})-(n_{o,\uparrow} + n_{e,\downarrow}).
%\]

\subsection{Properties of homogeneous ground state solutions}

We start by studying the dependence of different order parameters on the ratio $U_{12}/U$. This is shown in Fig.~\ref{fig:fig1_appendix} for the same parameters as used in Fig.~2(e-g) of the main text. Here, we do not consider phase separation or instabilities. The evolution of different order parameters as a function of $zt/U$ is similar with a slight increase in the critical tunnelling for the transition to supersolid or superfluid phases with increasing $U_{12}/U$. Such an increase is expected as effectively the short-range interaction energy increases at the fixed density. 

The order of various phase transitions either as a function of tunnelling or long-range interaction strength is summarized in the Tables \ref{table:1} and \ref{table:2}.

\begin{table}[h!]
	\centering
	\begin{tabular}{|m{2.1cm}| m{3cm}| m{3cm}|} 
		\hline
		Transition & $U_s/U_v$=$4.64$, $U_{12}/U=$[0.9,1,1.1]& $U_s/U_v$=$0.33$, $U_{12}/U=$[0.9,1,1.1] \\  
		\hline
		AFM$\rightarrow$SF & 1,1,1 & 1,1,1 \\ 
		AFM$\rightarrow$AF-SS & ~~- & 2,2,2  \\
		AFM$\rightarrow$SS & 1,1,1 & ~~- \\ 
		AF-SS$\rightarrow$SF & ~~- & 2,2,2 \\
		CDW$\rightarrow$SS & 1,2,1 & ~~- \\
		%		CDW$\rightarrow$SF & -,-,1 & ~~- \\ 
		SS$\rightarrow$SF & 2,2,2 & ~~- \\ 
		\hline
	\end{tabular}
	\caption{Order of various transitions as a function of increasing tunnelling. The notation phase 1$\rightarrow$phase 2 denotes the transition from phase 1 to phase 2 on increasing tunnelling. First (second) order transition is referred by 1(2). `-' implies that the corresponding transition is not found in the calculations.}
	\label{table:1}
\end{table}

\begin{table}[h!]
	\centering
	\begin{tabular}{|m{2.1cm}| m{3cm}| m{3cm}|} 
		\hline
		Transition & $U_s/U_v$=$4.64$, $U_{12}/U=$[0.9,1,1.1]& $U_s/U_v$=$0.33$, $U_{12}/U=$[0.9,1,1.1] \\  
		\hline
		AFM$\rightarrow$CDW & 1,1,1 & ~~- \\ 
		AFM$\rightarrow$SS & 1,1,1 & ~~- \\ 
		AF-SS$\rightarrow$AFM & ~~- & 2,2,2 \\
		SS$\rightarrow$CDW & 1,2,1 & ~~- \\ 
		SF$\rightarrow$AFM & 1,1,1 & 1,1,1 \\ 
		SF$\rightarrow$AF-SS & ~~- & 2,2,2  \\
		SF$\rightarrow$SS & 2,2,2 & ~~- \\ 
		%		SF$\rightarrow$CDW & -,-,1 & ~~- \\
		\hline
	\end{tabular}
	\caption{Order of various transitions as a function of increasing long-range interaction strength. The notation phase 1$\rightarrow$phase 2 denotes the transition from phase 1 to phase 2 on increasing long-range interaction strength. First (second) order transition is marked by 1(2).  `-' implies that the corresponding transition is not found in the calculations.}
	\label{table:2}
\end{table}

\begin{figure}[ht]
	\centering
	\includegraphics[width=\columnwidth]{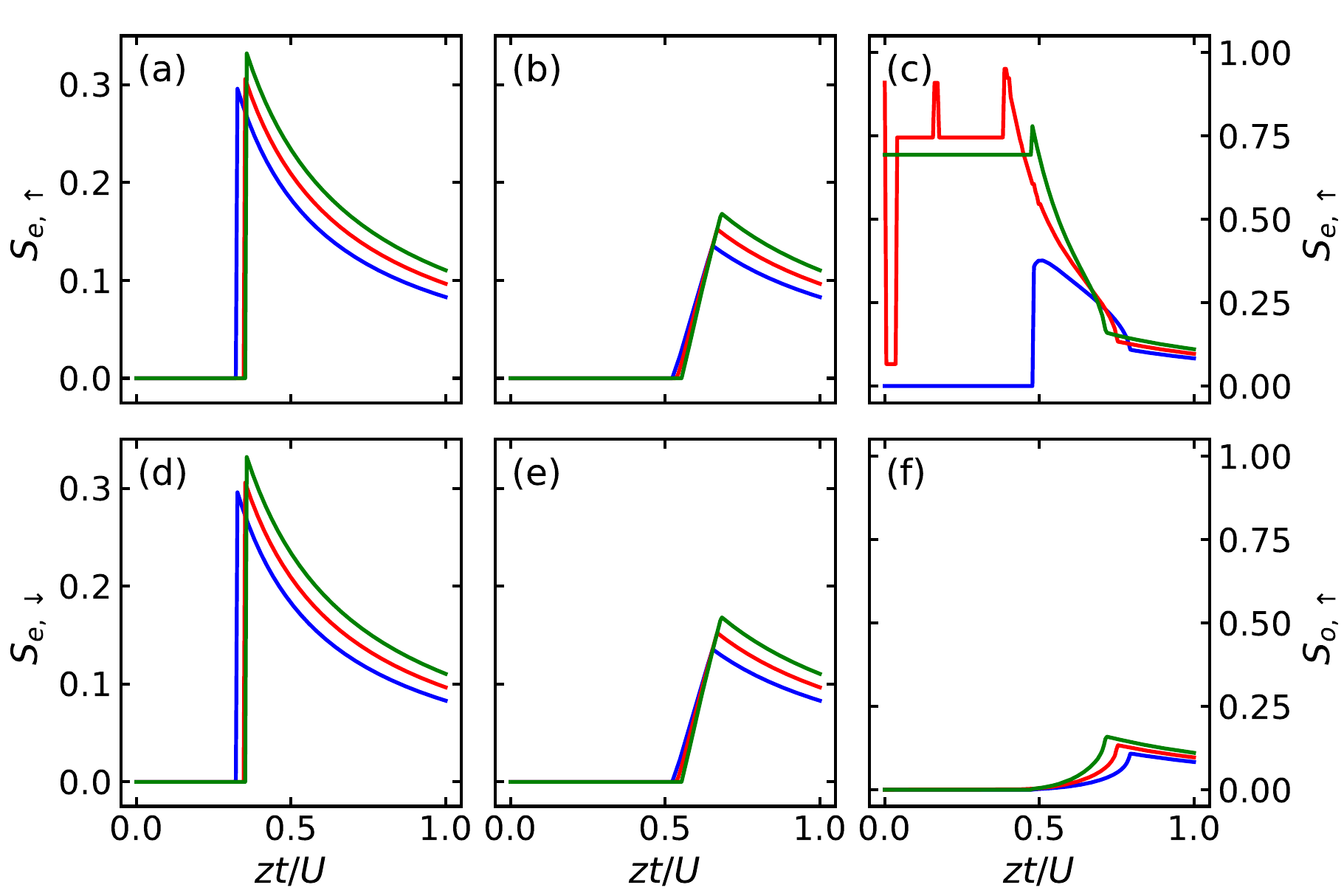}
	\caption{ Reduced von Neumann entropy as a function of normalized tunnelling strength $zt/U$. Different colours represent different values of $U_{12}/U$: 0.9 (blue), 1 (red) and 1.1 (green). (a,d) are plotted for $U_s/U_v\approx 4.64$ and $U_L/U=0.6$, (b,e) are plotted for $U_s/U_v\approx 0.33$ and $U_L/U=1.3$ and (c,f) are plotted for $U_s/U_v\approx 4.64$ and $U_L/U=1.41$. The y-axes on the right side are defined for panels (c,f).}
	\label{fig:fig2_appendix}
\end{figure}

Next, we look at the von Neumann entropy for different phases. We construct the density matrix with our variational ansatz for the optimum value of different parameters. We then trace out one of the spin states to calculate the reduced density matrix $\rho_{\mathrm{red}}$ of the other spin state either on the even or the odd site which is used to obtain the reduced von Neumann entropy via: $S = -\sum_i p_i\log{p_i}$, where the summation is over all the eigenvalues $p_i$ of $\rho_\text{red}$. By construction, the entropy $S_{j,\uparrow} = S_{j,\downarrow}$ for $j \epsilon\{e,o\}$ for uniform density order solutions and $S_{e,\uparrow} = S_{o,\downarrow}$, $S_{o,\uparrow} = S_{e,\downarrow}$ for uniform spin order solutions. So, we accordingly show only $S_{e,\uparrow}$, $S_{e,\downarrow}$ and $S_{e,\uparrow}$, $S_{o,\uparrow}$ for phases with density and spin order respectively, see Fig.~\ref{fig:fig2_appendix}.

We see that except for CDW phase in the $U_{12}/U \geq 1$ cases, the insulating phases are locally product states and hence have zero von Neumann entropy. For $U_{12}/U > 1$, the ground state is $(|2,0\rangle) + |0,2\rangle/\sqrt{2}$ on one site and $|0,0\rangle$ on the other which is the maximally entangled state in a two-dimensional subspace and hence has $\rho_{\mathrm{red}} = \ln{2}$. For $U_{12}/U = 1$, the ground state is an arbitrary superposition of $(|2,0\rangle) + |0,2\rangle/\sqrt{2}$ and $|1,1\rangle$ on one site and $|0,0\rangle$ on the other which gives finite S with jumps as different superpositions have different entropy. The jump and increase in $S_{e,\uparrow}$ on entering the corresponding AF-SS phase (for $U_{12}/U \geq 1$) is due to inclusion of more states in the ground state solution. AF-SS has lower values of entropy as compared to lattice SS. This is because AF-SS (lattice SS) correspond to the two spins primarily occupying sites with opposite (same) parity. In the superfluid phase, the entropy goes down with increasing tunneling which is expected as the kinetic energy term favors a product state as the solution. This further elucidates that such entanglement arises due to the presence of short-range interparticle interactions.

Besides the exceptions discussed above, there are jumps in the entropy as a function of tunnelling corresponding to the first-order transitions (revealed by the jumps in the order parameters). We also note that $S_{e,\uparrow}=S_{e,\downarrow}$ for states with $\theta_S \neq 0$ as well. Finally, we comment about experimental challenges in detecting the entanglement entropy. A system with populations in two different spin states is created by shining a resonant radio-frequency pulse of controlled duration on a single spin state. Subsequent spin-changing processes are experimentally suppressed by applying a strong external magnetic field which lifts the degeneracy of $\uparrow$ and $\downarrow$ spin states. Any experimental noise will convert such a state into an incoherent mixture and detection of generated entanglement has to be done locally while accounting for the Zeeman shift from the external magnetic field.

\subsection{Properties of phase-separated solutions}
\begin{figure}[ht]
	\centering
	\includegraphics[width=\columnwidth]{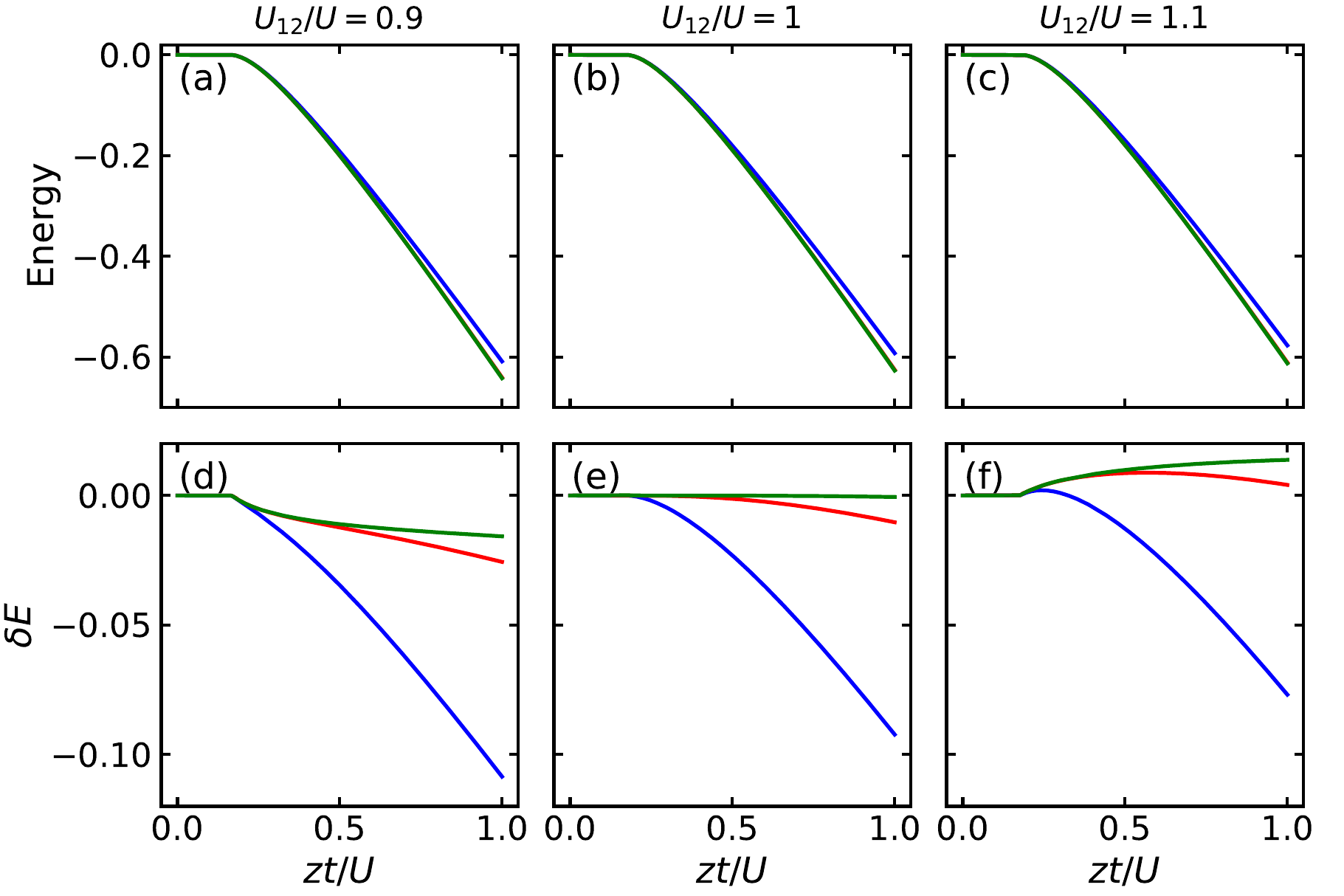}
	\caption{ (a-c) shows the minimum energy densities of the homogeneous solution for different values of $U_{12}/U$. (d-f) shows the corresponding energy density difference $\delta E$ of homogeneous solution and (fully) phase-separated solution. The blue, red and green colours correspond to $n_{\mathrm{max}}=2,3,4$ respectively. $U_L/U=0$ is considered.}
	\label{fig:fig4_appendix}
\end{figure}

We first show that increasing the maximum occupancy per spin per lattice site $n_{\mathrm{max}}$ and $m_{\mathrm{max}}$ in our ansatz Eq.~\eqref{eq:full_state} affects the regime where the fully phase-separated solutions have the lowest energy. This is illustrated in Fig.~\ref{fig:fig4_appendix}.

We note that for $U_{12}/U =1.1$, the phase-separated solution is the one with the lowest energy density in the SF regime. We observe such a behaviour only for $n_{\mathrm{max}} \geq 3$. Similarly, we expect homogeneous solution and phase-separated solution to be degenerate for $U_{12}/U =1,$ which is observed only for $n_{\mathrm{max}} = 4$ in the SF phase (for all $zt/U$). Such a limitation of our results with $n_{\mathrm{max}} = 3$ is visible in Fig. 3(b) of the main text.

\begin{figure}[ht]
	\centering
	\includegraphics[width=\columnwidth]{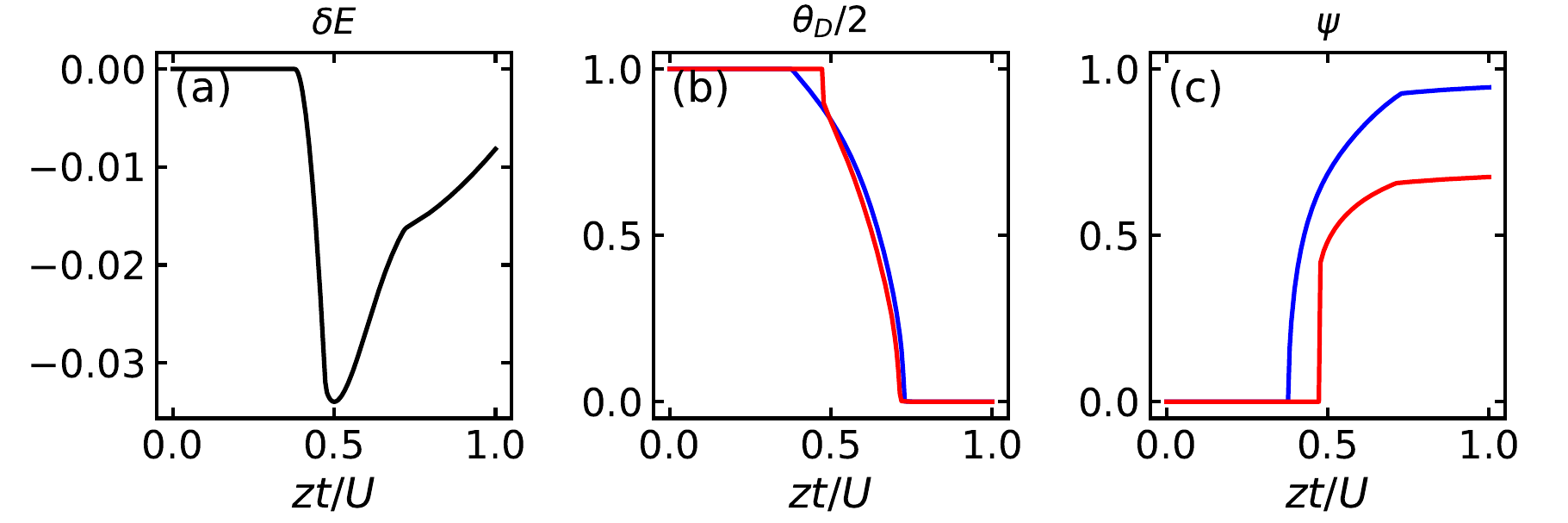}
	\caption{ (a) The energy difference between the homogeneous solution and the phase-separated solution $\delta E$ is plotted. The order parameters (b) $\theta_D$ and (c) $\psi$ are plotted for both homogeneous (red) and phase-separated (blue) solutions. $U_{12}/U =1.1$, $U_s/U_v\approx 4.64$ and $U_L/U=1.41$ is considered.}
	\label{fig:fig3_appendix}
\end{figure}

In Fig.~\ref{fig:fig3_appendix}, we compare the energy density and the order parameters for homogeneous and phase-separated solutions for $U_{12}/U =1.1$. We consider the regime where density order is favoured and find that the phase-separated lattice supersolid and superfluid become the only ground state solutions. This makes the transition from CDW to SS a second order one. In general, we always find that phase-separated solution with full separation (one spin component in one half of the system and the other spin component in the other) has the lowest energy and partial phase separation is never energetically favoured.

\subsection{Phase instability}
\begin{figure}[ht]
	\centering
	\includegraphics[width=\columnwidth]{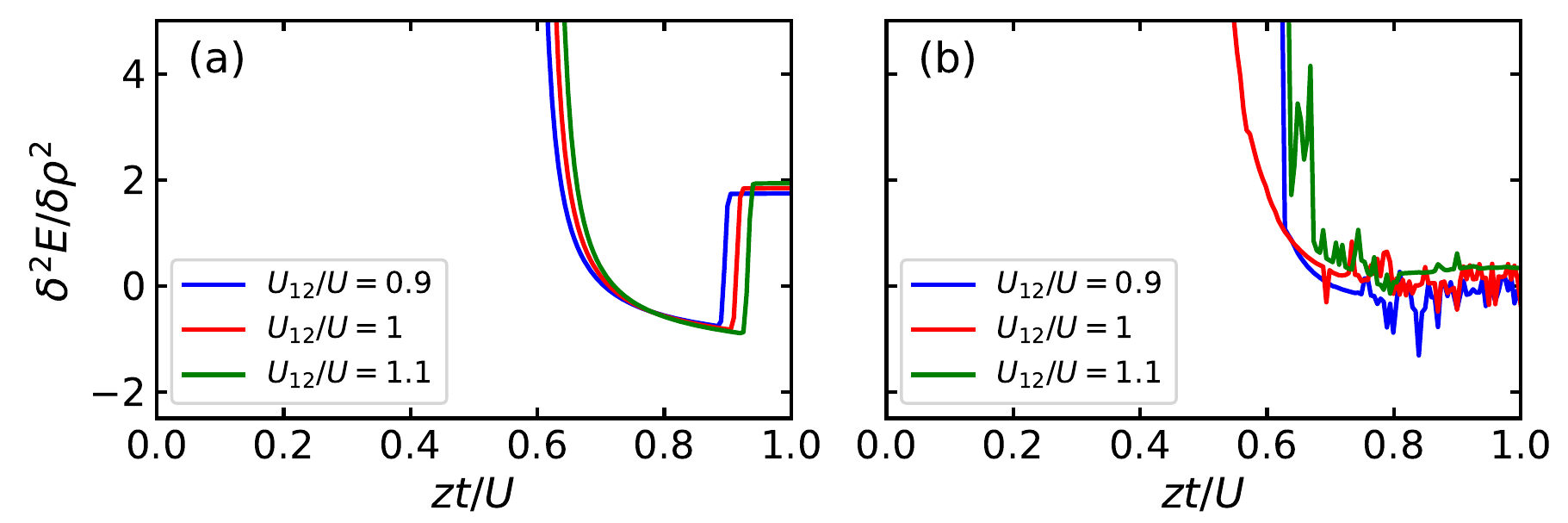}
	\caption{ Second derivative of energy density with respect to the total density $\rho.$ Different colours represent different values of $U_{12}/U$: 0.9 (blue), 1(red) and 1.1 (green). (a) is plotted for $U_s/U_v\approx 4.64$ and $U_L/U=1.93$ and (b) is plotted for $U_s/U_v\approx 0.33$ and $U_L/U=1.93$. For small $zt/U$, $\delta^2 \mathcal{E}/\delta \rho^2$ is positive and large and thus invisible in the plot.}
	\label{fig:fig6_appendix}
\end{figure}

To identify the regime of phase instability, we first determine the minimal energy density $\mathcal{E}(\rho)$ using the ansatzes $\ket{\phi_{e,o}}$ and $\ket{\phi_{e,o}^{D}}$ for a total density $\rho=\rho_{\up}+\rho_{\down}.$ Then, we calculate  the compressibility which is the second derivative of energy with respect to density. Numerically, the second derivative is approximated by 
\begin{align}
	\frac{\delta\mu}{\delta\rho}=\frac{\delta^2\mathcal{E}}{\delta \rho^2} = (\mathcal{E}(\rho + \Delta\rho) + \mathcal{E}(\rho - \Delta\rho) - 2\mathcal{E}(\rho))/(\Delta \rho)^2.
\end{align}
We evaluate the compressibility at unity filling $\rho = \rho_\uparrow + \rho_\downarrow = 2$ and approximate $\Delta \rho = 10^{-2}$ as the discrete change in the total density. The variation of compressibility is shown in Fig.~\ref{fig:fig6_appendix} for two different set of parameters.

We find phase instability in both SS and AF-SS. There is a well-defined regime of instability for the AF-SS (Fig.~\ref{fig:fig6_appendix}(a)) which starts in the AF-SS phase and ends at the transition to the SF phase. The compressibility exhibits a scattered behavior for the SS case (Fig.~\ref{fig:fig6_appendix}(b)), which is compatible with the scattered instability regions in Fig.3(a,b) of the main text. Its origin is very likely due to technical limitations of the numerical method, but goes beyond the scope of this work.

% \subsection{Dynamical Mean-Field}
%
% \begin{itemize}
% \item Description of the dynamical mean-field with chemical potential
% \item Decoupling of tunnelling
% \item Linearizing of the long-range interactions
% \item Discussion of the t=0 limit
% \item Phasediagrams?
% \end{itemize}
% Should we include the dynamical mean-field calculations in the appendix?
\section{Self-Consistent Mean-Field Method}
In this section, we give details on crosschecks we did using mean-field calculations in the grand canonical ensemble (GCE). In the GCE the system after the adiabatic elimination of the cavity field is described by 
\begin{equation}
	\hat{H}_{\mathrm{GCE}}=\hat{H}_{\mathrm{MB}}+\hat{H}_{\mathrm{Long}}-\mu_{\up}\hat{N}_{\up}-\mu_{\down}\hat{N}_{\down}
	\label{eq:H_GCE}
\end{equation}
where the total number of atoms $\braket{\hat{N}_{\up,(\down)}}$ in spin $\up (\down)$ are determined by the chemical potentials $\mu_{\up}$ and $\mu_{\down},$ $\hat{H}_{\mathrm{MB}}$ and $\hat{H}_{\mathrm{Long}}$ are defined in Eq.~(3) and Eq.~(4) of the main text. Starting from Eq.~\eqref{eq:H_GCE} we derive an effective two-site two-spin Hamiltonian by decoupling the tunnelling term \cite{Sheshadri1993superfluid,Oosten2001_quantum_phases} and linearising the long-range interactions \cite{dogra2016phase}. Concretely, we approximate the bosonic annihilation operators and the operators $\hat{\Theta}_{D,S}$ as their expectation values plus fluctuations  
\begin{align}
	\bpp_{e,m}&\rightarrow \psi_{e,m}+\delta\bpp_{e,m} \nonumber \\
	\bpp_{o,m}&\rightarrow \psi_{o,m}+\delta\bpp_{o,m}\\
	\hat{\Theta}_{D,S}&\rightarrow\braket{\hat{\Theta}_{D,S}}+\delta\hat{\Theta}_{D,S} \nonumber
\end{align}
where $\psi_{e(o),m}=\braket{\psi_G|\bpp_{e(o),m}|\psi_G}$ denote the mean-field superfluid order parameters as introduced in the main text. Here, $e$ and $o$ are neighbouring even and odd sites. We assume a homogeneous distribution of the atoms on the two super lattices defined by even and odd sites, which yields $\braket{\psi_G|\hat{\Theta}_{D,S}|\Psi_G}=\frac{K}{2}\theta_{D,S}.$ The density and spin order parameters $\theta_{D,S}$ are given by Eq.~\eqref{eq:theta_D} and Eq.~\eqref{eq:theta_S}. 
Neglecting fluctuations in 2nd order and higher yields an effective decoupled two-site Hamiltonian 
\begin{equation}
	\begin{split}
		&\hat{H}_{\mathrm{eff}}(\psi_{i,m},\theta_S,\theta_D)\\
		&=-zt\sum_{m\in \{\up,\down\}}\bigg( \psi_{o,m}\left(\bpp_{e,m}+\bp_{e,m}\right)\\
		&+\psi_{e,m}\left(\bpp_{o,m}+\bp_{o,m}\right)-2\psi_{e,m}\psi_{o,m}\bigg)\\
		&+\frac{U}{2}\sum_m \left\{\hat{n}_{e,m}\left(\hat{n}_{e,m}-1\right)+\hat{n}_{o,m}\left(\hat{n}_{o,m}-1\right)\right\}\\
		&+U_{12}\left(\hat{n}_{e,\up}\hat{n}_{e,\down}+\hat{n}_{o,\up}\hat{n}_{o,\down}\right)- \sum_{m\in\{\up,\down\}}\mu_m\sum_{i\in\{e,o\}}\hat{n}_{i,m}\\
		&-U_s(\phi)\left(\theta_D\hat{\theta}_D-\frac{\theta_D^2}{2}\right)-U_v(\phi)\left(\theta_S\hat{\theta}_S-\frac{\theta_S^2}{2}\right)
	\end{split}
	\label{eq:H_eff}
\end{equation}

where $z=2D$ denotes the coordination number in a $D$-dimensional lattice (e.g. for a 2-dimensional square-lattice, we get $z=4$).\\
Notably, evaluating $\hat{H}_{\mathrm{eff}}$ on the true ground state 
\begin{equation}
	\mathcal{E}_G=\braket{\psi_G|\hat{H}_{\mathrm{eff}}|\psi_G}
\end{equation}
yields the same energy density as given in Eq.~\eqref{eq:energy_density} (excluding the chemical potential term). Henceforth, both methods are equivalent.\\
The self-consistent Hamiltonian given in Eq.~\eqref{eq:H_eff} can be solved  numerically using an iterative algorithm \cite{dhar2011mean}: At step $i=0,$ we start with an initial guess of the order parameters $\mathbf{X}^{(0)}=\left(\psi^{(0)}_{i,m},\theta_D^{(0)},\theta_S^{(0)}\right).$ We diagonalize the effective Hamiltonian $\hat{H}_{\mathrm{eff}}(\mathbf{X}^{(0)})$ in the Hilbert space ${\mathcal{H}=\mathcal{H}_e\otimes \mathcal{H}_o}$ where $\mathcal{H}_i$ is the  local Hilbert space of site $i=e,o.$ These local Hilbert spaces are spanned by the Fock states $\ket{n,m}_i$ where $n=0,...,n_{\max}$ and $m=0,...,m_{\max}$ denote the number of atoms in spin $\up$ and spin $\down$ on site $i.$ The local basis is truncated to at most $n_{\max} (m_{\max})$ atoms on each site in spin $\up(\down).$ As the effective Hamiltonian, Eq.~\eqref{eq:H_eff}, can be written as ${\hat{H}_{\mathrm{eff}}(\mathbf{X}^{(0)})=\hat{H}_{\mathrm{eff},e}(\mathbf{X}^{(0)})+\hat{H}_{\mathrm{eff},o}(\mathbf{X}^{(0)})}$ with $\hat{H}_{\mathrm{eff},e(o)}(\mathbf{X}^{(0)})$ acting only on site $e(o),$ we can diagonalize each one separately.\\
The diagonalization yields an approximate ground state state $\ket{\psi_G^{(0)}}.$ We update the order parameters $\psi_{i,m}^{(1)}=\braket{ \psi_G^{(0)}|\bpp_{i,m}|\psi_G^{(0)}}$ and $\theta_{D,S}^{(1)}= \braket{ \psi_G^{(0)}|\hat{\theta}_{D,S}|\psi_G^{(0)}}.$
The above described procedure is repeated until $||\mathbf{X}^{(i+1)}-\mathbf{X}^{(i)}||_{\infty}<tol$, where $||\cdot||_{\infty}$ denotes the sup norm and $tol$ is some tolerance level (in our calculations $tol=10^{-8}$).\\
As the algorithm is sensitive to the choice of the initial conditions $\mathbf{X}^{(0)}$, we repeated the iterative procedure for several initial values, partially randomly chosen, and selected the solution with the minimal ground state energy. In our calculations, we chose equal chemical potentials for the two spin species. Furthermore, we restricted our analysis mostly to unity fillings to benchmark the Gutzwiller calculations. In this regime, we truncated the local basis to at most $n_{\max}=n_{\max}=10.$

\subsection{Perturbative Calculations}
The phase boundary displayed in Fig.~2 of the main part is obtained by employing a perturbation theory analysis in the GCE starting from the effective Hamiltonian Eq.~\eqref{eq:H_eff} \cite{Oosten2001_quantum_phases}. We rewrite Eq.~\eqref{eq:H_eff}, 
\begin{equation}
	\hat{H}_{\mathrm{eff}}=\hat{H}_0+\hat{H}_t
\end{equation} 
where 
\begin{align}
	\hat{H}_{t}&=-t\sum_m\left(\psi_{e,m}\left(\bpp_{e,m}+\bp_{e,m}\right)+\psi_{o,m}\left(\bpp_{o,m}+\bp_{o,m}\right)\right) \nonumber \\
	&+2t\sum_m\psi_{e,m}\psi_{o,m}.
\end{align}
denotes the kinetic part of the effective Hamiltonian. In the following we treat the kinetic part $H_t$ as a perturbation and expand the ground state $E_g$ at in the superfluid order parameters $\psi_{i,m}$ up to second order.\\
The eigenstates of $\hat{H}_0$ are Fock states of the form $\ket{\mathbf{n}}=\ket{n_{e,\up},n_{e,\down},n_{o,\up},n_{o,\down}}$ with $n_{i,m}$ denoting the occupation number of the state located at site $i$ and carrying spin $m.$ We denote the ground state at $t=0$ as ${\ket{\mathbf{g}}=\ket{g_{e,\up},g_{e,\down},g_{o,\up},g_{o,\down}}}.$ Its corresponding energy density reads
\begin{equation}
	\begin{split}
		\mathcal{E}_g^{(0)}&=\braket{\mathbf{g}|\hat{H}_0|\mathbf{g}}\\
		&=\frac{1}{2}\sum_m\left(g_{e,m}(g_{e,m}-1)+g_{o,m}(g_{o,m}-1)\right)\\
		&+U_{12}\left(n_{e,\up}n_{e,\down}+n_{o,\up}n_{o,\down}\right)\\
		&-\mu\sum_m\left( n_{e,m}+n_{o,m}\right)\\
		&-\frac{U_s}{2}\theta_D^2-\frac{U_v}{2}\theta_S^2\\
	\end{split}
\end{equation}
where we used that self-consistency requirements
\begin{equation}
	\theta_{D,S}=\braket{\mathbf{g}|\hat{\theta}_{D,S}|\mathbf{g}}.
\end{equation}
The second order expansion of the energy density reads
\begin{equation}
	\mathcal{E}^{(2)}=\mathcal{E}^{(0)}_g+\braket{\mathbf{g}|\hat{H}_t|\mathbf{g}}-\sum_{\mathbf{n}\neq \mathbf{g}}\frac{|\braket{\mathbf{n}|\hat{H}_t|\mathbf{g}}|^2}{\mathcal{E}^{(0)}(\mathbf{n},\theta_D,\theta_S)-\mathcal{E}^{(0)}_g}
\end{equation}
with $\mathcal{E}^{(0)}(\mathbf{n},\theta_D,\theta_S)=\braket{\mathbf{n}|\hat{H}_0|\mathbf{n}}.$
We obtain
\begin{equation}
	\mathcal{E}^{(2)}=E^{(0)}_g+2t\sum_m \psi_{e,m}\psi_{o,m}-t^2\sum_m\left( \psi_{e,m}^2 f_{o,m}+\psi_{o,m}^2f_{e,m}\right)
\end{equation}
where ${f_{i,m}=\frac{g_{i,m}}{\Delta E(g_{i,m}-1)}+\frac{g_{i,m}+1}{\Delta \mathcal{E}(g_{i,m}+1)}.}$ Here,
$${\Delta \mathcal{E}(g_{e,\up}\pm 1)=\mathcal{E}^{(0)}\left(\left(g_{e,\up}\pm 1,g_{e,\down},g_{o,\up},g_{o,\down}\right) ,\theta_D,\theta_S\right)-\mathcal{E}_g}$$ denotes the energy density difference when adding or removing one atom with spin $\up$ on site $e$. Similar, ${\Delta \mathcal{E}(g_{i,m}\pm 1)}$ is defined. To determine the insulating to superfluid transition we determine where the point $\psi_{i,m}=0$ changes from being a local minima to a local maximum of $\mathcal{E}^{(2)}$ when varying $zt/U.$ Hereby, we make additional assumptions on $\psi_{i,m}$ dependent on the underlying ground state $\mathbf{g}$ at $t/U=0.$ For the AFM phase  we set $\mathbf{g}_{\mathrm{AFM}}=\left(1,0,0,1\right)$ and $\theta_S=2$ and $\theta_D=0.$ Further, we assume $\psi_{e,\up}= \psi_{e,\down}.$ With the additional assumptions $E^{(2)}$ becomes a function of only two variables.  In this case, we have
\begin{align}
	f_{e,\up}(\mathrm{AFM})&=\frac{1}{\mu/U+2U_v/U}+\frac{2}{1-\mu/U-2U_v/U} \nonumber\\
	f_{e,\down}(\mathrm{AFM})&=\frac{1}{U_{12}/U-\mu+2U_v/U}. \
\end{align}
and $f_{o,\down}(\mathrm{AFM})=f_{e,\up}(\mathrm{AFM})$ and $f_{o,\up}(\mathrm{AFM})=f_{e,\down}(\mathrm{AFM}).$
The determinant of the Hessian $\mathrm{Hess}(\mathcal{E}^{(2)})$ changes its sign at 
\begin{equation}
	\frac{zt_c^{(\mathrm{AFM})}}{U}(\mu/U,U_v/U)=\frac{1}{\sqrt{f_{e,\up}(\mathrm{AFM})f_{o,\up}(\mathrm{AFM})}}.
\end{equation} Thus, 
$t_{c}^{(\mathrm{AFM})}(\mu/U,U_v/U)$ determines AFM to AF-SS phase boundary (see black line in Fig.~\ref{fig:6_appendix} (a)).\\
For the CDW to SS transition we find that boundary matches the numerical results when assuming a polarized state $\mathbf{g}_{\mathrm{CDW}}=(2,0,0,0)$ (and $\theta_D=2$ and $\theta_S=0$) for $U_{12}/U\geq 1$. In this case the problem reduces to a uniform mixture. The corresponding analysis is already discussed in \cite{dogra2016phase}. For completeness, we restate here the values $f_{e,(\up)}(\mathrm{CDW})$ and $f_{o,(\up)}(\mathrm{CDW})$
\begin{align}
	f_{e,(\up)}(\mathrm{CDW})&=\frac{2}{\mu/U+2U_s/U - 1}+\frac{3}{2-2U_s/U-\mu/U} \nonumber\\
	f_{o,(\up)}(\mathrm{CDW})&=\frac{1}{-\mu/U-2U_s/U}.
\end{align} 
Similar, as above the CDW to SS phase boundary is given by $zt_c^{(\mathrm{CDW})}/U=\frac{1}{\sqrt{f_{e,\up}(\mathrm{CDW})f_{o,\up}(\mathrm{CDW})}}.$ We find that the boundary, obtained by assuming a mixed CDW phase, $\ket{\up\down,0},$ does not match the numerical results.\\
To obtain the AFM to AF-SS and the CDW to SS phase boundary in the CE,  we maximize $zt_c^{(\mathrm{AFM})}/U$ and $zt_c^{(\mathrm{CDW})}/U$ as a function of $\mu/U.$

\subsection{Comparison with the Gutzwiller Ansatz}
\begin{figure}[H]
	\includegraphics[width=\columnwidth]{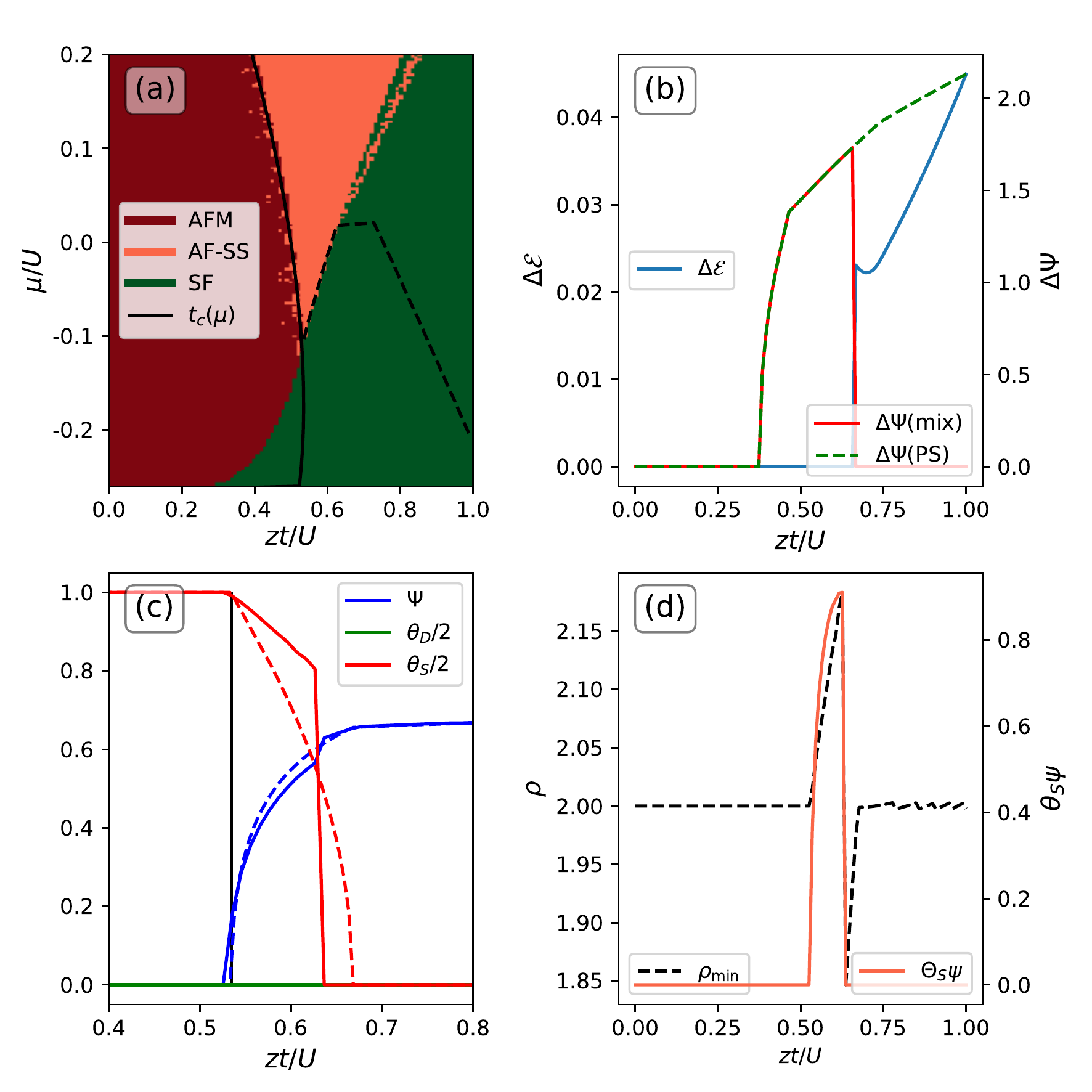}
	\caption{(a) Phase diagram at $U_L/U=1.3,$ $U_s/U_v=0.33$ and $U_{12}/U=1,$ obtained for $n_{\max}=m_{\max}=10.$ (b) Energy density difference $\Delta\mathcal{E}=\mathcal{E}(\mathrm{mix})-\mathcal{E}(PS)$ between homogeneous ($\mathcal{E}(\mathrm{mix}$) and polarized ($\mathcal{E}(\mathrm{PS}$) solutions along a line of constant chemical potential $\mu/U=0.1525$ at $U_L/U=0.7,$  $U_v/U=0$ and $U_{12}/U=1.1$ (left vertical axis). The former is obtained by assuming mixed starting values with $\Delta\rho=0,$ the latter by assuming PS starting values. Difference in the superfluid order parameters $\Delta \psi=|\psi_{e,\up}+\psi_{o,\up}-\psi_{e,\down}-\psi_{o,\down}|$ (left vertical axis) for the mixed solution (solid) and the phase separated solution (dotted). (c) Order parameters (solid) along a line of approximate constant density $\rho_{\min}\approx 2$ ( black dotted line in (a)) together with the solutions at fixed density from the Gutzwiller ansatz used in the main text (dotted). The vertical black line indicates the AFM to AF-SS  transition predicted from perturbation theory. (d) Closest density $\rho_{\min}$ to unity filling (left axis) and AF-SS order parameter $\theta_s\psi$ following the line of closest density (right axis).
	}
	\label{fig:6_appendix}
\end{figure}

We systematically compare the calculations in the GCE with the ones in the CE. Different than in the CE the density $\rho$ is not a priori constrained but determined by the chemical potential. We also do not make any a priori assumptions on the form of the local wave function and allow for a significant higher local occupation number. To compare to the results in the CE, we choose values of $\mu,$ which resulted in a density of (approximate) $\rho=2,$ if possible. Here, we give examples for $U_s/U_v=0.33$ and $U_L/U=1.3$ (Fig.~2 (g) in the main text) and $U_s/U_v=4.46$ and $U_L/U=1.4$ (Fig.~2(f)). For $U_s/U_v=0.33$ and $U_{12}/U=1$ the phase diagram in the regime of $\rho=2$ is displayed in Fig.~\ref{fig:6_appendix}(a) as $zt/U$ and $\mu/U$ are varied. At $zt/U\ll 1$ the system is in an incompressible AFM phase forming a characteristic lobe. With increasing $zt/U$ the system enters either a SF or a AF-SS phase depending on the value of $\mu/U.$  The AFM to AF-SS transition is 2nd-order and matches the prediction from the perturbative analysis discussed above. The AFM to SF transition is of 1st-order. Remarkably, outside the AFM lobe there is a region where the density for any value of $\mu$ is never conserved (Fig.~\ref{fig:fig6_appendix}(d)). This indicates that the AF-SS phase is nowhere stable in the GCE. The non-conserved density in the AF-SS phase makes a direct comparison with the Gutzwiller Ansatz difficult (Fig.~\ref{fig:fig6_appendix})(d)). The tip of the AFM lobe is at a slightly lower $zt/U$ than where the AFM to AF-SS transition in the CE occurs. However, the maximum point of the transition line from perturbation theory matches the results in the CE ensemble. In the superfluid phase, the order parameters of the two methods agree well with each other. The SF order parameters in the GCE are slightly larger due to the larger local Hilbert space dimensions. \\
For $U_s/U_v=4.64,$ $U_L/U=1.4$ and $U_{12}/U \in \{0.9,1,1.1\}$ there are density ordered phases (CDW and SS) in the CE. Since both $\theta_S$ and $\theta_D$ are simultaneous maximized for the polarized configuration $\mathbf{n}=(2,0,0,0),$ the self-consistent method yields phase separated solutions. In order to study also the possibility of mixture in this regime, we set $U_v=0.$ Similarly to the CE, we find a mixed CDW and a SS phase at $U_{12}/U<1$ and a PS CDW and a PS-SS phase at $U_{12}/U\geq 1.$ Additionally, for $U_{12}/U>1$ we obtain a PS SF. In Fig.~\ref{fig:fig6_appendix}(b), we show the energy densities along a slice of constant $\mu/U$ for $U_{12}/U$ and $n_{\max}=m_{\max}=10,$ obtained by assuming either homogeneous starting values with $\Delta\rho=0$ or phase separated starting values with $\Delta\rho>0.$ In the CDW and the SS phase both the initially homogeneous starting values and the PS ones converge to fully PS solutions, indicating that the PS SS phase remains the energetically favourable for $U_{12}/U>1$ even with increased Hilbert-space. In the SF, the homogeneous SF phase is a metastable state with a slightly higher energy than the PS SF.

\section{Effective theory for AFM excitations}

The ground state of the antiferromagnetic (AFM) phase is defined within a square lattice. The lattice vectors are given by ${\bf a}_1=\left({\bf e}_x+{\bf e}_z\right)\lambda/2$ and ${\bf a}_2=\left({\bf e}_x-{\bf e}_z\right)\lambda/2$, yielding the reciprocal lattice vectors ${\bf b}_1=\left({\bf e}_x+{\bf e}_z\right) \frac{2\pi}{a}$ and ${\bf b}_2=\left({\bf e}_x-{\bf e}_z\right) \frac{2\pi}{a}$ with $a=\lambda/2$. The unit cell contains two spins, either at $\left(R_\uparrow=0; R_\downarrow=a{\bf e}_x\right)$ or $\left(R_\downarrow=0; R_\uparrow=a{\bf e}_x\right)$, depending on the way the global even-odd site symmetry is broken. 

\renewcommand{\arraystretch}{1.7}
\begin{table}[!htb]
	\centering
	\begin{tabular}{c|c|c|c|}
		\hline
		\multicolumn{4}{|c|}{\textbf{Tunneling to nearest neighbors}}    	 					 														 	 \\ \hline
		& $E_{t=0}$					  & Energy shift                 			& Effective tunneling rate   					\\ \hline
		\multicolumn{1}{|c|}{P2} & $(1+4{\bar U}_v)$                 & $+\frac{t^2}{2U_v}$                     & $+\frac{t^2}{2U_v}$       					 \\ 
		\multicolumn{1}{|c|}{P1} & $1$                            				& $-\frac{t^2}{2U_v}$                     & $-\frac{t^2}{2U_v}$          					 \\ 
		\multicolumn{1}{|c|}{S}  & $4{\bar U}_v$                         & $-\frac{(3-4{\bar U}_v)t^2}{(1-4{\bar U}_v)(1+4{\bar U}_v)}$ & $-\frac{(1+12U_v)t^4}{2U_v(1-4{\bar U}_v)(1+4{\bar U}_v)^2}$ \\ 
		\multicolumn{1}{|c|}{H}  & $0$                            				& $-\frac{2t^2}{(1+4{\bar U}_v)}$                & $-\frac{2t^2}{(1+4{\bar U}_v)}$                       \\ \hline
		
		\multicolumn{4}{c}{} \\
		\hline
		\multicolumn{4}{|c|}{\textbf{Tunneling to next-nearest neighbors}}    	 					 														 \\ \hline
		& $E_{t=0}$					  & Energy shift                 			& Effective tunneling rate   					 \\ \hline
		\multicolumn{1}{|c|}{P2} & $(1+4{\bar U}_v)$                     & $+\frac{t^2}{4U_v}$                     & $+\frac{t^2}{4U_v}$       					 \\ 
		\multicolumn{1}{|c|}{P1} & $1$                            & $-\frac{t^2}{4U_v}$                     & $-\frac{t^2}{4U_v}$          					 \\ 
		\multicolumn{1}{|c|}{S}  & $4{\bar U}_v$                         & $-\frac{(3-4{\bar U}_v)t^2}{(1-4{\bar U}_v)(1+4{\bar U}_v)}$ & $-\frac{t^4}{2U_v(1-4{\bar U}_v)(1+4{\bar U}_v)}$			 \\ 
		\multicolumn{1}{|c|}{H}  & $0$                            & $-\frac{t^2}{(1+4{\bar U}_v)}$                 & $-\frac{t^2}{(1+4{\bar U}_v)}$                        \\ \hline
	\end{tabular}
	\caption{Effective quasiparticle energy shifts and tunneling rates, obtained via first order perturbation theory. Here ${\bar U}_v \equiv U_v/U$. The results are obtained by including up to forth-order processes and adiabatically eliminating all the intermediate states.}
	\label{table:3}
\end{table}
\renewcommand{\arraystretch}{1}

\begin{figure}[H]
	\centering
	\includegraphics[width=0.9\columnwidth]{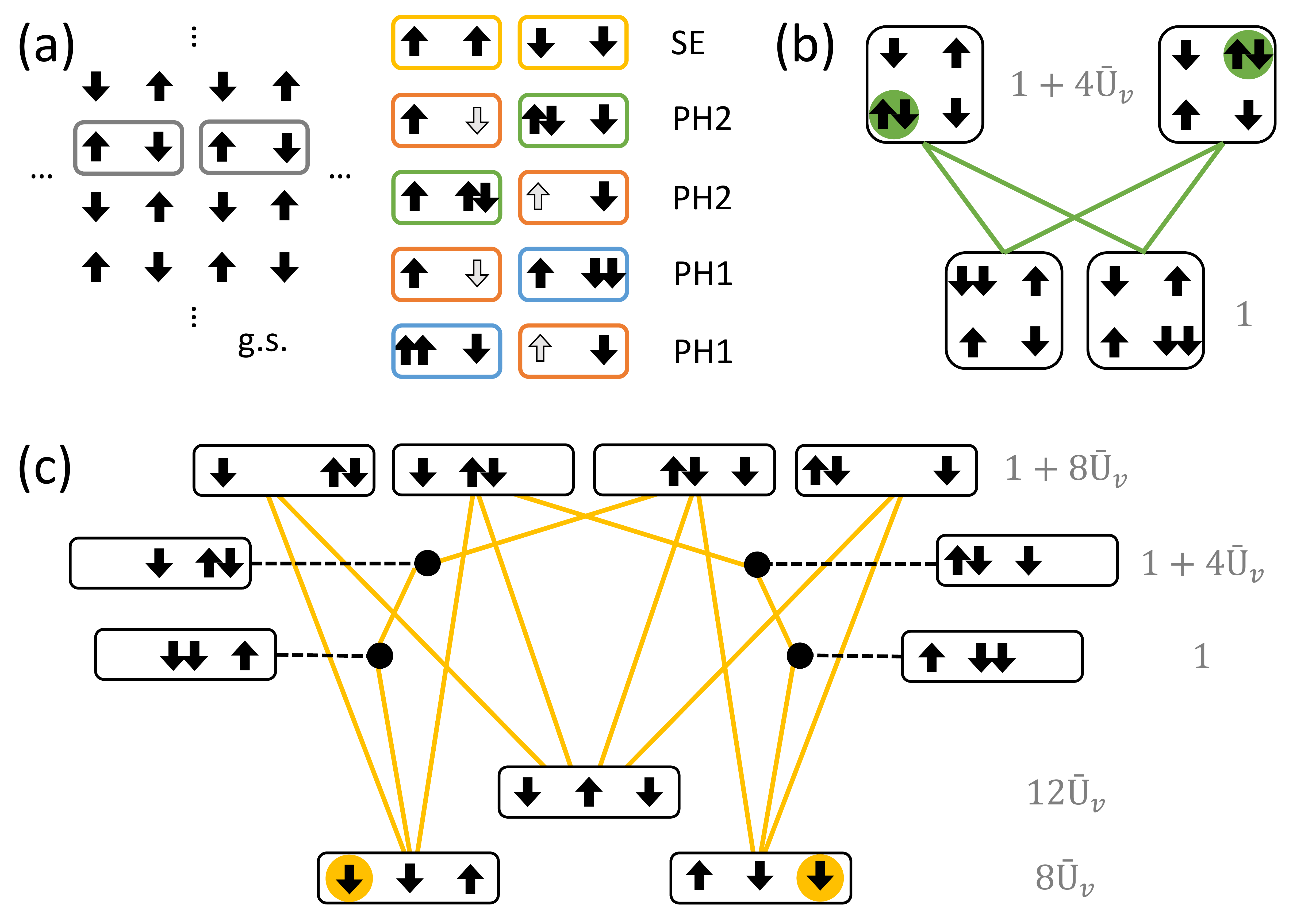}
	\caption{(a) Possible quasiparticle pairs, corresponding to the different low-lying excitations above the AFM ground state (g.s.), on a pair of superlattice unit cells (denoted with grey boxes on the g.s. lattice). The SE excitation is composed of two spin (S) quasiparticles, the PH2 excitation is composed of a particle (P2) and hole (H) quasiparticle, the PH1 excitation is composed of a particle (P1) and hole (H) quasiparticle. The color coding is consistent with the main part of the manuscript.
		The effective energy shifts and tunneling rates of the quasiparticle pairs are obtained via first order perturbation theory, by constructing graphs that connect the inital and final state through up-to-three intermediate virtual states. Examples of graphs for (b) hopping of particle excitation P2 to first neighbor, and (c) hopping of spin excitation S to second neighbor. }
	\label{fig:fig5_appendix}
\end{figure}

Each of the lowest excitations above the ground state (particle-hole PH1 and PH2, and spin-exchange SE) corresponds to a specific change in two of the unit cells of a square superlattice. An excitation can be seen as two quasiparticles that are always created in pair and are free to delocalize on the square superlattice defined for the ground state. The quasiparticle pairs for a pair of superlattice unit cells are shown in Fig.~\ref{fig:fig5_appendix}(a). The PH1 excitation is composed of a particle (P1) and corresponding hole (H) quasiparticle, the PH2 excitation is composed of a particle (P2) and corresponding hole (H) quasiparticle, while the SE excitation is composed of two aligned-spin (S) quasiparticles. The energies of the quasiparticles in the $t=0$ limit are shown in the first column of Tab.~\ref{table:3}.

The quasiparticles can independently tunnel through the square superlattice with different effective tunneling rates $t'$, which shifts their energy as compared to the $t=0$ limit, $E_{t=0}\rightarrow E_{t=0}+\Delta E$. The effective energies and tunneling rates of the quasiparticles to nearest and next-nearest neighbors on the square superlattice, obtained through first-order perturbation theory and by taking into account all processes up to forth order, are listed in Tab.~\ref{table:3}.
Examples of graphs used to evaluate the effective energy shifts and tunneling rates of quasiparticles are shown in~\ref{fig:fig5_appendix}(b,c). The results are calculated by adiabatically eliminating all states but the initial and final, i.e. by assuming no population change in all the other states. Such an approach is valid as long as the tunneling rate $t$ is small compared to the energy gaps between the initial/final states and each of the adiabatically eliminating virtual states, i.e. $t\ll 4U_v$ and $t\ll U$.

The transition elements between the excitation quasiparticles in the non-interacting theory are vanishing for all pairs except P1-P2: A P1 quasiparticle can transition to a P2 quasiparticle through a tunneling process of a spin within the unit cell (tunneling rate $t$). On the other hand, a P1 quasiparticle could only mix with an S quasiparticle through a process in which a P1 and H quasiparticles interact, after which to create two S quasiparticles. We do not consider the quasiparticle interaction terms in the effective Hamiltonian.

\begin{thebibliography}{68}%
	\makeatletter
	\providecommand \@ifxundefined [1]{%
		\@ifx{#1\undefined}
	}%
	\providecommand \@ifnum [1]{%
		\ifnum #1\expandafter \@firstoftwo
		\else \expandafter \@secondoftwo
		\fi
	}%
	\providecommand \@ifx [1]{%
		\ifx #1\expandafter \@firstoftwo
		\else \expandafter \@secondoftwo
		\fi
	}%
	\providecommand \natexlab [1]{#1}%
	\providecommand \enquote  [1]{``#1''}%
	\providecommand \bibnamefont  [1]{#1}%
	\providecommand \bibfnamefont [1]{#1}%
	\providecommand \citenamefont [1]{#1}%
	\providecommand \href@noop [0]{\@secondoftwo}%
	\providecommand \href [0]{\begingroup \@sanitize@url \@href}%
	\providecommand \@href[1]{\@@startlink{#1}\@@href}%
	\providecommand \@@href[1]{\endgroup#1\@@endlink}%
	\providecommand \@sanitize@url [0]{\catcode `\\12\catcode `\$12\catcode
		`\&12\catcode `\#12\catcode `\^12\catcode `\_12\catcode `\%12\relax}%
	\providecommand \@@startlink[1]{}%
	\providecommand \@@endlink[0]{}%
	\providecommand \url  [0]{\begingroup\@sanitize@url \@url }%
	\providecommand \@url [1]{\endgroup\@href {#1}{\urlprefix }}%
	\providecommand \urlprefix  [0]{URL }%
	\providecommand \Eprint [0]{\href }%
	\providecommand \doibase [0]{https://doi.org/}%
	\providecommand \selectlanguage [0]{\@gobble}%
	\providecommand \bibinfo  [0]{\@secondoftwo}%
	\providecommand \bibfield  [0]{\@secondoftwo}%
	\providecommand \translation [1]{[#1]}%
	\providecommand \BibitemOpen [0]{}%
	\providecommand \bibitemStop [0]{}%
	\providecommand \bibitemNoStop [0]{.\EOS\space}%
	\providecommand \EOS [0]{\spacefactor3000\relax}%
	\providecommand \BibitemShut  [1]{\csname bibitem#1\endcsname}%
	\let\auto@bib@innerbib\@empty
	%</preamble>
	\bibitem [{\citenamefont {Jaksch}\ \emph {et~al.}(1998)\citenamefont {Jaksch},
		\citenamefont {Bruder}, \citenamefont {Cirac}, \citenamefont {Gardiner},\
		and\ \citenamefont {Zoller}}]{jaksch1998cold}%
	\BibitemOpen
	\bibfield  {author} {\bibinfo {author} {\bibfnamefont {D.}~\bibnamefont
			{Jaksch}}, \bibinfo {author} {\bibfnamefont {C.}~\bibnamefont {Bruder}},
		\bibinfo {author} {\bibfnamefont {J.~I.}\ \bibnamefont {Cirac}}, \bibinfo
		{author} {\bibfnamefont {C.~W.}\ \bibnamefont {Gardiner}},\ and\ \bibinfo
		{author} {\bibfnamefont {P.}~\bibnamefont {Zoller}},\ }\bibfield  {title}
	{\bibinfo {title} {{Cold Bosonic Atoms in Optical Lattices}},\ }\href
	{https://doi.org/10.1103/physrevlett.81.3108} {\bibfield  {journal} {\bibinfo
			{journal} {Physical Review Letters}\ }\textbf {\bibinfo {volume} {81}},\
		\bibinfo {pages} {3108} (\bibinfo {year} {1998})}\BibitemShut {NoStop}%
	\bibitem [{\citenamefont {Greiner}\ \emph {et~al.}(2002)\citenamefont
		{Greiner}, \citenamefont {Mandel}, \citenamefont {Esslinger}, \citenamefont
		{H{\"{a}}nsch},\ and\ \citenamefont {Bloch}}]{greiner2002quantum}%
	\BibitemOpen
	\bibfield  {author} {\bibinfo {author} {\bibfnamefont {M.}~\bibnamefont
			{Greiner}}, \bibinfo {author} {\bibfnamefont {O.}~\bibnamefont {Mandel}},
		\bibinfo {author} {\bibfnamefont {T.}~\bibnamefont {Esslinger}}, \bibinfo
		{author} {\bibfnamefont {T.~W.}\ \bibnamefont {H{\"{a}}nsch}},\ and\ \bibinfo
		{author} {\bibfnamefont {I.}~\bibnamefont {Bloch}},\ }\bibfield  {title}
	{\bibinfo {title} {{Quantum phase transition from a superfluid to a Mott
				insulator in a gas of ultracold atoms}},\ }\href
	{https://doi.org/10.1038/415039a} {\bibfield  {journal} {\bibinfo  {journal}
			{Nature}\ }\textbf {\bibinfo {volume} {415}},\ \bibinfo {pages} {39}
		(\bibinfo {year} {2002})}\BibitemShut {NoStop}%
	\bibitem [{\citenamefont {Gross}\ and\ \citenamefont
		{Bloch}(2017)}]{gross2017quantum}%
	\BibitemOpen
	\bibfield  {author} {\bibinfo {author} {\bibfnamefont {C.}~\bibnamefont
			{Gross}}\ and\ \bibinfo {author} {\bibfnamefont {I.}~\bibnamefont {Bloch}},\
	}\bibfield  {title} {\bibinfo {title} {{Quantum simulations with ultracold
				atoms in optical lattices}},\ }\href@noop {} {\bibfield  {journal} {\bibinfo
			{journal} {Science}\ }\textbf {\bibinfo {volume} {357}},\ \bibinfo {pages}
		{995} (\bibinfo {year} {2017})}\BibitemShut {NoStop}%
	\bibitem [{\citenamefont {Sch{\"{a}}fer}\ \emph {et~al.}(2020)\citenamefont
		{Sch{\"{a}}fer}, \citenamefont {Fukuhara}, \citenamefont {Sugawa},
		\citenamefont {Takasu},\ and\ \citenamefont {Takahashi}}]{schaefer2020tools}%
	\BibitemOpen
	\bibfield  {author} {\bibinfo {author} {\bibfnamefont {F.}~\bibnamefont
			{Sch{\"{a}}fer}}, \bibinfo {author} {\bibfnamefont {T.}~\bibnamefont
			{Fukuhara}}, \bibinfo {author} {\bibfnamefont {S.}~\bibnamefont {Sugawa}},
		\bibinfo {author} {\bibfnamefont {Y.}~\bibnamefont {Takasu}},\ and\ \bibinfo
		{author} {\bibfnamefont {Y.}~\bibnamefont {Takahashi}},\ }\bibfield  {title}
	{\bibinfo {title} {{Tools for quantum simulation with ultracold atoms in
				optical lattices}},\ }\href {https://doi.org/10.1038/s42254-020-0195-3}
	{\bibfield  {journal} {\bibinfo  {journal} {Nature Reviews Physics}\ }\textbf
		{\bibinfo {volume} {2}},\ \bibinfo {pages} {411} (\bibinfo {year}
		{2020})}\BibitemShut {NoStop}%
	\bibitem [{\citenamefont {Vuletic}\ \emph {et~al.}(1999)\citenamefont
		{Vuletic}, \citenamefont {Kerman}, \citenamefont {Chin},\ and\ \citenamefont
		{Chu}}]{Vuletic1999}%
	\BibitemOpen
	\bibfield  {author} {\bibinfo {author} {\bibfnamefont {V.}~\bibnamefont
			{Vuletic}}, \bibinfo {author} {\bibfnamefont {A.~J.}\ \bibnamefont {Kerman}},
		\bibinfo {author} {\bibfnamefont {C.}~\bibnamefont {Chin}},\ and\ \bibinfo
		{author} {\bibfnamefont {S.}~\bibnamefont {Chu}},\ }\bibfield  {title}
	{\bibinfo {title} {{Observation of low-field feshbach resonances in
				collisions of cesium atoms}},\ }\href
	{https://doi.org/10.1103/PhysRevLett.82.1406} {\bibfield  {journal} {\bibinfo
			{journal} {Physical Review Letters}\ }\textbf {\bibinfo {volume} {82}},\
		\bibinfo {pages} {1406} (\bibinfo {year} {1999})}\BibitemShut {NoStop}%
	\bibitem [{\citenamefont {Moses}\ \emph {et~al.}(2017)\citenamefont {Moses},
		\citenamefont {Covey}, \citenamefont {Miecnikowski}, \citenamefont {Jin},\
		and\ \citenamefont {Ye}}]{Moses2017}%
	\BibitemOpen
	\bibfield  {author} {\bibinfo {author} {\bibfnamefont {S.~A.}\ \bibnamefont
			{Moses}}, \bibinfo {author} {\bibfnamefont {J.~P.}\ \bibnamefont {Covey}},
		\bibinfo {author} {\bibfnamefont {M.~T.}\ \bibnamefont {Miecnikowski}},
		\bibinfo {author} {\bibfnamefont {D.~S.}\ \bibnamefont {Jin}},\ and\ \bibinfo
		{author} {\bibfnamefont {J.}~\bibnamefont {Ye}},\ }\bibfield  {title}
	{\bibinfo {title} {{New frontiers for quantum gases of polar molecules}},\
	}\href {https://doi.org/10.1038/nphys3985} {\bibfield  {journal} {\bibinfo
			{journal} {Nature Physics}\ }\textbf {\bibinfo {volume} {13}},\ \bibinfo
		{pages} {13} (\bibinfo {year} {2017})},\ \Eprint
	{https://arxiv.org/abs/1610.07711} {1610.07711} \BibitemShut {NoStop}%
	\bibitem [{\citenamefont {Lahaye}\ \emph {et~al.}(2009)\citenamefont {Lahaye},
		\citenamefont {Menotti}, \citenamefont {Santos}, \citenamefont {Lewenstein},\
		and\ \citenamefont {Pfau}}]{Lahaye2009}%
	\BibitemOpen
	\bibfield  {author} {\bibinfo {author} {\bibfnamefont {T.}~\bibnamefont
			{Lahaye}}, \bibinfo {author} {\bibfnamefont {C.}~\bibnamefont {Menotti}},
		\bibinfo {author} {\bibfnamefont {L.}~\bibnamefont {Santos}}, \bibinfo
		{author} {\bibfnamefont {M.}~\bibnamefont {Lewenstein}},\ and\ \bibinfo
		{author} {\bibfnamefont {T.}~\bibnamefont {Pfau}},\ }\bibfield  {title}
	{\bibinfo {title} {{The physics of dipolar bosonic quantum gases}},\ }\href
	{https://doi.org/10.1088/0034-4885/72/12/126401} {\bibfield  {journal}
		{\bibinfo  {journal} {Reports on Progress in Physics}\ }\textbf {\bibinfo
			{volume} {72}},\ \bibinfo {pages} {126401} (\bibinfo {year} {2009})},\
	\Eprint {https://arxiv.org/abs/0905.0386} {0905.0386} \BibitemShut {NoStop}%
	\bibitem [{\citenamefont {Browaeys}\ \emph {et~al.}(2016)\citenamefont
		{Browaeys}, \citenamefont {Barredo},\ and\ \citenamefont
		{Lahaye}}]{Browaeys2016}%
	\BibitemOpen
	\bibfield  {author} {\bibinfo {author} {\bibfnamefont {A.}~\bibnamefont
			{Browaeys}}, \bibinfo {author} {\bibfnamefont {D.}~\bibnamefont {Barredo}},\
		and\ \bibinfo {author} {\bibfnamefont {T.}~\bibnamefont {Lahaye}},\
	}\bibfield  {title} {\bibinfo {title} {{Experimental investigations of
				dipole–dipole interactions between a few Rydberg atoms}},\ }\href
	{https://doi.org/10.1088/0953-4075/49/15/152001} {\bibfield  {journal}
		{\bibinfo  {journal} {Journal of Physics B: Atomic, Molecular and Optical
				Physics}\ }\textbf {\bibinfo {volume} {49}},\ \bibinfo {pages} {152001}
		(\bibinfo {year} {2016})}\BibitemShut {NoStop}%
	\bibitem [{\citenamefont {Ritsch}\ \emph {et~al.}(2013)\citenamefont {Ritsch},
		\citenamefont {Domokos}, \citenamefont {Brennecke},\ and\ \citenamefont
		{Esslinger}}]{ritsch2013cold}%
	\BibitemOpen
	\bibfield  {author} {\bibinfo {author} {\bibfnamefont {H.}~\bibnamefont
			{Ritsch}}, \bibinfo {author} {\bibfnamefont {P.}~\bibnamefont {Domokos}},
		\bibinfo {author} {\bibfnamefont {F.}~\bibnamefont {Brennecke}},\ and\
		\bibinfo {author} {\bibfnamefont {T.}~\bibnamefont {Esslinger}},\ }\bibfield
	{title} {\bibinfo {title} {Cold atoms in cavity-generated dynamical optical
			potentials},\ }\href {https://doi.org/10.1103/RevModPhys.85.553} {\bibfield
		{journal} {\bibinfo  {journal} {Rev. Mod. Phys.}\ }\textbf {\bibinfo {volume}
			{85}},\ \bibinfo {pages} {553} (\bibinfo {year} {2013})}\BibitemShut
	{NoStop}%
	\bibitem [{\citenamefont {Mivehvar}\ \emph {et~al.}(2021)\citenamefont
		{Mivehvar}, \citenamefont {Piazza}, \citenamefont {Donner},\ and\
		\citenamefont {Ritsch}}]{mivehvar2021cavityqed}%
	\BibitemOpen
	\bibfield  {author} {\bibinfo {author} {\bibfnamefont {F.}~\bibnamefont
			{Mivehvar}}, \bibinfo {author} {\bibfnamefont {F.}~\bibnamefont {Piazza}},
		\bibinfo {author} {\bibfnamefont {T.}~\bibnamefont {Donner}},\ and\ \bibinfo
		{author} {\bibfnamefont {H.}~\bibnamefont {Ritsch}},\ }\bibfield  {title}
	{\bibinfo {title} {{Cavity QED with quantum gases: new paradigms in many-body
				physics}},\ }\href {https://doi.org/10.1080/00018732.2021.1969727} {\bibfield
		{journal} {\bibinfo  {journal} {Advances in Physics}\ }\textbf {\bibinfo
			{volume} {70}},\ \bibinfo {pages} {1} (\bibinfo {year} {2021})}\BibitemShut
	{NoStop}%
	\bibitem [{\citenamefont {Li}\ \emph {et~al.}(2013)\citenamefont {Li},
		\citenamefont {He},\ and\ \citenamefont {Hofstetter}}]{li2013lattice}%
	\BibitemOpen
	\bibfield  {author} {\bibinfo {author} {\bibfnamefont {Y.}~\bibnamefont
			{Li}}, \bibinfo {author} {\bibfnamefont {L.}~\bibnamefont {He}},\ and\
		\bibinfo {author} {\bibfnamefont {W.}~\bibnamefont {Hofstetter}},\ }\bibfield
	{title} {\bibinfo {title} {Lattice-supersolid phase of strongly correlated
			bosons in an optical cavity},\ }\href
	{https://doi.org/10.1103/PhysRevA.87.051604} {\bibfield  {journal} {\bibinfo
			{journal} {Phys. Rev. A}\ }\textbf {\bibinfo {volume} {87}},\ \bibinfo
		{pages} {051604} (\bibinfo {year} {2013})}\BibitemShut {NoStop}%
	\bibitem [{\citenamefont {Bakhtiari}\ \emph {et~al.}(2015)\citenamefont
		{Bakhtiari}, \citenamefont {Hemmerich}, \citenamefont {Ritsch},\ and\
		\citenamefont {Thorwart}}]{Bakhtiari2015nonequilibrium}%
	\BibitemOpen
	\bibfield  {author} {\bibinfo {author} {\bibfnamefont {M.~R.}\ \bibnamefont
			{Bakhtiari}}, \bibinfo {author} {\bibfnamefont {A.}~\bibnamefont
			{Hemmerich}}, \bibinfo {author} {\bibfnamefont {H.}~\bibnamefont {Ritsch}},\
		and\ \bibinfo {author} {\bibfnamefont {M.}~\bibnamefont {Thorwart}},\
	}\bibfield  {title} {\bibinfo {title} {Nonequilibrium phase transition of
			interacting bosons in an intra-cavity optical lattice},\ }\href
	{https://doi.org/10.1103/PhysRevLett.114.123601} {\bibfield  {journal}
		{\bibinfo  {journal} {Phys. Rev. Lett.}\ }\textbf {\bibinfo {volume} {114}},\
		\bibinfo {pages} {123601} (\bibinfo {year} {2015})}\BibitemShut {NoStop}%
	\bibitem [{\citenamefont {Caballero-Benitez}\ and\ \citenamefont
		{Mekhov}(2015)}]{CaballeroBenitez2015Quantum}%
	\BibitemOpen
	\bibfield  {author} {\bibinfo {author} {\bibfnamefont {S.~F.}\ \bibnamefont
			{Caballero-Benitez}}\ and\ \bibinfo {author} {\bibfnamefont {I.~B.}\
			\bibnamefont {Mekhov}},\ }\bibfield  {title} {\bibinfo {title} {Quantum
			optical lattices for emergent many-body phases of ultracold atoms},\ }\href
	{https://doi.org/10.1103/PhysRevLett.115.243604} {\bibfield  {journal}
		{\bibinfo  {journal} {Phys. Rev. Lett.}\ }\textbf {\bibinfo {volume} {115}},\
		\bibinfo {pages} {243604} (\bibinfo {year} {2015})}\BibitemShut {NoStop}%
	\bibitem [{\citenamefont {Dogra}\ \emph {et~al.}(2016)\citenamefont {Dogra},
		\citenamefont {Brennecke}, \citenamefont {Huber},\ and\ \citenamefont
		{Donner}}]{dogra2016phase}%
	\BibitemOpen
	\bibfield  {author} {\bibinfo {author} {\bibfnamefont {N.}~\bibnamefont
			{Dogra}}, \bibinfo {author} {\bibfnamefont {F.}~\bibnamefont {Brennecke}},
		\bibinfo {author} {\bibfnamefont {S.~D.}\ \bibnamefont {Huber}},\ and\
		\bibinfo {author} {\bibfnamefont {T.}~\bibnamefont {Donner}},\ }\bibfield
	{title} {\bibinfo {title} {{Phase transitions in a Bose-Hubbard model with
				cavity-mediated global-range interactions}},\ }\href
	{https://doi.org/10.1103/PhysRevA.94.023632} {\bibfield  {journal} {\bibinfo
			{journal} {Phys. Rev. A}\ }\textbf {\bibinfo {volume} {94}},\ \bibinfo
		{pages} {023632} (\bibinfo {year} {2016})}\BibitemShut {NoStop}%
	\bibitem [{\citenamefont {Chen}\ \emph {et~al.}(2016)\citenamefont {Chen},
		\citenamefont {Yu},\ and\ \citenamefont {Zhai}}]{chen2016quantum}%
	\BibitemOpen
	\bibfield  {author} {\bibinfo {author} {\bibfnamefont {Y.}~\bibnamefont
			{Chen}}, \bibinfo {author} {\bibfnamefont {Z.}~\bibnamefont {Yu}},\ and\
		\bibinfo {author} {\bibfnamefont {H.}~\bibnamefont {Zhai}},\ }\bibfield
	{title} {\bibinfo {title} {{Quantum phase transitions of the Bose-Hubbard
				model inside a cavity}},\ }\href {https://doi.org/10.1103/PhysRevA.93.041601}
	{\bibfield  {journal} {\bibinfo  {journal} {Phys. Rev. A}\ }\textbf {\bibinfo
			{volume} {93}},\ \bibinfo {pages} {041601} (\bibinfo {year}
		{2016})}\BibitemShut {NoStop}%
	\bibitem [{\citenamefont {Sundar}\ and\ \citenamefont
		{Mueller}(2016)}]{sundar2016latticeboson}%
	\BibitemOpen
	\bibfield  {author} {\bibinfo {author} {\bibfnamefont {B.}~\bibnamefont
			{Sundar}}\ and\ \bibinfo {author} {\bibfnamefont {E.~J.}\ \bibnamefont
			{Mueller}},\ }\bibfield  {title} {\bibinfo {title} {Lattice bosons with
			infinite-range checkerboard interactions},\ }\href
	{https://doi.org/10.1103/PhysRevA.94.033631} {\bibfield  {journal} {\bibinfo
			{journal} {Phys. Rev. A}\ }\textbf {\bibinfo {volume} {94}},\ \bibinfo
		{pages} {033631} (\bibinfo {year} {2016})}\BibitemShut {NoStop}%
	\bibitem [{\citenamefont {Flottat}\ \emph {et~al.}(2017)\citenamefont
		{Flottat}, \citenamefont {{De Parny}}, \citenamefont {H{\'{e}}bert},
		\citenamefont {Rousseau}, \citenamefont {Batrouni}, \citenamefont {{de Forges
				de Parny}}, \citenamefont {H{\'{e}}bert}, \citenamefont {Rousseau},\ and\
		\citenamefont {Batrouni}}]{Flottat2017_phasediagram}%
	\BibitemOpen
	\bibfield  {author} {\bibinfo {author} {\bibfnamefont {T.}~\bibnamefont
			{Flottat}}, \bibinfo {author} {\bibfnamefont {L.~D.~F.}\ \bibnamefont {{De
					Parny}}}, \bibinfo {author} {\bibfnamefont {F.}~\bibnamefont {H{\'{e}}bert}},
		\bibinfo {author} {\bibfnamefont {V.~G.}\ \bibnamefont {Rousseau}}, \bibinfo
		{author} {\bibfnamefont {G.~G.}\ \bibnamefont {Batrouni}}, \bibinfo {author}
		{\bibfnamefont {L.}~\bibnamefont {{de Forges de Parny}}}, \bibinfo {author}
		{\bibfnamefont {F.}~\bibnamefont {H{\'{e}}bert}}, \bibinfo {author}
		{\bibfnamefont {V.~G.}\ \bibnamefont {Rousseau}},\ and\ \bibinfo {author}
		{\bibfnamefont {G.~G.}\ \bibnamefont {Batrouni}},\ }\bibfield  {title}
	{\bibinfo {title} {{Phase diagram of bosons in a two-dimensional optical
				lattice with infinite-range cavity-mediated interactions}},\ }\href
	{https://doi.org/10.1103/PhysRevB.95.144501} {\bibfield  {journal} {\bibinfo
			{journal} {Physical Review B}\ }\textbf {\bibinfo {volume} {95}},\ \bibinfo
		{pages} {1} (\bibinfo {year} {2017})}\BibitemShut {NoStop}%
	\bibitem [{\citenamefont {Liao}\ \emph {et~al.}(2018)\citenamefont {Liao},
		\citenamefont {Chen}, \citenamefont {Zheng},\ and\ \citenamefont
		{Huang}}]{liao2018theoretical}%
	\BibitemOpen
	\bibfield  {author} {\bibinfo {author} {\bibfnamefont {R.}~\bibnamefont
			{Liao}}, \bibinfo {author} {\bibfnamefont {H.-J.}\ \bibnamefont {Chen}},
		\bibinfo {author} {\bibfnamefont {D.-C.}\ \bibnamefont {Zheng}},\ and\
		\bibinfo {author} {\bibfnamefont {Z.-G.}\ \bibnamefont {Huang}},\ }\bibfield
	{title} {\bibinfo {title} {{Theoretical exploration of competing phases of
				lattice Bose gases in a cavity}},\ }\href
	{https://doi.org/10.1103/PhysRevA.97.013624} {\bibfield  {journal} {\bibinfo
			{journal} {Phys. Rev. A}\ }\textbf {\bibinfo {volume} {97}},\ \bibinfo
		{pages} {013624} (\bibinfo {year} {2018})}\BibitemShut {NoStop}%
	\bibitem [{\citenamefont {Chen}\ \emph {et~al.}(2020)\citenamefont {Chen},
		\citenamefont {Yu}, \citenamefont {Zheng},\ and\ \citenamefont
		{Liao}}]{Chen2020extended}%
	\BibitemOpen
	\bibfield  {author} {\bibinfo {author} {\bibfnamefont {H.-J.}\ \bibnamefont
			{Chen}}, \bibinfo {author} {\bibfnamefont {Y.-Q.}\ \bibnamefont {Yu}},
		\bibinfo {author} {\bibfnamefont {D.-C.}\ \bibnamefont {Zheng}},\ and\
		\bibinfo {author} {\bibfnamefont {R.}~\bibnamefont {Liao}},\ }\bibfield
	{title} {\bibinfo {title} {{Extended Bose-Hubbard Model with Cavity-Mediated
				Infinite-Range Interactions at Finite Temperatures}},\ }\href
	{https://doi.org/10.1038/s41598-020-66054-1} {\bibfield  {journal} {\bibinfo
			{journal} {Scientific Reports}\ }\textbf {\bibinfo {volume} {10}},\ \bibinfo
		{pages} {9076} (\bibinfo {year} {2020})}\BibitemShut {NoStop}%
	\bibitem [{\citenamefont {Klinder}\ \emph {et~al.}(2015)\citenamefont
		{Klinder}, \citenamefont {Ke\ss{}ler}, \citenamefont {Bakhtiari},
		\citenamefont {Thorwart},\ and\ \citenamefont
		{Hemmerich}}]{Klinder2015observation}%
	\BibitemOpen
	\bibfield  {author} {\bibinfo {author} {\bibfnamefont {J.}~\bibnamefont
			{Klinder}}, \bibinfo {author} {\bibfnamefont {H.}~\bibnamefont {Ke\ss{}ler}},
		\bibinfo {author} {\bibfnamefont {M.~R.}\ \bibnamefont {Bakhtiari}}, \bibinfo
		{author} {\bibfnamefont {M.}~\bibnamefont {Thorwart}},\ and\ \bibinfo
		{author} {\bibfnamefont {A.}~\bibnamefont {Hemmerich}},\ }\bibfield  {title}
	{\bibinfo {title} {{Observation of a Superradiant Mott Insulator in the
				Dicke-Hubbard Model}},\ }\href
	{https://doi.org/10.1103/PhysRevLett.115.230403} {\bibfield  {journal}
		{\bibinfo  {journal} {Phys. Rev. Lett.}\ }\textbf {\bibinfo {volume} {115}},\
		\bibinfo {pages} {230403} (\bibinfo {year} {2015})}\BibitemShut {NoStop}%
	\bibitem [{\citenamefont {Landig}\ \emph {et~al.}(2016)\citenamefont {Landig},
		\citenamefont {Hruby}, \citenamefont {Dogra}, \citenamefont {Landini},
		\citenamefont {Mottl}, \citenamefont {Donner},\ and\ \citenamefont
		{Esslinger}}]{landig2016quantum}%
	\BibitemOpen
	\bibfield  {author} {\bibinfo {author} {\bibfnamefont {R.}~\bibnamefont
			{Landig}}, \bibinfo {author} {\bibfnamefont {L.}~\bibnamefont {Hruby}},
		\bibinfo {author} {\bibfnamefont {N.}~\bibnamefont {Dogra}}, \bibinfo
		{author} {\bibfnamefont {M.}~\bibnamefont {Landini}}, \bibinfo {author}
		{\bibfnamefont {R.}~\bibnamefont {Mottl}}, \bibinfo {author} {\bibfnamefont
			{T.}~\bibnamefont {Donner}},\ and\ \bibinfo {author} {\bibfnamefont
			{T.}~\bibnamefont {Esslinger}},\ }\bibfield  {title} {\bibinfo {title}
		{{Quantum phases from competing short- and long-range interactions in an
				optical lattice}},\ }\href {https://doi.org/10.1038/nature17409} {\bibfield
		{journal} {\bibinfo  {journal} {Nature}\ }\textbf {\bibinfo {volume} {532}},\
		\bibinfo {pages} {476} (\bibinfo {year} {2016})}\BibitemShut {NoStop}%
	\bibitem [{\citenamefont {Hruby}\ \emph {et~al.}(2018)\citenamefont {Hruby},
		\citenamefont {Dogra}, \citenamefont {Landini}, \citenamefont {Donner},\ and\
		\citenamefont {Esslinger}}]{hruby2018metastability}%
	\BibitemOpen
	\bibfield  {author} {\bibinfo {author} {\bibfnamefont {L.}~\bibnamefont
			{Hruby}}, \bibinfo {author} {\bibfnamefont {N.}~\bibnamefont {Dogra}},
		\bibinfo {author} {\bibfnamefont {M.}~\bibnamefont {Landini}}, \bibinfo
		{author} {\bibfnamefont {T.}~\bibnamefont {Donner}},\ and\ \bibinfo {author}
		{\bibfnamefont {T.}~\bibnamefont {Esslinger}},\ }\bibfield  {title} {\bibinfo
		{title} {{Metastability and avalanche dynamics in strongly correlated gases
				with long-range interactions}},\ }\href
	{https://doi.org/10.1073/pnas.1720415115} {\bibfield  {journal} {\bibinfo
			{journal} {Proceedings of the National Academy of Sciences}\ }\textbf
		{\bibinfo {volume} {115}},\ \bibinfo {pages} {3279} (\bibinfo {year}
		{2018})},\ \Eprint {https://arxiv.org/abs/1708.02229} {1708.02229}
	\BibitemShut {NoStop}%
	\bibitem [{\citenamefont {Landini}\ \emph {et~al.}(2018)\citenamefont
		{Landini}, \citenamefont {Dogra}, \citenamefont {Kroeger}, \citenamefont
		{Hruby}, \citenamefont {Donner},\ and\ \citenamefont
		{Esslinger}}]{landini2018formation}%
	\BibitemOpen
	\bibfield  {author} {\bibinfo {author} {\bibfnamefont {M.}~\bibnamefont
			{Landini}}, \bibinfo {author} {\bibfnamefont {N.}~\bibnamefont {Dogra}},
		\bibinfo {author} {\bibfnamefont {K.}~\bibnamefont {Kroeger}}, \bibinfo
		{author} {\bibfnamefont {L.}~\bibnamefont {Hruby}}, \bibinfo {author}
		{\bibfnamefont {T.}~\bibnamefont {Donner}},\ and\ \bibinfo {author}
		{\bibfnamefont {T.}~\bibnamefont {Esslinger}},\ }\bibfield  {title} {\bibinfo
		{title} {Formation of a spin texture in a quantum gas coupled to a cavity},\
	}\href {https://doi.org/10.1103/PhysRevLett.120.223602} {\bibfield  {journal}
		{\bibinfo  {journal} {Phys. Rev. Lett.}\ }\textbf {\bibinfo {volume} {120}},\
		\bibinfo {pages} {223602} (\bibinfo {year} {2018})}\BibitemShut {NoStop}%
	\bibitem [{\citenamefont {Kroeze}\ \emph {et~al.}(2018)\citenamefont {Kroeze},
		\citenamefont {Guo}, \citenamefont {Vaidya}, \citenamefont {Keeling},\ and\
		\citenamefont {Lev}}]{Kroeze2018spinor}%
	\BibitemOpen
	\bibfield  {author} {\bibinfo {author} {\bibfnamefont {R.~M.}\ \bibnamefont
			{Kroeze}}, \bibinfo {author} {\bibfnamefont {Y.}~\bibnamefont {Guo}},
		\bibinfo {author} {\bibfnamefont {V.~D.}\ \bibnamefont {Vaidya}}, \bibinfo
		{author} {\bibfnamefont {J.}~\bibnamefont {Keeling}},\ and\ \bibinfo {author}
		{\bibfnamefont {B.~L.}\ \bibnamefont {Lev}},\ }\bibfield  {title} {\bibinfo
		{title} {{Spinor Self-Ordering of a Quantum Gas in a Cavity}},\ }\href
	{https://doi.org/10.1103/PhysRevLett.121.163601} {\bibfield  {journal}
		{\bibinfo  {journal} {Phys. Rev. Lett.}\ }\textbf {\bibinfo {volume} {121}},\
		\bibinfo {pages} {163601} (\bibinfo {year} {2018})}\BibitemShut {NoStop}%
	\bibitem [{\citenamefont {Ferri}\ \emph {et~al.}(2021)\citenamefont {Ferri},
		\citenamefont {Rosa-Medina}, \citenamefont {Finger}, \citenamefont {Dogra},
		\citenamefont {Soriente}, \citenamefont {Zilberberg}, \citenamefont
		{Donner},\ and\ \citenamefont {Esslinger}}]{ferri2021emerging}%
	\BibitemOpen
	\bibfield  {author} {\bibinfo {author} {\bibfnamefont {F.}~\bibnamefont
			{Ferri}}, \bibinfo {author} {\bibfnamefont {R.}~\bibnamefont {Rosa-Medina}},
		\bibinfo {author} {\bibfnamefont {F.}~\bibnamefont {Finger}}, \bibinfo
		{author} {\bibfnamefont {N.}~\bibnamefont {Dogra}}, \bibinfo {author}
		{\bibfnamefont {M.}~\bibnamefont {Soriente}}, \bibinfo {author}
		{\bibfnamefont {O.}~\bibnamefont {Zilberberg}}, \bibinfo {author}
		{\bibfnamefont {T.}~\bibnamefont {Donner}},\ and\ \bibinfo {author}
		{\bibfnamefont {T.}~\bibnamefont {Esslinger}},\ }\bibfield  {title} {\bibinfo
		{title} {{Emerging Dissipative Phases in a Superradiant Quantum Gas with
				Tunable Decay}},\ }\href {https://doi.org/10.1103/PhysRevX.11.041046}
	{\bibfield  {journal} {\bibinfo  {journal} {Phys. Rev. X}\ }\textbf {\bibinfo
			{volume} {11}},\ \bibinfo {pages} {041046} (\bibinfo {year}
		{2021})}\BibitemShut {NoStop}%
	\bibitem [{\citenamefont {Simon}\ \emph {et~al.}(2007)\citenamefont {Simon},
		\citenamefont {Tanji}, \citenamefont {Ghosh},\ and\ \citenamefont
		{Vuleti{\'{c}}}}]{Simon2007single}%
	\BibitemOpen
	\bibfield  {author} {\bibinfo {author} {\bibfnamefont {J.}~\bibnamefont
			{Simon}}, \bibinfo {author} {\bibfnamefont {H.}~\bibnamefont {Tanji}},
		\bibinfo {author} {\bibfnamefont {S.}~\bibnamefont {Ghosh}},\ and\ \bibinfo
		{author} {\bibfnamefont {V.}~\bibnamefont {Vuleti{\'{c}}}},\ }\bibfield
	{title} {\bibinfo {title} {{Single-photon bus connecting spin-wave quantum
				memories}},\ }\href {https://doi.org/10.1038/nphys726} {\bibfield  {journal}
		{\bibinfo  {journal} {Nature Physics}\ }\textbf {\bibinfo {volume} {3}},\
		\bibinfo {pages} {765} (\bibinfo {year} {2007})}\BibitemShut {NoStop}%
	\bibitem [{\citenamefont {Strack}\ and\ \citenamefont
		{Sachdev}(2011)}]{Strack2011Dicke}%
	\BibitemOpen
	\bibfield  {author} {\bibinfo {author} {\bibfnamefont {P.}~\bibnamefont
			{Strack}}\ and\ \bibinfo {author} {\bibfnamefont {S.}~\bibnamefont
			{Sachdev}},\ }\bibfield  {title} {\bibinfo {title} {{Dicke Quantum Spin Glass
				of Atoms and Photons}},\ }\href
	{https://doi.org/10.1103/PhysRevLett.107.277202} {\bibfield  {journal}
		{\bibinfo  {journal} {Phys. Rev. Lett.}\ }\textbf {\bibinfo {volume} {107}},\
		\bibinfo {pages} {277202} (\bibinfo {year} {2011})}\BibitemShut {NoStop}%
	\bibitem [{\citenamefont {Gopalakrishnan}\ \emph {et~al.}(2011)\citenamefont
		{Gopalakrishnan}, \citenamefont {Lev},\ and\ \citenamefont
		{Goldbart}}]{Gopalakrishnan2011frustration}%
	\BibitemOpen
	\bibfield  {author} {\bibinfo {author} {\bibfnamefont {S.}~\bibnamefont
			{Gopalakrishnan}}, \bibinfo {author} {\bibfnamefont {B.~L.}\ \bibnamefont
			{Lev}},\ and\ \bibinfo {author} {\bibfnamefont {P.~M.}\ \bibnamefont
			{Goldbart}},\ }\bibfield  {title} {\bibinfo {title} {{Frustration and
				Glassiness in Spin Models with Cavity-Mediated Interactions}},\ }\href
	{https://doi.org/10.1103/PhysRevLett.107.277201} {\bibfield  {journal}
		{\bibinfo  {journal} {Phys. Rev. Lett.}\ }\textbf {\bibinfo {volume} {107}},\
		\bibinfo {pages} {277201} (\bibinfo {year} {2011})}\BibitemShut {NoStop}%
	\bibitem [{\citenamefont {Buchhold}\ \emph {et~al.}()\citenamefont {Buchhold},
		\citenamefont {Strack}, \citenamefont {Sachdev},\ and\ \citenamefont
		{Diehl}}]{Buchhold2013dicke}%
	\BibitemOpen
	\bibfield  {author} {\bibinfo {author} {\bibfnamefont {M.}~\bibnamefont
			{Buchhold}}, \bibinfo {author} {\bibfnamefont {P.}~\bibnamefont {Strack}},
		\bibinfo {author} {\bibfnamefont {S.}~\bibnamefont {Sachdev}},\ and\ \bibinfo
		{author} {\bibfnamefont {S.}~\bibnamefont {Diehl}},\ }\bibfield  {title}
	{\bibinfo {title} {{Dicke-model quantum spin and photon glass in optical
				cavities: Nonequilibrium theory and experimental signatures}},\ }\href@noop
	{} {\ }\BibitemShut {NoStop}%
	\bibitem [{\citenamefont {Zhiqiang}\ \emph {et~al.}(2017)\citenamefont
		{Zhiqiang}, \citenamefont {Lee}, \citenamefont {Kumar}, \citenamefont
		{Arnold}, \citenamefont {Masson}, \citenamefont {Parkins},\ and\
		\citenamefont {Barrett}}]{Zhiqiang2017nonequilibrium}%
	\BibitemOpen
	\bibfield  {author} {\bibinfo {author} {\bibfnamefont {Z.}~\bibnamefont
			{Zhiqiang}}, \bibinfo {author} {\bibfnamefont {C.~H.}\ \bibnamefont {Lee}},
		\bibinfo {author} {\bibfnamefont {R.}~\bibnamefont {Kumar}}, \bibinfo
		{author} {\bibfnamefont {K.~J.}\ \bibnamefont {Arnold}}, \bibinfo {author}
		{\bibfnamefont {S.~J.}\ \bibnamefont {Masson}}, \bibinfo {author}
		{\bibfnamefont {A.~S.}\ \bibnamefont {Parkins}},\ and\ \bibinfo {author}
		{\bibfnamefont {M.~D.}\ \bibnamefont {Barrett}},\ }\bibfield  {title}
	{\bibinfo {title} {{Nonequilibrium phase transition in a spin-1 Dicke
				model}},\ }\href {https://doi.org/10.1364/optica.4.000424} {\bibfield
		{journal} {\bibinfo  {journal} {Optica}\ }\textbf {\bibinfo {volume} {4}},\
		\bibinfo {pages} {424} (\bibinfo {year} {2017})}\BibitemShut {NoStop}%
	\bibitem [{\citenamefont {Mivehvar}\ \emph {et~al.}(2017)\citenamefont
		{Mivehvar}, \citenamefont {Piazza},\ and\ \citenamefont
		{Ritsch}}]{Mivehvar2017disorder}%
	\BibitemOpen
	\bibfield  {author} {\bibinfo {author} {\bibfnamefont {F.}~\bibnamefont
			{Mivehvar}}, \bibinfo {author} {\bibfnamefont {F.}~\bibnamefont {Piazza}},\
		and\ \bibinfo {author} {\bibfnamefont {H.}~\bibnamefont {Ritsch}},\
	}\bibfield  {title} {\bibinfo {title} {{Disorder-Driven Density and Spin
				Self-Ordering of a Bose-Einstein Condensate in a Cavity}},\ }\href
	{https://doi.org/10.1103/PhysRevLett.119.063602} {\bibfield  {journal}
		{\bibinfo  {journal} {Phys. Rev. Lett.}\ }\textbf {\bibinfo {volume} {119}},\
		\bibinfo {pages} {063602} (\bibinfo {year} {2017})}\BibitemShut {NoStop}%
	\bibitem [{\citenamefont {Lewis-Swan}\ \emph {et~al.}(2018)\citenamefont
		{Lewis-Swan}, \citenamefont {Norcia}, \citenamefont {Cline}, \citenamefont
		{Thompson},\ and\ \citenamefont {Rey}}]{LewisSwan2018robust}%
	\BibitemOpen
	\bibfield  {author} {\bibinfo {author} {\bibfnamefont {R.~J.}\ \bibnamefont
			{Lewis-Swan}}, \bibinfo {author} {\bibfnamefont {M.~A.}\ \bibnamefont
			{Norcia}}, \bibinfo {author} {\bibfnamefont {J.~R.~K.}\ \bibnamefont
			{Cline}}, \bibinfo {author} {\bibfnamefont {J.~K.}\ \bibnamefont
			{Thompson}},\ and\ \bibinfo {author} {\bibfnamefont {A.~M.}\ \bibnamefont
			{Rey}},\ }\bibfield  {title} {\bibinfo {title} {{Robust Spin Squeezing via
				Photon-Mediated Interactions on an Optical Clock Transition}},\ }\href
	{https://doi.org/10.1103/PhysRevLett.121.070403} {\bibfield  {journal}
		{\bibinfo  {journal} {Phys. Rev. Lett.}\ }\textbf {\bibinfo {volume} {121}},\
		\bibinfo {pages} {070403} (\bibinfo {year} {2018})}\BibitemShut {NoStop}%
	\bibitem [{\citenamefont {Davis}\ \emph {et~al.}(2019)\citenamefont {Davis},
		\citenamefont {Bentsen}, \citenamefont {Homeier}, \citenamefont {Li},\ and\
		\citenamefont {Schleier-Smith}}]{davis2019photon}%
	\BibitemOpen
	\bibfield  {author} {\bibinfo {author} {\bibfnamefont {E.~J.}\ \bibnamefont
			{Davis}}, \bibinfo {author} {\bibfnamefont {G.}~\bibnamefont {Bentsen}},
		\bibinfo {author} {\bibfnamefont {L.}~\bibnamefont {Homeier}}, \bibinfo
		{author} {\bibfnamefont {T.}~\bibnamefont {Li}},\ and\ \bibinfo {author}
		{\bibfnamefont {M.~H.}\ \bibnamefont {Schleier-Smith}},\ }\bibfield  {title}
	{\bibinfo {title} {{Photon-Mediated Spin-Exchange Dynamics of Spin-1
				Atoms}},\ }\href {https://doi.org/10.1103/PhysRevLett.122.010405} {\bibfield
		{journal} {\bibinfo  {journal} {Phys. Rev. Lett.}\ }\textbf {\bibinfo
			{volume} {122}},\ \bibinfo {pages} {010405} (\bibinfo {year}
		{2019})}\BibitemShut {NoStop}%
	\bibitem [{\citenamefont {Mivehvar}\ \emph {et~al.}(2019)\citenamefont
		{Mivehvar}, \citenamefont {Ritsch},\ and\ \citenamefont
		{Piazza}}]{mivehvar2019cavity}%
	\BibitemOpen
	\bibfield  {author} {\bibinfo {author} {\bibfnamefont {F.}~\bibnamefont
			{Mivehvar}}, \bibinfo {author} {\bibfnamefont {H.}~\bibnamefont {Ritsch}},\
		and\ \bibinfo {author} {\bibfnamefont {F.}~\bibnamefont {Piazza}},\
	}\bibfield  {title} {\bibinfo {title} {Cavity-quantum-electrodynamical
			toolbox for quantum magnetism},\ }\href
	{https://doi.org/10.1103/PhysRevLett.122.113603} {\bibfield  {journal}
		{\bibinfo  {journal} {Phys. Rev. Lett.}\ }\textbf {\bibinfo {volume} {122}},\
		\bibinfo {pages} {113603} (\bibinfo {year} {2019})}\BibitemShut {NoStop}%
	\bibitem [{\citenamefont {Muniz}\ \emph {et~al.}(2020)\citenamefont {Muniz},
		\citenamefont {Barberena}, \citenamefont {Lewis-Swan}, \citenamefont {Young},
		\citenamefont {Cline}, \citenamefont {Rey},\ and\ \citenamefont
		{Thompson}}]{Muniz2020exploring}%
	\BibitemOpen
	\bibfield  {author} {\bibinfo {author} {\bibfnamefont {J.~A.}\ \bibnamefont
			{Muniz}}, \bibinfo {author} {\bibfnamefont {D.}~\bibnamefont {Barberena}},
		\bibinfo {author} {\bibfnamefont {R.~J.}\ \bibnamefont {Lewis-Swan}},
		\bibinfo {author} {\bibfnamefont {D.~J.}\ \bibnamefont {Young}}, \bibinfo
		{author} {\bibfnamefont {J.~R.~K.}\ \bibnamefont {Cline}}, \bibinfo {author}
		{\bibfnamefont {A.~M.}\ \bibnamefont {Rey}},\ and\ \bibinfo {author}
		{\bibfnamefont {J.~K.}\ \bibnamefont {Thompson}},\ }\bibfield  {title}
	{\bibinfo {title} {{Exploring dynamical phase transitions with cold atoms in
				an optical cavity}},\ }\href {https://doi.org/10.1038/s41586-020-2224-x}
	{\bibfield  {journal} {\bibinfo  {journal} {Nature}\ }\textbf {\bibinfo
			{volume} {580}},\ \bibinfo {pages} {602} (\bibinfo {year}
		{2020})}\BibitemShut {NoStop}%
	\bibitem [{\citenamefont {Stitely}\ \emph {et~al.}(2020)\citenamefont
		{Stitely}, \citenamefont {Giraldo}, \citenamefont {Krauskopf},\ and\
		\citenamefont {Parkins}}]{Stitely2020nonlinear}%
	\BibitemOpen
	\bibfield  {author} {\bibinfo {author} {\bibfnamefont {K.~C.}\ \bibnamefont
			{Stitely}}, \bibinfo {author} {\bibfnamefont {A.}~\bibnamefont {Giraldo}},
		\bibinfo {author} {\bibfnamefont {B.}~\bibnamefont {Krauskopf}},\ and\
		\bibinfo {author} {\bibfnamefont {S.}~\bibnamefont {Parkins}},\ }\bibfield
	{title} {\bibinfo {title} {{Nonlinear semiclassical dynamics of the
				unbalanced, open Dicke model}},\ }\href
	{https://doi.org/10.1103/PhysRevResearch.2.033131} {\bibfield  {journal}
		{\bibinfo  {journal} {Phys. Rev. Research}\ }\textbf {\bibinfo {volume}
			{2}},\ \bibinfo {pages} {033131} (\bibinfo {year} {2020})}\BibitemShut
	{NoStop}%
	\bibitem [{\citenamefont {Guan}\ \emph {et~al.}(2019)\citenamefont {Guan},
		\citenamefont {Fan}, \citenamefont {Zhou}, \citenamefont {Chen},\ and\
		\citenamefont {Jia}}]{Guan2019two}%
	\BibitemOpen
	\bibfield  {author} {\bibinfo {author} {\bibfnamefont {X.}~\bibnamefont
			{Guan}}, \bibinfo {author} {\bibfnamefont {J.}~\bibnamefont {Fan}}, \bibinfo
		{author} {\bibfnamefont {X.}~\bibnamefont {Zhou}}, \bibinfo {author}
		{\bibfnamefont {G.}~\bibnamefont {Chen}},\ and\ \bibinfo {author}
		{\bibfnamefont {S.}~\bibnamefont {Jia}},\ }\bibfield  {title} {\bibinfo
		{title} {{Two-component lattice bosons with cavity-mediated long-range
				interaction}},\ }\href {https://doi.org/10.1103/PhysRevA.100.013617}
	{\bibfield  {journal} {\bibinfo  {journal} {Physical Review A}\ }\textbf
		{\bibinfo {volume} {100}},\ \bibinfo {pages} {13617} (\bibinfo {year}
		{2019})}\BibitemShut {NoStop}%
	\bibitem [{\citenamefont {Lozano-M\'endez}\ \emph {et~al.}(2022)\citenamefont
		{Lozano-M\'endez}, \citenamefont {C\'asares},\ and\ \citenamefont
		{Caballero-Ben\'{\i}tez}}]{Lozano2020Spin}%
	\BibitemOpen
	\bibfield  {author} {\bibinfo {author} {\bibfnamefont {K.}~\bibnamefont
			{Lozano-M\'endez}}, \bibinfo {author} {\bibfnamefont {A.~H.}\ \bibnamefont
			{C\'asares}},\ and\ \bibinfo {author} {\bibfnamefont {S.~F.}\ \bibnamefont
			{Caballero-Ben\'{\i}tez}},\ }\bibfield  {title} {\bibinfo {title} {Spin
			entanglement and magnetic competition via long-range interactions in spinor
			quantum optical lattices},\ }\href
	{https://doi.org/10.1103/PhysRevLett.128.080601} {\bibfield  {journal}
		{\bibinfo  {journal} {Phys. Rev. Lett.}\ }\textbf {\bibinfo {volume} {128}},\
		\bibinfo {pages} {080601} (\bibinfo {year} {2022})}\BibitemShut {NoStop}%
	\bibitem [{\citenamefont {Fan}\ \emph {et~al.}(2018)\citenamefont {Fan},
		\citenamefont {Zhou}, \citenamefont {Zheng}, \citenamefont {Yi},
		\citenamefont {Chen},\ and\ \citenamefont {Jia}}]{Fan2018magnetic}%
	\BibitemOpen
	\bibfield  {author} {\bibinfo {author} {\bibfnamefont {J.}~\bibnamefont
			{Fan}}, \bibinfo {author} {\bibfnamefont {X.}~\bibnamefont {Zhou}}, \bibinfo
		{author} {\bibfnamefont {W.}~\bibnamefont {Zheng}}, \bibinfo {author}
		{\bibfnamefont {W.}~\bibnamefont {Yi}}, \bibinfo {author} {\bibfnamefont
			{G.}~\bibnamefont {Chen}},\ and\ \bibinfo {author} {\bibfnamefont
			{S.}~\bibnamefont {Jia}},\ }\bibfield  {title} {\bibinfo {title} {{Magnetic
				order in a Fermi gas induced by cavity-field fluctuations}},\ }\href
	{https://doi.org/10.1103/PhysRevA.98.043613} {\bibfield  {journal} {\bibinfo
			{journal} {Physical Review A}\ }\textbf {\bibinfo {volume} {98}},\ \bibinfo
		{pages} {043613} (\bibinfo {year} {2018})}\BibitemShut {NoStop}%
	\bibitem [{\citenamefont {Camacho-Guardian}\ \emph {et~al.}(2017)\citenamefont
		{Camacho-Guardian}, \citenamefont {Paredes},\ and\ \citenamefont
		{Caballero-Ben{\'{i}}tez}}]{CamachoGuardian2017quantum}%
	\BibitemOpen
	\bibfield  {author} {\bibinfo {author} {\bibfnamefont {A.}~\bibnamefont
			{Camacho-Guardian}}, \bibinfo {author} {\bibfnamefont {R.}~\bibnamefont
			{Paredes}},\ and\ \bibinfo {author} {\bibfnamefont {S.~F.}\ \bibnamefont
			{Caballero-Ben{\'{i}}tez}},\ }\bibfield  {title} {\bibinfo {title} {{Quantum
				simulation of competing orders with fermions in quantum optical lattices}},\
	}\href {https://doi.org/10.1103/PhysRevA.96.051602} {\bibfield  {journal}
		{\bibinfo  {journal} {Physical Review A}\ }\textbf {\bibinfo {volume} {96}},\
		\bibinfo {pages} {051602} (\bibinfo {year} {2017})}\BibitemShut {NoStop}%
	\bibitem [{\citenamefont {Kuklov}\ and\ \citenamefont
		{Svistunov}(2003)}]{kuklov2003_counterflow}%
	\BibitemOpen
	\bibfield  {author} {\bibinfo {author} {\bibfnamefont {A.~B.}\ \bibnamefont
			{Kuklov}}\ and\ \bibinfo {author} {\bibfnamefont {B.~V.}\ \bibnamefont
			{Svistunov}},\ }\bibfield  {title} {\bibinfo {title} {{Counterflow
				Superfluidity of Two-Species Ultracold Atoms in a Commensurate Optical
				Lattice}},\ }\href {https://doi.org/10.1103/PhysRevLett.90.100401} {\bibfield
		{journal} {\bibinfo  {journal} {Phys. Rev. Lett.}\ }\textbf {\bibinfo
			{volume} {90}},\ \bibinfo {pages} {100401} (\bibinfo {year}
		{2003})}\BibitemShut {NoStop}%
	\bibitem [{\citenamefont {Altman}\ \emph {et~al.}(2003)\citenamefont {Altman},
		\citenamefont {Hofstetter}, \citenamefont {Demler},\ and\ \citenamefont
		{Lukin}}]{Altman2003_phase_diagram}%
	\BibitemOpen
	\bibfield  {author} {\bibinfo {author} {\bibfnamefont {E.}~\bibnamefont
			{Altman}}, \bibinfo {author} {\bibfnamefont {W.}~\bibnamefont {Hofstetter}},
		\bibinfo {author} {\bibfnamefont {E.}~\bibnamefont {Demler}},\ and\ \bibinfo
		{author} {\bibfnamefont {M.~D.}\ \bibnamefont {Lukin}},\ }\bibfield  {title}
	{\bibinfo {title} {{Phase diagram of two-component bosons on an optical
				lattice}},\ }\href {https://doi.org/10.1088/1367-2630/5/1/113} {\bibfield
		{journal} {\bibinfo  {journal} {New Journal of Physics}\ }\textbf {\bibinfo
			{volume} {5}},\ \bibinfo {pages} {113} (\bibinfo {year} {2003})}\BibitemShut
	{NoStop}%
	\bibitem [{\citenamefont {Lingua}\ \emph {et~al.}(2015)\citenamefont {Lingua},
		\citenamefont {Guglielmino}, \citenamefont {Penna},\ and\ \citenamefont
		{Capogrosso~Sansone}}]{Lingua2015_demixing_effects}%
	\BibitemOpen
	\bibfield  {author} {\bibinfo {author} {\bibfnamefont {F.}~\bibnamefont
			{Lingua}}, \bibinfo {author} {\bibfnamefont {M.}~\bibnamefont {Guglielmino}},
		\bibinfo {author} {\bibfnamefont {V.}~\bibnamefont {Penna}},\ and\ \bibinfo
		{author} {\bibfnamefont {B.}~\bibnamefont {Capogrosso~Sansone}},\ }\bibfield
	{title} {\bibinfo {title} {{Demixing effects in mixtures of two bosonic
				species}},\ }\href {https://doi.org/10.1103/PhysRevA.92.053610} {\bibfield
		{journal} {\bibinfo  {journal} {Phys. Rev. A}\ }\textbf {\bibinfo {volume}
			{92}},\ \bibinfo {pages} {053610} (\bibinfo {year} {2015})}\BibitemShut
	{NoStop}%
	\bibitem [{\citenamefont {Bai}\ \emph {et~al.}(2020)\citenamefont {Bai},
		\citenamefont {Gaur}, \citenamefont {Sable}, \citenamefont {Bandyopadhyay},
		\citenamefont {Suthar},\ and\ \citenamefont
		{Angom}}]{Bai2020_segregated_quantum}%
	\BibitemOpen
	\bibfield  {author} {\bibinfo {author} {\bibfnamefont {R.}~\bibnamefont
			{Bai}}, \bibinfo {author} {\bibfnamefont {D.}~\bibnamefont {Gaur}}, \bibinfo
		{author} {\bibfnamefont {H.}~\bibnamefont {Sable}}, \bibinfo {author}
		{\bibfnamefont {S.}~\bibnamefont {Bandyopadhyay}}, \bibinfo {author}
		{\bibfnamefont {K.}~\bibnamefont {Suthar}},\ and\ \bibinfo {author}
		{\bibfnamefont {D.}~\bibnamefont {Angom}},\ }\bibfield  {title} {\bibinfo
		{title} {{Segregated quantum phases of dipolar bosonic mixtures in
				two-dimensional optical lattices}},\ }\href
	{https://doi.org/10.1103/PhysRevA.102.043309} {\bibfield  {journal} {\bibinfo
			{journal} {Physical Review A}\ }\textbf {\bibinfo {volume} {102}},\ \bibinfo
		{pages} {43309} (\bibinfo {year} {2020})}\BibitemShut {NoStop}%
	\bibitem [{\citenamefont {Zhang}\ \emph {et~al.}(2022)\citenamefont {Zhang},
		\citenamefont {Feng},\ and\ \citenamefont {Yang}}]{Zhang2022quantumphases}%
	\BibitemOpen
	\bibfield  {author} {\bibinfo {author} {\bibfnamefont {D.-C.}\ \bibnamefont
			{Zhang}}, \bibinfo {author} {\bibfnamefont {S.-P.}\ \bibnamefont {Feng}},\
		and\ \bibinfo {author} {\bibfnamefont {S.-J.}\ \bibnamefont {Yang}},\
	}\bibfield  {title} {\bibinfo {title} {{Quantum phases of two-component
				bosons in the extended Bose-Hubbard model}},\ }\href
	{https://doi.org/10.1016/j.physleta.2021.127912} {\bibfield  {journal}
		{\bibinfo  {journal} {Physics Letters A}\ }\textbf {\bibinfo {volume}
			{427}},\ \bibinfo {pages} {127912} (\bibinfo {year} {2022})}\BibitemShut
	{NoStop}%
	\bibitem [{\citenamefont {Batrouni}\ and\ \citenamefont
		{Scalettar}(2000)}]{Batrouni2000_phaseseparation}%
	\BibitemOpen
	\bibfield  {author} {\bibinfo {author} {\bibfnamefont {G.~G.}\ \bibnamefont
			{Batrouni}}\ and\ \bibinfo {author} {\bibfnamefont {R.~T.}\ \bibnamefont
			{Scalettar}},\ }\bibfield  {title} {\bibinfo {title} {{Phase Separation in
				Supersolids}},\ }\href {https://doi.org/10.1103/physrevlett.84.1599}
	{\bibfield  {journal} {\bibinfo  {journal} {Physical Review Letters}\
		}\textbf {\bibinfo {volume} {84}},\ \bibinfo {pages} {1599} (\bibinfo {year}
		{2000})}\BibitemShut {NoStop}%
	\bibitem [{SI()}]{SI}%
	\BibitemOpen
	\href@noop {} {}\bibinfo {note} {See Supplemental Material}\BibitemShut
	{NoStop}%
	\bibitem [{\citenamefont {Kien}\ \emph {et~al.}(2013)\citenamefont {Kien},
		\citenamefont {Schneeweiss}, \citenamefont {Rauschenbeutel}, \citenamefont
		{{Le Kien}}, \citenamefont {Schneeweiss},\ and\ \citenamefont
		{Rauschenbeutel}}]{kien2013dynamical}%
	\BibitemOpen
	\bibfield  {author} {\bibinfo {author} {\bibfnamefont {F.~L.}\ \bibnamefont
			{Kien}}, \bibinfo {author} {\bibfnamefont {P.}~\bibnamefont {Schneeweiss}},
		\bibinfo {author} {\bibfnamefont {A.}~\bibnamefont {Rauschenbeutel}},
		\bibinfo {author} {\bibfnamefont {F.}~\bibnamefont {{Le Kien}}}, \bibinfo
		{author} {\bibfnamefont {P.}~\bibnamefont {Schneeweiss}},\ and\ \bibinfo
		{author} {\bibfnamefont {A.}~\bibnamefont {Rauschenbeutel}},\ }\bibfield
	{title} {\bibinfo {title} {{Dynamical polarizability of atoms in arbitrary
				light fields: general theory and application to cesium}},\ }\href
	{https://doi.org/10.1140/epjd/e2013-30729-x} {\bibfield  {journal} {\bibinfo
			{journal} {The European Physical Journal D}\ }\textbf {\bibinfo {volume}
			{67}},\ \bibinfo {pages} {92} (\bibinfo {year} {2013})}\BibitemShut {NoStop}%
	\bibitem [{\citenamefont {Cohen-Tannoudji}\ \emph {et~al.}(1998)\citenamefont
		{Cohen-Tannoudji}, \citenamefont {Dupont-Roc},\ and\ \citenamefont
		{Grynberg}}]{cohentannoudji1998atom}%
	\BibitemOpen
	\bibfield  {author} {\bibinfo {author} {\bibfnamefont {C.}~\bibnamefont
			{Cohen-Tannoudji}}, \bibinfo {author} {\bibfnamefont {J.}~\bibnamefont
			{Dupont-Roc}},\ and\ \bibinfo {author} {\bibfnamefont {G.}~\bibnamefont
			{Grynberg}},\ }\href
	{https://www.ebook.de/de/product/3861563/claude_cohen_tannoudji_jacques_dupont_roc_gilbert_grynberg_atom_photon_interactions_basic_processes_and_applications.html}
	{\emph {\bibinfo {title} {{Atom-Photon Interactions: Basic Processes and
					Applications}}}}\ (\bibinfo  {publisher} {VCH PUBN},\ \bibinfo {year}
	{1998})\BibitemShut {NoStop}%
	\bibitem [{\citenamefont {Zwerger}(2003)}]{Zwerger2003mott}%
	\BibitemOpen
	\bibfield  {author} {\bibinfo {author} {\bibfnamefont {W.}~\bibnamefont
			{Zwerger}},\ }\bibfield  {title} {\bibinfo {title} {{Mott-Hubbard transition
				of cold atoms in optical lattices}},\ }\href
	{https://doi.org/10.1088/1464-4266/5/2/352} {\bibfield  {journal} {\bibinfo
			{journal} {Journal of Optics B: Quantum and Semiclassical Optics}\ }\textbf
		{\bibinfo {volume} {5}},\ \bibinfo {pages} {S9} (\bibinfo {year}
		{2003})}\BibitemShut {NoStop}%
	\bibitem [{\citenamefont {Fisher}\ \emph {et~al.}(1989)\citenamefont {Fisher},
		\citenamefont {Weichman}, \citenamefont {Grinstein},\ and\ \citenamefont
		{Fisher}}]{fisher1989boson}%
	\BibitemOpen
	\bibfield  {author} {\bibinfo {author} {\bibfnamefont {M.~P.~A.}\
			\bibnamefont {Fisher}}, \bibinfo {author} {\bibfnamefont {P.~B.}\
			\bibnamefont {Weichman}}, \bibinfo {author} {\bibfnamefont {G.}~\bibnamefont
			{Grinstein}},\ and\ \bibinfo {author} {\bibfnamefont {D.~S.}\ \bibnamefont
			{Fisher}},\ }\bibfield  {title} {\bibinfo {title} {{Boson localization and
				the superfluid-insulator transition}},\ }\href
	{https://doi.org/10.1103/physrevb.40.546} {\bibfield  {journal} {\bibinfo
			{journal} {Physical Review B}\ }\textbf {\bibinfo {volume} {40}},\ \bibinfo
		{pages} {546} (\bibinfo {year} {1989})}\BibitemShut {NoStop}%
	\bibitem [{\citenamefont {Stamper-Kurn}\ and\ \citenamefont
		{Ueda}(2013)}]{stamperkurn2013spinorbosegases}%
	\BibitemOpen
	\bibfield  {author} {\bibinfo {author} {\bibfnamefont {D.~M.}\ \bibnamefont
			{Stamper-Kurn}}\ and\ \bibinfo {author} {\bibfnamefont {M.}~\bibnamefont
			{Ueda}},\ }\bibfield  {title} {\bibinfo {title} {{Spinor Bose gases:
				Symmetries, magnetism, and quantum dynamics}},\ }\href
	{https://doi.org/10.1103/revmodphys.85.1191} {\bibfield  {journal} {\bibinfo
			{journal} {Reviews of Modern Physics}\ }\textbf {\bibinfo {volume} {85}},\
		\bibinfo {pages} {1191} (\bibinfo {year} {2013})}\BibitemShut {NoStop}%
	\bibitem [{\citenamefont {Rokhsar}\ and\ \citenamefont
		{Kotliar}(1991)}]{rokhsar1991gutzwiller}%
	\BibitemOpen
	\bibfield  {author} {\bibinfo {author} {\bibfnamefont {D.~S.}\ \bibnamefont
			{Rokhsar}}\ and\ \bibinfo {author} {\bibfnamefont {B.~G.}\ \bibnamefont
			{Kotliar}},\ }\bibfield  {title} {\bibinfo {title} {{Gutzwiller projection
				for bosons}},\ }\href {https://doi.org/10.1103/physrevb.44.10328} {\bibfield
		{journal} {\bibinfo  {journal} {Physical Review B}\ }\textbf {\bibinfo
			{volume} {44}},\ \bibinfo {pages} {10328} (\bibinfo {year}
		{1991})}\BibitemShut {NoStop}%
	\bibitem [{\citenamefont {Altman}\ and\ \citenamefont
		{Auerbach}(2002)}]{altman2002oscillating}%
	\BibitemOpen
	\bibfield  {author} {\bibinfo {author} {\bibfnamefont {E.}~\bibnamefont
			{Altman}}\ and\ \bibinfo {author} {\bibfnamefont {A.}~\bibnamefont
			{Auerbach}},\ }\bibfield  {title} {\bibinfo {title} {{Oscillating
				Superfluidity of Bosons in Optical Lattices}},\ }\href
	{https://doi.org/10.1103/PhysRevLett.89.250404} {\bibfield  {journal}
		{\bibinfo  {journal} {Phys. Rev. Lett.}\ }\textbf {\bibinfo {volume} {89}},\
		\bibinfo {pages} {250404} (\bibinfo {year} {2002})}\BibitemShut {NoStop}%
	\bibitem [{\citenamefont {Huber}\ \emph {et~al.}(2007)\citenamefont {Huber},
		\citenamefont {Altman}, \citenamefont {B\"uchler},\ and\ \citenamefont
		{Blatter}}]{huber2007dynamical}%
	\BibitemOpen
	\bibfield  {author} {\bibinfo {author} {\bibfnamefont {S.~D.}\ \bibnamefont
			{Huber}}, \bibinfo {author} {\bibfnamefont {E.}~\bibnamefont {Altman}},
		\bibinfo {author} {\bibfnamefont {H.~P.}\ \bibnamefont {B\"uchler}},\ and\
		\bibinfo {author} {\bibfnamefont {G.}~\bibnamefont {Blatter}},\ }\bibfield
	{title} {\bibinfo {title} {Dynamical properties of ultracold bosons in an
			optical lattice},\ }\href {https://doi.org/10.1103/PhysRevB.75.085106}
	{\bibfield  {journal} {\bibinfo  {journal} {Phys. Rev. B}\ }\textbf {\bibinfo
			{volume} {75}},\ \bibinfo {pages} {085106} (\bibinfo {year}
		{2007})}\BibitemShut {NoStop}%
	\bibitem [{\citenamefont {van Oosten}\ \emph {et~al.}(2001)\citenamefont {van
			Oosten}, \citenamefont {van~der Straten},\ and\ \citenamefont
		{Stoof}}]{Oosten2001_quantum_phases}%
	\BibitemOpen
	\bibfield  {author} {\bibinfo {author} {\bibfnamefont {D.}~\bibnamefont {van
				Oosten}}, \bibinfo {author} {\bibfnamefont {P.}~\bibnamefont {van~der
				Straten}},\ and\ \bibinfo {author} {\bibfnamefont {H.~T.~C.}\ \bibnamefont
			{Stoof}},\ }\bibfield  {title} {\bibinfo {title} {{Quantum phases in an
				optical lattice}},\ }\href {https://doi.org/10.1103/PhysRevA.63.053601}
	{\bibfield  {journal} {\bibinfo  {journal} {Phys. Rev. A}\ }\textbf {\bibinfo
			{volume} {63}},\ \bibinfo {pages} {053601} (\bibinfo {year}
		{2001})}\BibitemShut {NoStop}%
	\bibitem [{\citenamefont {Chen}\ and\ \citenamefont {Wu}(2003)}]{Chen2003}%
	\BibitemOpen
	\bibfield  {author} {\bibinfo {author} {\bibfnamefont {G.-H.}\ \bibnamefont
			{Chen}}\ and\ \bibinfo {author} {\bibfnamefont {Y.-S.}\ \bibnamefont {Wu}},\
	}\bibfield  {title} {\bibinfo {title} {{Quantum phase transition in a
				multicomponent Bose-Einstein condensate in optical lattices}},\ }\href
	{https://doi.org/10.1103/PhysRevA.67.013606} {\bibfield  {journal} {\bibinfo
			{journal} {Phys. Rev. A}\ }\textbf {\bibinfo {volume} {67}},\ \bibinfo
		{pages} {013606} (\bibinfo {year} {2003})}\BibitemShut {NoStop}%
	\bibitem [{\citenamefont {Zhao}\ \emph {et~al.}(2014)\citenamefont {Zhao},
		\citenamefont {Hu}, \citenamefont {Chang}, \citenamefont {Zhang},\ and\
		\citenamefont {Wang}}]{Zhao2014_ferromagnetism}%
	\BibitemOpen
	\bibfield  {author} {\bibinfo {author} {\bibfnamefont {J.}~\bibnamefont
			{Zhao}}, \bibinfo {author} {\bibfnamefont {S.}~\bibnamefont {Hu}}, \bibinfo
		{author} {\bibfnamefont {J.}~\bibnamefont {Chang}}, \bibinfo {author}
		{\bibfnamefont {P.}~\bibnamefont {Zhang}},\ and\ \bibinfo {author}
		{\bibfnamefont {X.}~\bibnamefont {Wang}},\ }\bibfield  {title} {\bibinfo
		{title} {{Ferromagnetism in a two-component Bose-Hubbard model with synthetic
				spin-orbit coupling}},\ }\href {https://doi.org/10.1103/PhysRevA.89.043611}
	{\bibfield  {journal} {\bibinfo  {journal} {Phys. Rev. A}\ }\textbf {\bibinfo
			{volume} {89}},\ \bibinfo {pages} {043611} (\bibinfo {year}
		{2014})}\BibitemShut {NoStop}%
	\bibitem [{\citenamefont {Dhar}\ \emph {et~al.}(2011)\citenamefont {Dhar},
		\citenamefont {Singh}, \citenamefont {Pai},\ and\ \citenamefont
		{Das}}]{dhar2011mean}%
	\BibitemOpen
	\bibfield  {author} {\bibinfo {author} {\bibfnamefont {A.}~\bibnamefont
			{Dhar}}, \bibinfo {author} {\bibfnamefont {M.}~\bibnamefont {Singh}},
		\bibinfo {author} {\bibfnamefont {R.~V.}\ \bibnamefont {Pai}},\ and\ \bibinfo
		{author} {\bibfnamefont {B.~P.}\ \bibnamefont {Das}},\ }\bibfield  {title}
	{\bibinfo {title} {Mean-field analysis of quantum phase transitions in a
			periodic optical superlattice},\ }\href
	{https://doi.org/10.1103/PhysRevA.84.033631} {\bibfield  {journal} {\bibinfo
			{journal} {Phys. Rev. A}\ }\textbf {\bibinfo {volume} {84}},\ \bibinfo
		{pages} {033631} (\bibinfo {year} {2011})}\BibitemShut {NoStop}%
	\bibitem [{\citenamefont {Cheneau}\ \emph {et~al.}(2012)\citenamefont
		{Cheneau}, \citenamefont {Barmettler}, \citenamefont {Poletti}, \citenamefont
		{Endres}, \citenamefont {Schau{\ss}}, \citenamefont {Fukuhara}, \citenamefont
		{Gross}, \citenamefont {Bloch}, \citenamefont {Kollath},\ and\ \citenamefont
		{Kuhr}}]{cheneau2011light-cone}%
	\BibitemOpen
	\bibfield  {author} {\bibinfo {author} {\bibfnamefont {M.}~\bibnamefont
			{Cheneau}}, \bibinfo {author} {\bibfnamefont {P.}~\bibnamefont {Barmettler}},
		\bibinfo {author} {\bibfnamefont {D.}~\bibnamefont {Poletti}}, \bibinfo
		{author} {\bibfnamefont {M.}~\bibnamefont {Endres}}, \bibinfo {author}
		{\bibfnamefont {P.}~\bibnamefont {Schau{\ss}}}, \bibinfo {author}
		{\bibfnamefont {T.}~\bibnamefont {Fukuhara}}, \bibinfo {author}
		{\bibfnamefont {C.}~\bibnamefont {Gross}}, \bibinfo {author} {\bibfnamefont
			{I.}~\bibnamefont {Bloch}}, \bibinfo {author} {\bibfnamefont
			{C.}~\bibnamefont {Kollath}},\ and\ \bibinfo {author} {\bibfnamefont
			{S.}~\bibnamefont {Kuhr}},\ }\bibfield  {title} {\bibinfo {title}
		{{Light-cone-like spreading of correlations in a quantum many-body system}},\
	}\href {https://doi.org/10.1038/nature10748} {\bibfield  {journal} {\bibinfo
			{journal} {Nature}\ }\textbf {\bibinfo {volume} {481}},\ \bibinfo {pages}
		{484} (\bibinfo {year} {2012})}\BibitemShut {NoStop}%
	\bibitem [{\citenamefont {Hauke}\ and\ \citenamefont
		{Tagliacozzo}(2013)}]{hauke2013spread}%
	\BibitemOpen
	\bibfield  {author} {\bibinfo {author} {\bibfnamefont {P.}~\bibnamefont
			{Hauke}}\ and\ \bibinfo {author} {\bibfnamefont {L.}~\bibnamefont
			{Tagliacozzo}},\ }\bibfield  {title} {\bibinfo {title} {{Spread of
				Correlations in Long-Range Interacting Quantum Systems}},\ }\href
	{https://doi.org/10.1103/PhysRevLett.111.207202} {\bibfield  {journal}
		{\bibinfo  {journal} {Phys. Rev. Lett.}\ }\textbf {\bibinfo {volume} {111}},\
		\bibinfo {pages} {207202} (\bibinfo {year} {2013})}\BibitemShut {NoStop}%
	\bibitem [{\citenamefont {Dogra}\ \emph {et~al.}(2019)\citenamefont {Dogra},
		\citenamefont {Landini}, \citenamefont {Kroeger}, \citenamefont {Hruby},
		\citenamefont {Donner},\ and\ \citenamefont
		{Esslinger}}]{dogra2019dissipation}%
	\BibitemOpen
	\bibfield  {author} {\bibinfo {author} {\bibfnamefont {N.}~\bibnamefont
			{Dogra}}, \bibinfo {author} {\bibfnamefont {M.}~\bibnamefont {Landini}},
		\bibinfo {author} {\bibfnamefont {K.}~\bibnamefont {Kroeger}}, \bibinfo
		{author} {\bibfnamefont {L.}~\bibnamefont {Hruby}}, \bibinfo {author}
		{\bibfnamefont {T.}~\bibnamefont {Donner}},\ and\ \bibinfo {author}
		{\bibfnamefont {T.}~\bibnamefont {Esslinger}},\ }\bibfield  {title} {\bibinfo
		{title} {Dissipation-induced structural instability and chiral dynamics in a
			quantum gas},\ }\href {https://doi.org/10.1126/science.aaw4465} {\bibfield
		{journal} {\bibinfo  {journal} {Science}\ }\textbf {\bibinfo {volume}
			{366}},\ \bibinfo {pages} {1496} (\bibinfo {year} {2019})}\BibitemShut
	{NoStop}%
	\bibitem [{\citenamefont {Chiacchio}\ and\ \citenamefont
		{Nunnenkamp}(2019)}]{chiacchio2019dissipation}%
	\BibitemOpen
	\bibfield  {author} {\bibinfo {author} {\bibfnamefont {E.~I.~R.}\
			\bibnamefont {Chiacchio}}\ and\ \bibinfo {author} {\bibfnamefont
			{A.}~\bibnamefont {Nunnenkamp}},\ }\bibfield  {title} {\bibinfo {title}
		{{Dissipation-Induced Instabilities of a Spinor Bose-Einstein Condensate
				Inside an Optical Cavity}},\ }\href
	{https://doi.org/10.1103/PhysRevLett.122.193605} {\bibfield  {journal}
		{\bibinfo  {journal} {Phys. Rev. Lett.}\ }\textbf {\bibinfo {volume} {122}},\
		\bibinfo {pages} {193605} (\bibinfo {year} {2019})}\BibitemShut {NoStop}%
	\bibitem [{\citenamefont {Bu\ifmmode~\check{c}\else \v{c}\fi{}a}\ and\
		\citenamefont {Jaksch}(2019)}]{buca2019dissipation}%
	\BibitemOpen
	\bibfield  {author} {\bibinfo {author} {\bibfnamefont {B.}~\bibnamefont
			{Bu\ifmmode~\check{c}\else \v{c}\fi{}a}}\ and\ \bibinfo {author}
		{\bibfnamefont {D.}~\bibnamefont {Jaksch}},\ }\bibfield  {title} {\bibinfo
		{title} {{Dissipation Induced Nonstationarity in a Quantum Gas}},\ }\href
	{https://doi.org/10.1103/PhysRevLett.123.260401} {\bibfield  {journal}
		{\bibinfo  {journal} {Phys. Rev. Lett.}\ }\textbf {\bibinfo {volume} {123}},\
		\bibinfo {pages} {260401} (\bibinfo {year} {2019})}\BibitemShut {NoStop}%
	\bibitem [{\citenamefont {Rosa-Medina}\ \emph {et~al.}(2022)\citenamefont
		{Rosa-Medina}, \citenamefont {Ferri}, \citenamefont {Finger}, \citenamefont
		{Dogra}, \citenamefont {Kroeger}, \citenamefont {Lin}, \citenamefont
		{Chitra}, \citenamefont {Donner},\ and\ \citenamefont
		{Esslinger}}]{Rosa-Medina2022}%
	\BibitemOpen
	\bibfield  {author} {\bibinfo {author} {\bibfnamefont {R.}~\bibnamefont
			{Rosa-Medina}}, \bibinfo {author} {\bibfnamefont {F.}~\bibnamefont {Ferri}},
		\bibinfo {author} {\bibfnamefont {F.}~\bibnamefont {Finger}}, \bibinfo
		{author} {\bibfnamefont {N.}~\bibnamefont {Dogra}}, \bibinfo {author}
		{\bibfnamefont {K.}~\bibnamefont {Kroeger}}, \bibinfo {author} {\bibfnamefont
			{R.}~\bibnamefont {Lin}}, \bibinfo {author} {\bibfnamefont {R.}~\bibnamefont
			{Chitra}}, \bibinfo {author} {\bibfnamefont {T.}~\bibnamefont {Donner}},\
		and\ \bibinfo {author} {\bibfnamefont {T.}~\bibnamefont {Esslinger}},\
	}\bibfield  {title} {\bibinfo {title} {{Observing Dynamical Currents in a
				Non-Hermitian Momentum Lattice}},\ }\href
	{https://doi.org/10.1103/PhysRevLett.128.143602} {\bibfield  {journal}
		{\bibinfo  {journal} {Phys. Rev. Lett.}\ }\textbf {\bibinfo {volume} {128}},\
		\bibinfo {pages} {143602} (\bibinfo {year} {2022})}\BibitemShut {NoStop}%
	\bibitem [{\citenamefont {Goldman}\ \emph {et~al.}(2014)\citenamefont
		{Goldman}, \citenamefont {Juzeli{\={u}}nas}, \citenamefont {Öhberg},\ and\
		\citenamefont {Spielman}}]{Goldman_2014}%
	\BibitemOpen
	\bibfield  {author} {\bibinfo {author} {\bibfnamefont {N.}~\bibnamefont
			{Goldman}}, \bibinfo {author} {\bibfnamefont {G.}~\bibnamefont
			{Juzeli{\={u}}nas}}, \bibinfo {author} {\bibfnamefont {P.}~\bibnamefont
			{Öhberg}},\ and\ \bibinfo {author} {\bibfnamefont {I.~B.}\ \bibnamefont
			{Spielman}},\ }\bibfield  {title} {\bibinfo {title} {Light-induced gauge
			fields for ultracold atoms},\ }\href
	{https://doi.org/10.1088/0034-4885/77/12/126401} {\bibfield  {journal}
		{\bibinfo  {journal} {Reports on Progress in Physics}\ }\textbf {\bibinfo
			{volume} {77}},\ \bibinfo {pages} {126401} (\bibinfo {year}
		{2014})}\BibitemShut {NoStop}%
	\bibitem [{\citenamefont {Dogra}(2019)}]{dogra2019phd}%
	\BibitemOpen
	\bibfield  {author} {\bibinfo {author} {\bibfnamefont {N.}~\bibnamefont
			{Dogra}},\ }\emph {\bibinfo {title} {Interaction- and dissipation-induced
			phenomena in a quantum gas coupled to a cavity}},\ \href
	{https://doi.org/10.3929/ETHZ-B-000383084} {Ph.D. thesis},\ \bibinfo
	{school} {ETH Zurich} (\bibinfo {year} {2019})\BibitemShut {NoStop}%
	\bibitem [{\citenamefont {Sheshadri}\ \emph {et~al.}(1993)\citenamefont
		{Sheshadri}, \citenamefont {Krishnamurthy}, \citenamefont {Pandit},\ and\
		\citenamefont {Ramakrishnan}}]{Sheshadri1993superfluid}%
	\BibitemOpen
	\bibfield  {author} {\bibinfo {author} {\bibfnamefont {K.}~\bibnamefont
			{Sheshadri}}, \bibinfo {author} {\bibfnamefont {H.~R.}\ \bibnamefont
			{Krishnamurthy}}, \bibinfo {author} {\bibfnamefont {R.}~\bibnamefont
			{Pandit}},\ and\ \bibinfo {author} {\bibfnamefont {T.~V.}\ \bibnamefont
			{Ramakrishnan}},\ }\bibfield  {title} {\bibinfo {title} {Superfluid and
			insulating phases in an interacting-boson model: Mean-field theory and the
			{RPA}},\ }\href {https://doi.org/10.1209/0295-5075/22/4/004} {\bibfield
		{journal} {\bibinfo  {journal} {Europhysics Letters ({EPL})}\ }\textbf
		{\bibinfo {volume} {22}},\ \bibinfo {pages} {257} (\bibinfo {year}
		{1993})}\BibitemShut {NoStop}%
\end{thebibliography}
\end{document}